\documentclass[10pt,a4paper,final]{iopart}

\usepackage{iopams}
\usepackage{stackrel}
\usepackage{fullpage}
\usepackage{graphicx}
\usepackage[breaklinks=true,colorlinks=true,linkcolor=blue,urlcolor=blue,citecolor=blue]{hyperref}
\usepackage[capitalize,nameinlink,noabbrev]{cleveref}
\usepackage[utf8]{inputenc}
\usepackage[english]{babel}
\usepackage{enumitem}
\usepackage[toc,page]{appendix}
\usepackage{bm}
\usepackage{array}
\usepackage{mdframed}
\usepackage{braket}
\renewcommand\bra[1]{{\langle{#1}|}}
\renewcommand\ket[1]{{|{#1}\rangle}}

\usepackage{comment}
\usepackage{multirow}
\usepackage[format=hang,font=small,labelfont=bf,textfont=sf,figurename=Fig.]{caption}[2013/02/03]
\usepackage[margin=0.2cm]{subcaption}
\usepackage[usenames,dvipsnames,svgnames,table]{xcolor}
\usepackage{breqn}
\usepackage{cite}

\eqnobysec

\begin{document}
	\renewcommand{\equationautorefname}{Eq.} 
	\renewcommand{\figureautorefname}{figure} 
	\renewcommand{\chapterautorefname}{Ch.} 
	\renewcommand{\sectionautorefname}{Sec.} 
	\renewcommand{\subsectionautorefname}{Subsec.} 
	\newcommand{\Mod}[1]{\ (\mathrm{mod}\ #1)}

\title{Practical decoy-state method for twin-field quantum key distribution}

\author{Federico Grasselli$^1$, Marcos Curty$^2$}
\address{$^1$Institut für Theoretische Physik III, Heinrich-Heine-Universität Düsseldorf, Universitätsstraße 1, D-40225 Düsseldorf, Germany \\ $^2$Escuela de Ingenier\'ia de Telecomunicaci\'on, Dept. of Signal Theory and Communications, University of Vigo, E-36310 Vigo, Spain}
\ead{federico.grasselli@hhu.de} 
\begin{abstract}
Twin-Field (TF) quantum key distribution (QKD) represents a novel QKD approach whose principal merit is to beat the point-to-point private capacity of a lossy quantum channel, thanks to performing single-photon interference in an untrusted node. Indeed, recent security proofs of various TF-QKD type protocols have confirmed that the secret key rate of these schemes scales essentially as the square root of the transmittance of the channel. Here, we focus on the TF-QKD protocol introduced by Curty \textit{et al.}, whose secret key rate is nearly an order of magnitude higher than previous solutions. Its security relies on the estimation of the detection probabilities associated to various photon-number states through the decoy-state method. We derive analytical bounds on these quantities assuming that each party uses either two, three or four decoy intensity settings, and we investigate the protocol’s performance in this scenario. Our simulations show that two decoy intensity settings are enough to beat the point-to-point private capacity of the channel, and that the use of four decoys is already basically optimal, in the sense that it almost reproduces the ideal scenario of infinite decoys. We also observe that the protocol seems to be quite robust against intensity fluctuations of the optical pulses prepared by the parties.
\end{abstract}


The last few decades have witnessed major advancements in the field of quantum communication \cite{quantum-communication-review,Kimble}, with quantum key distribution (QKD) \cite{BB84,E91,Scarani-review,Curty-MDIQKD,DIQKD1,Curty-review,Diamanti-review,Epping,DIQKD2,Ribeiro,Grasselli} being its most developed application. Recent experiments over about 400 km of optical fibers \cite{404km,421km} and over about 1000 km of satellite-to-ground links \cite{free-space-QKD1,free-space-QKD2} demonstrated that QKD over long distances is possible. Despite such remarkable experimental achievements, the private capacity of point-to-point QKD is intrinsically limited by fundamental bounds \cite{Takeoka, PLOB}.
These bounds state that in the high-loss regime the key rate scales basically linearly with the transmittance of the channel connecting the end-users Alice and Bob, i.e. it decreases exponentially with the total channel length. This imposes strict practical constraints on the possibility of achieving point-to-point QKD over arbitrary long distances.\\
A way to overcome this limitation is to employ one or more intermediate nodes in the quantum channel connecting the parties. For instance, the use of quantum repeaters \cite{review-quantum-repeaters} yields a polynomial scaling of the communication efficiency with the distance \cite{DLCZ}. Moreover, a quantum repeater scheme can be arbitrarily iterated along the quantum channel, thus increasing in principle the total communication distance between Alice and Bob as much as desired. Unfortunately, however, quantum repeaters are very challenging to build in practice with current technology: they either require quantum memories \cite{review-quantum-repeaters, DLCZ, Grudka-quantumrep} or quantum error correction \cite{No-quantum-memories, all-photonic-repeaters}. Of course, technology is improving, and quantum repeaters may become viable in the future.\\
Other solutions, which attain a square-root improvement in the scaling of the key rate with respect to the transmittance of the channel, are obtained by placing a single untrusted relay between Alice and Bob. Such protocols include, for instance, Measurement-Device-Independent-QKD \cite{Curty-MDIQKD} (MDI-QKD) with quantum memories \cite{Abruzzo-MDIQKD, mem-assisited-MDIQKD} and adaptive MDI-QKD featuring quantum non-demolition measurements \cite{Azuma-intercityQKD}. The philosophy behind both types of protocols is that the central relay is able to adapt the pairings of photons received from Alice and Bob to the photon losses. In this way, for every signal sent by Alice and Bob to the central relay, just one of the two signals is required to arrive, leading to the mentioned square-root improvement in the key rate scaling. However, both protocols still require two-photon interference in the central node, as in the original MDI-QKD scheme \cite{Curty-MDIQKD}. More recently, \cite{Lucamarini-TF} proposed the Twin-Field (TF) QKD protocol, still characterized by an untrusted central node, and conjectured a square-root improvement in the key rate scaling. This scaling has been later on confirmed in~\cite{Tamaki-security-TF,Ma-security-TF} for two variants of the original scheme. The advantage of TF-QKD lies in the fact that it is designed to generate key bits from single-photon interference in the central node, thus naturally retaining the scaling with the square-root of the transmittance without the need to adapt to photon losses via sophisticated devices.\\
Since the original proposal, there has been an intense research activity to develop different versions of TF-QKD protocols equipped with their security proofs
\cite{Tamaki-security-TF, Ma-security-TF, Cui-security-TF, Lutkenhaus-security-TF, Curty-security-TF} as well as to investigate their experimental feasibility~\cite{experiment-Toshiba,experiment-chinese,experiment-Toronto}. Among these protocols, the one that seems to deliver the higher secret ket rate \cite{comparison-TF} is that introduced in \cite{Curty-security-TF}. Its security relies on the ability to estimate the detection statistics (usually called yields) of various Fock states sent by Alice and Bob through the decoy-state method \cite{decoy-state-method-Hwang, decoy-state-method-Lo, decoy-state-method-Wang}. The key-rate simulations provided in \cite{Curty-security-TF} indeed exhibit an improved scaling with the loss, but the estimation of the yields is only carried out by means of {\it numerical} tools based on linear programming and considering only the case of three decoy intensity settings.\\
In this paper, we derive {\it analytical} bounds on the yields which are required to evaluate the key rate formula of \cite{Curty-security-TF}, assuming two, three and four decoy intensity settings. In so doing, we are able to show, for instance, that the use of two decoy intensity settings is already enough to beat the point-to-point private capacity bound reported in \cite{PLOB}. Also, we show that the use of four decoys is basically optimal in the sense that the resulting secret key rate is already very close to the ideal scenario which assumes infinite decoy intensity settings. Analytical bounds imply a fully-analytical expression for the protocol's secret key rate, which could be very convenient for performance optimization in scenarios where the number of parameters is high, like for instance in finite-key security analyses. In addition, we study how the performance of TF-QKD is affected under intensity fluctuations, which are inevitable in practice, and we demonstrate that the protocol in \cite{Curty-security-TF} seems to be actually quite robust against such fluctuations.\\

Like in \cite{Curty-security-TF}, for simplicity, here we focus on the asymptotic-key rate scenario. However, we remark that by using the techniques reported in \cite{Curty-finitekey}, it is cumbersome but straightforward to adapt our analytical methods also to the finite-key rate scenario, where, as mentioned above, it becomes particularly useful to have analytical bounds for the main quantities that enter the key rate formula.\\
The article is structured as follows. In \autoref{TF-protocol} we present the TF protocol from \cite{Curty-security-TF} and highlight the main yields that need to be bounded. In \autoref{yields-bounds-2decoys} we provide the analytical bounds on the yields for the case of two decoys (the cases of three and four decoys are treated in \ref{yields-bounds-3decoys} and \ref{yields-bounds-4decoys}, respectively). In \autoref{simulations} we provide simulations of the secret key rate versus the loss for a typical channel model (briefly described in \ref{channel-model}), and we also evaluate the effect of intensity fluctuations. We conclude the paper in \autoref{conclusions}.

\section{The TF-QKD protocol} \label{TF-protocol}  
\begin{figure}[!htb]
	\centering
	\includegraphics[width=0.8\linewidth,keepaspectratio]{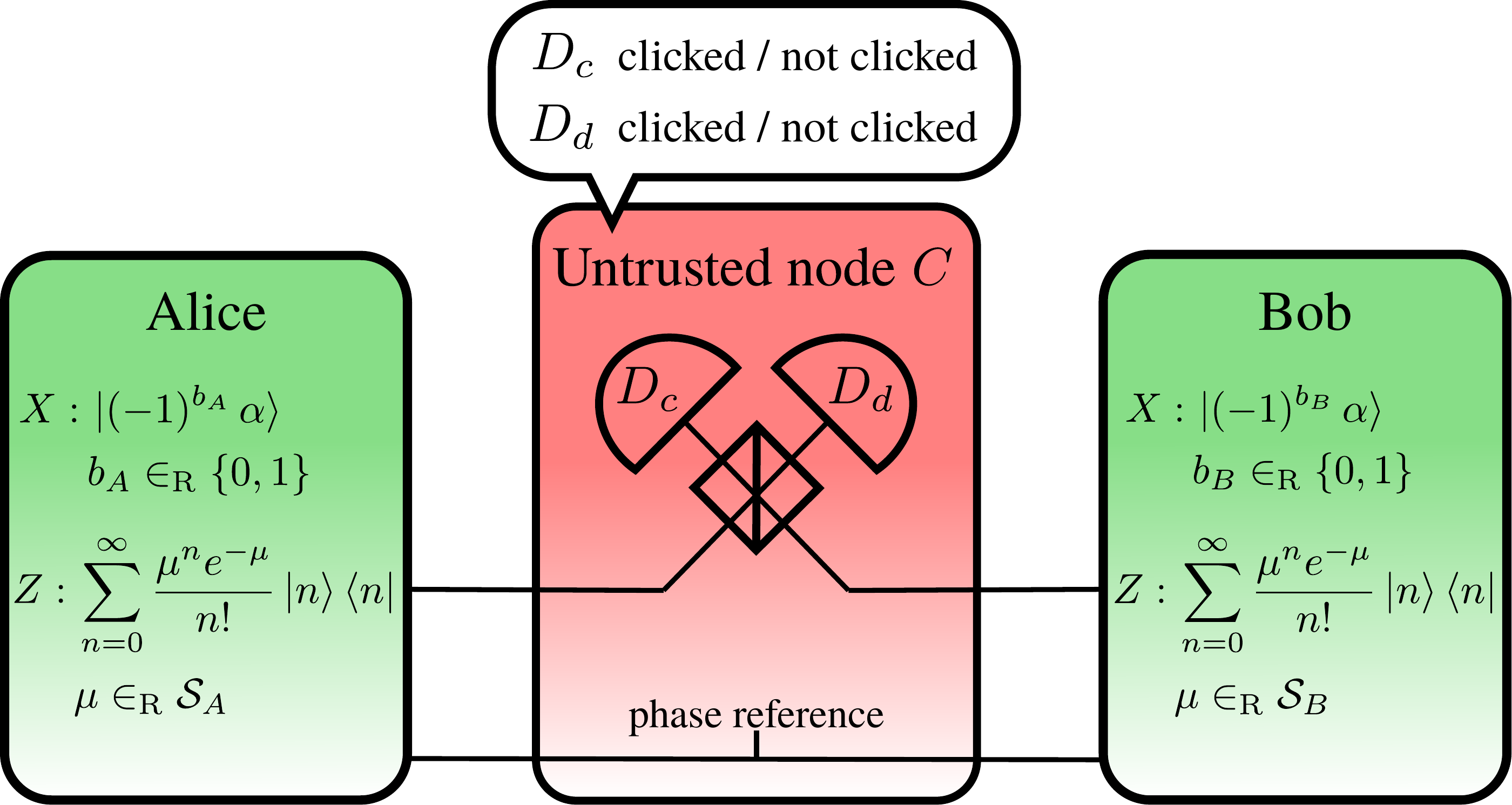}
	\caption{The Twin-Field QKD protocol introduced in \cite{Curty-security-TF}.}
	\label{TF-scheme}
\end{figure}
As discussed above, we consider the TF-QKD protocol presented in \cite{Curty-security-TF} and sketched in \autoref{TF-scheme}. Alice and Bob establish a secret shared key by sending optical pulses to a central untrusted node, $C$. It is assumed that the node $C$ shares a phase reference with Alice and Bob, which can be achieved by the transmission of strong optical pulses. The protocol is composed of the following five steps:
\begin{enumerate}
\item Alice (Bob) chooses the $X$-basis with probability $p_X$ and the $Z$-basis with probability $p_Z=1-p_X$. Upon choosing the $X$-basis, Alice (Bob) prepares an optical pulse in a coherent state $\ket{\alpha}$ or $\ket{-\alpha}$ at random, corresponding to the key bit $b_A=0$ ($b_B=0$) or $b_A=1$ ($b_B=1$), respectively. Upon choosing the $Z$-basis, she (he) prepares an optical pulse in a phase-randomized coherent state:
\begin{equation}
	\hat{\rho}_{\beta_A}=\frac{1}{2\pi} \int_{0}^{2\pi} d\theta \ket{\beta_A e^{i\theta}}\bra{\beta_A e^{i\theta}}= \sum_{n=0}^{\infty} \frac{(\beta_A^2)^n e^{-\beta_A^2}}{n!} \ket{n}\bra{n}  \label{phase-randomized-coherent-state}
\end{equation}
($\hat{\rho}_{\beta_B}$) whose intensity $\beta_A^2$ ($\beta^2_B$) is drawn randomly from a set $\mathcal{S}_A=\{\beta^2_i\}_i$ ($\mathcal{S}_B=\{\beta^2_j\}_j$) of real nonnegative numbers.
\item Both parties send their optical pulses to the untrusted node $C$ via optical channels in a synchronized manner.
\item The central node $C$ applies a balanced beamsplitter to the incoming pulses and features two threshold detectors at its output ports. The detector placed at the output port associated to constructive (destructive) interference is denoted by $D_c$ ($D_d$).
\item The node $C$ announces the measurement outcome $k_c$ ($k_d$) of detector $D_c$ ($D_d$), with $k_c=0$ and $k_c=1$ ($k_d=0$ and $k_d=1$) corresponding to a no-click and a click event, respectively.
\item Alice and Bob form their raw keys with the bits $b_A$ and $b_B$ collected when both parties chose the $X$-basis and node $C$ reported a click in only one detector ($k_c+k_d =1$). Bob flips his bits $b_B$ for which the click occurred in $D_d$.
\end{enumerate}

\subsection{Secret key rate formula}
The security analysis performed in \cite{Curty-security-TF} yields the following lower bound on the asymptotic key rate $R$:
\begin{equation}
 R \geq \max \{R_{10},0 \} + \max \{R_{01},0 \}  \label{key-rate} \,\,,
\end{equation}
where the terms $R_{k_c k_d}$, for $(k_c,k_d) \in \{ (1,0), (0,1)\}$, are defined as:
\begin{equation}
 R_{k_c k_d} = p_X^2 \,p(k_c,k_d) \left[1-f \,h(e_{k_c k_d}) -h(e^{\mathrm{ph}}_{k_c k_d}) \right]  \label{R_kckd} \,\,,
\end{equation}
with $h(x)=-x \log_2 x  -(1-x) \log_2 (1-x)$ being the binary entropy function, $f$ the inefficiency function associated to error correction, and $p(k_c,k_d)$ the conditional probability that node $C$ announces the outcome $(k_c,k_d)$ when both parties selected the $X$-basis. The probability $p(k_c,k_d)$ can be expressed as:
\begin{equation}
 p(k_c,k_d)= \sum_{b_A,b_B=0}^{1} p(b_A,b_B) p(k_c,k_d|b_A,b_B) \label{prob-kckd} \,\,,
\end{equation}
where $p(b_A,b_B)$ is the joint probability of Alice and Bob preparing the coherent states $\ket{(-1)^{b_A} \alpha}$ and $\ket{(-1)^{b_B} \alpha}$, respectively. According to the protocol description above, we have: $p(b_A,b_B)=1/4 \,\,\forall\, b_A,b_B$. $p(k_c,k_d|b_A,b_B)$ instead denotes the conditional probability that node C announced $(k_c,k_d)$ given that Alice and Bob sent the coherent states $\ket{(-1)^{b_A} \alpha}$ and $\ket{(-1)^{b_B} \alpha}$, respectively. Since we consider the asymptotic key-rate scenario, we assume that $p(k_c,k_d|b_A,b_B)$ coincides with the correspondent distribution observed by the parties.\\
Finally, the terms $e_{k_c k_d}$ and $e^{\mathrm{ph}}_{k_c k_d}$ in (\ref{R_kckd}) represent the bit-error rate in the $X$-basis and an upper bound on the phase-error rate, respectively. The former is defined as:
\begin{eqnarray}
 e_{10} &= \frac{\sum_{i,j=0 | i\oplus j=1}^{1} p(b_A=i,b_B=j)\, p(k_c=1,k_d=0|b_A=i,b_B=j)}{p(k_c=1,k_d=0)}  \,\,,\label{e10} \\
 e_{01} &= \frac{\sum_{i=0}^{1} p(b_A=i,b_B=i)\, p(k_c=0,k_d=1|b_A=i,b_B=i)}{p(k_c=0,k_d=1)}  \label{e01} \,\,,
\end{eqnarray}
and the latter as:
\begin{eqnarray}
\fl e^{\mathrm{ph}}_{k_c k_d} = \frac{1}{p(k_c,k_d)} \left[\left(\sum_{n,m=0}^{\infty} c_{2 n} c_{2 m} \sqrt{Y^{k_c,k_d}_{2n\,2m}}\right)^2 + \left(\sum_{n,m=0}^{\infty} c_{2 n+1} c_{2 m+1} \sqrt{Y^{k_c,k_d}_{2n+1\,2m+1}}\right)^2 \right]   \label{phase-error-rate} \,\,,
\end{eqnarray}
where the coefficients $c_n$ are defined as $c_n=e^{\frac{-\alpha^2}{2}} \alpha^n/\sqrt{n!}$ and the yields $Y^{k_c,k_d}_{nm}$ are the conditional probabilities that node $C$ announces the outcome $(k_c,k_d)$ given that Alice and Bob emitted an $n$-photon state and an $m$-photon state, respectively. Note that the only yields contributing to (\ref{phase-error-rate}) are those $Y^{k_c,k_d}_{nm}$ such that $n+m$ is an even number.\\
The yields $Y^{k_c,k_d}_{nm}$ are quantities that are not directly observed by the parties, however they can be estimated either numerically or analytically with techniques based on the decoy-state method \cite{decoy-state-method-Hwang,decoy-state-method-Lo,decoy-state-method-Wang}. Here we consider the analytical approach. In particular, we assume that Alice and Bob have at their disposal either two, three or four decoy intensity settings when choosing the $Z$-basis.
To each further decoy intensity correspond additional linear constraints on the yields, leading to tighter estimations of $Y^{k_c,k_d}_{nm}$ and thus to a higher key rate. However, a finite number of decoys only allows to derive non-trivial upper bounds\footnote{Every yield is a probability, thus it is trivially bounded by 1.} on a limited number of yields in (\ref{phase-error-rate}), whereas the other yields are set to 1. Nevertheless, even bounding just four yields in a non-trivial way is enough for the secret key rate to beat the point-to-point private capacity bound (PLOB bound) \cite{PLOB} at high losses (see \autoref{simulations}). Also, as we show below, with four decoy intensity settings one can already obtain a secret key rate very close to that achievable with infinite decoy intensity settings.\\
We remark that standard decoy-state-based QKD protocols require to \textit{lower} bound the value of a few yields (typically those associated to vacuum and single-photon pulses) \cite{decoy-state-method-Ma}, while the TF-QKD protocol considered here upper bounds the value of the phase-error rate (\ref{phase-error-rate}) by \textit{upper} bounding several yields. In particular, we upper bound the yields $Y^{k_c,k_d}_{nm}$ for $(n,m)\in \mathcal{I}$, where $\mathcal{I}$ is a certain subset of $\{(n,m)| \,n,m \in \mathbb{N}_0\}$ which depends on the number of decoys. Thanks to the derived upper bounds on the yields (which we shall denote by $Y^{U,\, k_c,k_d}_{nm}$) we are able to estimate the phase error rate (\ref{phase-error-rate}) as follows:
\begin{eqnarray}
	\fl e^{\mathrm{ph}}_{k_c k_d} \leq \frac{1}{p(k_c,k_d)} &\left[\left(\sum_{(2n,2m)\in\mathcal{I}} c_{2 n} c_{2 m} \sqrt{Y^{U,\, k_c,k_d}_{2n\,2m}} +\sum_{(2n,2m)\notin\mathcal{I}} c_{2 n} c_{2 m} \right)^2 \right. \nonumber\\
	 \fl &\left. + \left(\sum_{(2n+1,2m+1)\in\mathcal{I}} c_{2 n+1} c_{2 m+1} \sqrt{Y^{U,\, k_c,k_d}_{2n+1\,2m+1}} +\sum_{(2n+1,2m+1)\notin\mathcal{I}} c_{2 n+1} c_{2 m+1} \right)^2 \right]   \label{phase-error-rate-bound} \,\,.
\end{eqnarray}

\section{Yields estimation} \label{yields-bounds-2decoys}
When both Alice and Bob choose the $Z$-basis in the first step of the TF-QKD protocol, they prepare phase-randomized coherent states with intensities $\beta^2_A$ and $\beta^2_B$, respectively, and send them to $C$. From Eve's viewpoint, she cannot distinguish this scenario from the case in which the parties prepared number states $\ket{n}$ and $\ket{m}$ according to the Poissonian distributions $P_{\beta_A^2} (n)$ and $P_{\beta_B^2} (m)$ (see \autoref{phase-randomized-coherent-state}), where $P_{\mu}(n)=e^{-\mu}\mu^n/n!$. Therefore Eve's attack can only depend on the number states $\ket{n}$ and $\ket{m}$ but not on the signals' intensities $\beta_A^2$ and $\beta_B^2$. As a consequence, the probability that Eve announces outcomes $(k_c,k_d)$ only depends on the number of photons $(n,m)$ she received from Alice and Bob, i.e. the yields $Y^{k_c,k_d}_{n m}$ are independent of the decoy intensities chosen by the parties.\\
For this reason, one can derive a set of linear constraints on the yields $Y^{k_c,k_d}_{n m}$ by expressing the experimentally observed gains $Q^{\beta_A^2,\beta_B^2}_{k_c,k_d}$ --which are defined as the conditional probabilities that node $C$ announced the outcome $(k_c,k_d)$ given that Alice and Bob sent phase-randomized coherent states of intensities  $\beta_A^2$ and $\beta_B^2$, respectively-- in terms of the yields:
\begin{equation}
 Q^{\beta_A^2,\beta_B^2}_{k_c,k_d} = \sum_{n,m=0}^{\infty} e^{-\beta_A^2-\beta_B^2} \frac{(\beta_A^2)^n(\beta_B^2)^m}{n!m!} Y^{k_c,k_d}_{n m}  \label{gains} \,\,.
\end{equation}
As it is clear from (\ref{gains}), to every distinct pair of decoy intensities $(\beta_A^2,\beta_B^2)$ corresponds a new constraint on the set of infinite yields $\{Y^{k_c,k_d}_{n m}\}_{n,m}$, which leads to tighter upper bounds and thus to a higher secret key rate. On the other hand, having a large number of decoy intensities is experimentally demanding, hence the need to derive the tightest possible bounds on the yields with a limited number of decoys.\bigskip\\
In this Section we present a simple analytical method to obtain tight bounds on the yields of largest contribution\footnote{The same method can --in principle-- be applied to any yield, however the limited number of decoy settings prevents from obtaining a non-trivial bound on every yield.} in (\ref{phase-error-rate}) --i.e. relative to the largest coefficients $c_n$-- when the parties use two intensity settings in the $Z$-basis. It is basically a Gaussian elimination-type technique but involving infinite-size coefficient matrices. In particular, the guiding principle that we use is to combine the constraints (\ref{gains}) so that in the resulting expression the yield to be bounded is the one with the largest coefficient, while the yields which had larger coefficients in the initial constraints have been removed in the combination. However, in some cases it turns out that is not possible to remove all the yields with larger coefficients than the one to be bounded, due to a lack of decoy intensity settings (i.e. constraints). In other cases, we manage to remove from the resulting expression even some yields which had a smaller coefficient than the one to be bounded. Such a procedure can be readily extended to the case of three and four decoy intensity settings. The results for these last two cases are presented in \ref{yields-bounds-3decoys} and \ref{yields-bounds-4decoys}, respectively.\\
From now on, we assume that both optical channels linking the parties to the central node $C$ have the same transmittance $\sqrt{\eta}$. Therefore the set of optimal decoy intensities $\beta_A^2$ and $\beta_B^2$ is the same for both parties \cite{asymmetric-MDI-QKD} and we define it as: $\{\mu_0,\mu_1\}$.
In order to simplify the notation, we also omit the measurement outcome $(k_c,k_d)$ from the constraints given by (\ref{gains}). Hence the yields are subjected to the following four equality constraints:
\begin{equation}
	\tilde{Q}^{k,l} \equiv e^{\mu_k + \mu_l} Q^{k,l} =\sum_{n,m=0}^{\infty} \frac{Y_{nm}}{n!m!} {\mu_k}^n {\mu_l}^m \quad k,l \in \{0,1\} \,\,, \label{constr-2decoys}
\end{equation}
and to the inequality constraints:
\begin{equation}
	 0 \leq Y_{nm} \leq 1 \quad \forall\, n,m  \label{ineq-constr} \,\,.
\end{equation}
Below we derive upper bounds on the yields: $Y_{00},Y_{11},Y_{02}$ and $Y_{20}$.

\subsection{Upper bound on $Y_{11}$}
Consider the following combination of gains:
\begin{eqnarray}
	G_{11} &= \tilde{Q}^{0,0} + \tilde{Q}^{1,1} - (\tilde{Q}^{0,1} +\tilde{Q}^{1,0})  \nonumber \\
	&= \sum_{n,m=0}^{\infty} \frac{Y_{nm}}{n!m!} \left(\mu^n_0 -\mu^n_1\right) \left(\mu^m_0 - \mu^m_1\right) \,\,. \label{G-11}
\end{eqnarray}
The subscript in $G_{11}$ indicates the yield that is going to be bounded with this combination of gains. In (\ref{G-11}) the coefficients of the yields $Y_{0m}$ and 
$Y_{n0}$, for any $n$ and $m$, are identically zero. Thus (\ref{G-11}) can be rewritten as:
\begin{eqnarray}
	G_{11}= Y_{11} (\mu_0 - \mu_1)^2 + \sum_{\stackrel[n+m>2]{n,m=1}{}}^{\infty} \frac{Y_{nm}}{n!m!} \left(\mu^n_0 -\mu^n_1\right) \left(\mu^m_0 - \mu^m_1\right)
	\label{G-11_1} \,\,. 
\end{eqnarray}
We observe that the coefficients that multiply the yields $Y_{n m}$ are always positive, being the product of two factors of equal sign.
A valid upper bound for $Y_{11}$ is obtained considering the worst-case scenario for the other yields, taking into account that (\ref{ineq-constr}) holds. 
Since all the yields' coefficients carry the same sign in (\ref{G-11_1}) --regardless of the relation between $\mu_0$ and $\mu_1$--, the yield $Y_{11}$ is maximal when all the other yields are minimal. Thus the upper bound on $Y_{11}$ is extracted by setting all the other yields to zero in (\ref{G-11_1}):
\begin{equation}
	Y^U_{11} = \frac{G_{11}}{(\mu_0 - \mu_1)^2} \,\,,   \label{Y11-upperbound}
\end{equation}
where $G_{11}$ is defined in (\ref{G-11}).\medskip\\
We remark that by combining the gains as in (\ref{G-11}), we manage to obtain a closed expression for $Y_{11}$ in which the contribution of all the yields $Y_{0m}$ and $Y_{n0}$ is removed. Additionally, $Y_{11}$ is now the yield with the ``highest weight'' in (\ref{G-11_1}) since it has the largest coefficient. All the yields' bounds presented in this work follow the same philosophy.

\subsection{Upper bound on $Y_{02}$}
Consider the following combination of gains:
\begin{eqnarray}
	G_{02} &= \mu_1 \tilde{Q}^{0,0} + \mu_0 \tilde{Q}^{1,1} - \mu_1 \tilde{Q}^{0,1} -\mu_0 \tilde{Q}^{1,0}  \nonumber\\
	&= \sum_{n,m=0}^{\infty} \frac{Y_{nm}}{n!m!} \left(\mu_1 \mu^n_0 -\mu_0 \mu^n_1\right) \left(\mu^m_0 - \mu^m_1\right) \,\,. \label{G-02}
\end{eqnarray}
In (\ref{G-02}) the coefficients of the yields $Y_{n0}$ and $Y_{1m}$ are identically zero. Thus (\ref{G-02}) can be rewritten as:
\begin{eqnarray}
	 G_{02} = &-Y_{01} (\mu_0 - \mu_1)^2 -\frac{Y_{02}}{2}(\mu_0 + \mu_1)(\mu_0 - \mu_1)^2 -\sum_{m=3}^{\infty}\frac{Y_{0m}}{m!}(\mu_0 - \mu_1)(\mu_0^m - \mu_1^m) 
	\nonumber\\  
	 &+\sum_{\stackrel[m=1]{n=2}{}}^{\infty} \frac{Y_{nm}}{n!m!} \mu_0 \mu_1 \left(\mu^{n-1}_0 -\mu^{n-1}_1\right) \left(\mu^m_0 - \mu^m_1\right)
	\label{G-02_1} \,\,. 
\end{eqnarray}
Like in the derivation of $Y_{11}$'s bound given by (\ref{Y11-upperbound}), a valid upper bound for $Y_{02}$ is obtained by considering the worst-case scenario for the remaining yields in (\ref{G-02_1}). More specifically, $Y_{02}$ is maximal when the yields whose coefficient has the same sign as $Y_{02}$'s coefficient are minimal, and the yields whose coefficient has opposite sign to $Y_{02}$'s are maximal. Recalling constraint (\ref{ineq-constr}), this means setting $Y_{01}$ and $Y_{0m}$ to zero and $Y_{nm}$ with $n\geq 2$ and $m\geq 1$, to 1 in (\ref{G-02_1}). In so doing, after rearranging the terms we obtain:
\begin{eqnarray}
	 Y^U_{02} = \frac{2}{(\mu_0 + \mu_1)(\mu_0 - \mu_1)^2}\left[-G_{02} + \left(\sum_{m=1}^{\infty}\frac{\mu_0^m}{m!}-\frac{\mu_1^m}{m!}\right)
	\left(\sum_{n=2}^{\infty}\mu_1 \frac{\mu_0^n}{n!}-\mu_0 \frac{\mu_1^n}{n!}\right) \right] \,\,,
\end{eqnarray}
which leads to the following upper bound on $Y_{02}$:
\begin{eqnarray}
	Y^U_{02} = \frac{2\left(e^{\mu_0}-e^{\mu_1}\right)\left(\mu_0-\mu_1 +\mu_1 e^{\mu_0} -\mu_0 e^{\mu_1} \right) -2G_{02}}{(\mu_0 + \mu_1)(\mu_0 - \mu_1)^2}
	\,\,. \label{Y02-upperbound}
\end{eqnarray}

\subsection{Upper bound on $Y_{20}$}
Consider the following combination of gains:
\begin{eqnarray}
G_{20} &= \mu_1 \tilde{Q}^{0,0} + \mu_0 \tilde{Q}^{1,1} - \mu_0 \tilde{Q}^{0,1} -\mu_1 \tilde{Q}^{1,0}  \nonumber\\
&= \sum_{n,m=0}^{\infty} \frac{Y_{nm}}{n!m!} \left(\mu^n_0 - \mu^n_1\right) \left(\mu_1 \mu^m_0 -\mu_0 \mu^m_1\right)  \,\,. \label{G-20}
\end{eqnarray}
In (\ref{G-20}) the coefficients of the yields $Y_{n1}$ and $Y_{0m}$ are identically zero. Thus (\ref{G-20}) can be rewritten as:
\begin{eqnarray}
 G_{20} = &-Y_{10} (\mu_0 - \mu_1)^2 -\frac{Y_{20}}{2}(\mu_0 + \mu_1)(\mu_0 - \mu_1)^2 -\sum_{n=3}^{\infty}\frac{Y_{n0}}{n!}(\mu_0 - \mu_1)(\mu_0^n - \mu_1^n) 
\nonumber\\  
 &+\sum_{\stackrel[m=2]{n=1}{}}^{\infty} \frac{Y_{nm}}{n!m!} \mu_0 \mu_1 \left(\mu^{n}_0 -\mu^{n}_1\right) \left(\mu^{m-1}_0 - \mu^{m-1}_1\right)
\label{G-20_1} \,\,. 
\end{eqnarray}
A valid upper bound for $Y_{20}$ is obtained by setting to zero the yields whose coefficient has the same sign as $Y_{20}$'s coefficient, and by setting to 1
the yields whose coefficient has opposite sign to $Y_{20}$'s. In the case of (\ref{G-20_1}) this means setting $Y_{10}$ and $Y_{n0}$ to zero and $Y_{nm}$ with $n\geq 1$ and $m\geq 2$, to 1. In this way we obtain:
\begin{eqnarray}
 Y^U_{20} = \frac{2}{(\mu_0 + \mu_1)(\mu_0 - \mu_1)^2}\left[-G_{20} + \left(\sum_{n=1}^{\infty}\frac{\mu_0^n}{n!}-\frac{\mu_1^n}{n!}\right)
\left(\sum_{m=2}^{\infty}\mu_1 \frac{\mu_0^m}{m!}-\mu_0 \frac{\mu_1^m}{m!}\right) \right] \,\,,
\end{eqnarray}
which leads to the following upper bound on $Y_{20}$:
\begin{eqnarray}
Y^U_{20} = \frac{2\left(e^{\mu_0}-e^{\mu_1}\right)\left(\mu_0-\mu_1 +\mu_1 e^{\mu_0} -\mu_0 e^{\mu_1} \right) -2G_{20}}{(\mu_0 + \mu_1)(\mu_0 - \mu_1)^2}
\,\,. \label{Y20-upperbound}
\end{eqnarray}

\subsection{Upper bound on $Y_{00}$}
Consider the following combination of gains:
\begin{eqnarray}
	G_{00} &= \mu^2_1 \tilde{Q}^{0,0}+\mu^2_0 \tilde{Q}^{1,1} -\mu_0 \mu_1 (\tilde{Q}^{0,1} +\tilde{Q}^{1,0})  \nonumber\\
	&= \sum_{n,m=0}^{\infty} \frac{Y_{nm}}{n!m!} \left(\mu^n_0 \mu_1 -\mu_0 \mu^n_1\right) \left(\mu^m_0 \mu_1 -\mu_0 \mu^m_1\right) \,\,. \label{G-00}
\end{eqnarray}
In (\ref{G-00}) the coefficients of the yields $Y_{1m}$ and $Y_{n1}$, for any $n$ and $m$, are identically zero. Thus (\ref{G-00}) can be rewritten as:
\begin{eqnarray}
\fl	G_{00}= &Y_{00}(\mu_0-\mu_1)^2 -\mu_0 \mu_1(\mu_0 -\mu_1)\left[\sum_{m=2}^{\infty} \frac{Y_{0m}}{m!}(\mu^{m-1}_0-\mu^{m-1}_1)+\sum_{n=2}^{\infty} \frac{Y_{n0}}{n!} 
	(\mu^{n-1}_0-\mu^{n-1}_1) \right] \nonumber\\
\fl	&+ \mu^2_0 \mu^2_1 \sum_{n,m=2}^{\infty} \frac{Y_{nm}}{n! m!} (\mu^{n-1}_0-\mu^{n-1}_1)(\mu^{m-1}_0-\mu^{m-1}_1)  \label{G-00_1} \,\,.
\end{eqnarray}
As usual we extract an upper bound on $Y_{00}$ by setting to their lowest value the yields whose coefficient has the same sign as $Y_{00}$'s coefficient (which correspond to the $Y_{nm}$ with $n,m\geq 2$), and by setting to their maximum value the yields whose coefficient has opposite sign to $Y_{00}$'s coefficient (which correspond to $Y_{0m}$ and $Y_{n0}$). We know that every yield is trivially bounded by (\ref{ineq-constr}). However, in order to derive a tighter bound on $Y_{00}$, we employ non-trivial bounds for all the yields $Y_{nm}$ with $n+m \leq 4$ in (\ref{G-00_1}). The upper bound on $Y_{00}$ thus satisfies:
\begin{eqnarray}
\fl	G_{00} = &Y^U_{00}(\mu_0-\mu_1)^2 -\mu_0 \mu_1(\mu_0 -\mu_1)\left[\frac{(\mu_0 -\mu_1)}{2}(Y^U_{02}+Y^U_{20}) +\frac{(\mu^2_0 -\mu^2_1)}{6}(Y^U_{03}+Y^U_{30}) 
	  \right.  \nonumber\\
\fl	  &\left. +\frac{(\mu^3_0 -\mu^3_1)}{24}(Y^U_{04}+Y^U_{40}) + 2\sum_{n=5}^{\infty} \frac{(\mu^{n-1}_0-\mu^{n-1}_1)}{n!} \right] 
	+ \frac{\mu^2_0 \mu^2_1 (\mu_0 -\mu_1)^2}{4} Y^L_{22}  \label{G-00_2} \,\,.
\end{eqnarray}
In this equation $Y^U_{ij}$ are upper bounds and $Y^L_{ij}$ are lower bounds. From (\ref{G-00_2}) we obtain the following upper bound on $Y_{00}$:
\begin{eqnarray}
\fl	Y^U_{00} =  &\frac{G_{00}}{(\mu_0-\mu_1)^2} + \frac{\mu_0 \mu_1}{\mu_0-\mu_1}\left[\frac{(\mu_0 -\mu_1)}{2}(Y^U_{02}+Y^U_{20}) 
	+\frac{(\mu^2_0-\mu^2_1)}{6}(Y^U_{03}+Y^U_{30}) + \frac{(\mu^3_0 -\mu^3_1)}{24}(Y^U_{04}+Y^U_{40})\right] \nonumber\\
\fl	&+ \frac{2}{\mu_0 - \mu_1} \left[\mu_1 \left(e^{\mu_0}-1-\frac{\mu_0^2}{2}-\frac{\mu_0^3}{6}-\frac{\mu_0^4}{24}\right) -
	   \mu_0 \left(e^{\mu_1}-1-\frac{\mu_1^2}{2}-\frac{\mu_1^3}{6}-\frac{\mu_1^4}{24}\right)\right]	-\frac{\mu_0^2 \mu_1^2}{4} Y_{22}^L \,\,.  \label{Y00-upperbound}
\end{eqnarray}
where $Y^U_{02}$ and $Y^U_{20}$ are given in (\ref{Y02-upperbound}) and (\ref{Y20-upperbound}), respectively. The expressions for $Y^U_{03}$ and $Y^U_{04}$ in (\ref{Y00-upperbound}) can be found by starting from the same expression (\ref{G-02_1}) that we used to derive $Y^U_{02}$, i.e.:
\begin{eqnarray}
\fl	G_{02} = & -\sum_{m=1}^{\infty}\frac{Y_{0m}}{m!}(\mu_0 - \mu_1)(\mu_0^m - \mu_1^m)+\sum_{\stackrel[m=1]{n=2}{}}^{\infty} \frac{Y_{nm}}{n!m!} \mu_0 \mu_1
	\left(\mu^{n-1}_0 -\mu^{n-1}_1\right) \left(\mu^m_0 - \mu^m_1\right)  \,\,.
\end{eqnarray}
From this expression we can extract an upper bound on any generic $Y_{0m}$ as follows:
\begin{eqnarray}
\fl	Y^U_{0m} = \min\left\{ \frac{m!}{(\mu_0 - \mu_1)(\mu_0^m - \mu_1^m)} \left[-G_{02} + (e^{\mu_0}-e^{\mu_1})(\mu_0 -\mu_1 +\mu_1 e^{\mu_0}-\mu_0 e^{\mu_1})\right]
	 \,\,,\,\,1 \right\}  \label{Y0m-upperbound} \,\,,
\end{eqnarray}
where we employ the constraint (\ref{ineq-constr}). Similarly, the expressions for $Y^U_{30}$ and $Y^U_{40}$  are obtained starting from (\ref{G-20_1}) and deriving an upper bound on a generic $Y_{n0}$ as follows:
\begin{eqnarray}
\fl Y^U_{n0} = \min\left\{ \frac{n!}{(\mu_0 - \mu_1)(\mu_0^n - \mu_1^n)} \left[-G_{20} + (e^{\mu_0}-e^{\mu_1})(\mu_0 -\mu_1 +\mu_1 e^{\mu_0}-\mu_0 e^{\mu_1})\right]
\,\,,\,\,1 \right\}  \label{Yn0-upperbound} \,\,.
\end{eqnarray}
At last, the expression for $Y^L_{22}$ can be derived from the same combination of yields which led to $Y^U_{11}$. In particular, from (\ref{G-11_1}) we have that:
\begin{eqnarray}
	G_{11}= \sum_{n,m=1}^{\infty} \frac{Y_{nm}}{n!m!} \left(\mu^n_0 -\mu^n_1\right) \left(\mu^m_0 - \mu^m_1\right) \nonumber \,\,.
\end{eqnarray}
Then, by setting to 1 all the yields whose coefficient has equal sign to $Y_{22}$'s we obtain:
\begin{eqnarray}
	G_{11}= \sum_{n,m=1}^{\infty} \frac{\mu^n_0 -\mu^n_1}{n!} \frac{\mu^m_0 - \mu^m_1}{m!} -\frac{(\mu_0^2 -\mu_1^2)^2}{4} +\frac{(\mu_0^2 -\mu_1^2)^2}{4} Y_{22}^L
	\,\,,
\end{eqnarray}
which yields:
\begin{eqnarray}
	Y_{22}^L =  \max \left\{ \frac{4}{(\mu_0 -\mu_1)^2 (\mu_0 +\mu_1)^2} \left[G_{11} - (e^{\mu_0}-e^{\mu_1})^2\right] +1 \,\,,\,\,0\right\} \label{Y22-lowerbound} \,\,.
\end{eqnarray}
Note that the upper bounds derived on $Y_{04}$ and $Y_{40}$ in this Section could be used to improve the estimation of the phase error rate given by (\ref{phase-error-rate-bound}). However, the resulting improvement in the secret key rate would be extremely small in this case and we neglect it for simplicity.

\section{Simulations}  \label{simulations}
In this Section we provide plots of the secret key rate given by (\ref{key-rate}) against the overall loss ($-10\log_{10} \eta$) measured in dB of the two optical channels linking Alice and Bob to node $C$. The channel model we use to simulate the quantities that would be observed experimentally --i.e. the gains $p(k_c,k_d|b_A,b_B)$ and $Q^{\beta_A^2,\beta_B^2}_{k_c,k_d}$-- is given in \ref{channel-model} \cite{Curty-security-TF}. It accounts for: the loss in the optical channels together with the non-unity detection efficiency of $D_c$ and $D_d$ (altogether described by the parameter $\eta$), the polarization and phase misalignments introduced by the channel and a dark count probability $p_d$ in each detector.
For concreteness, in all the plots below we assume fixed polarization and phase misalignments of 2\%, independently of the channel loss. Note that, as pointed out in \cite{Curty-security-TF}, the TF-QKD protocol analyzed in this work is quite robust against phase mismatch. This is so because phase misalignment only affects the quantum bit error rate but not the phase error rate.\\
For illustration purposes every plot is obtained for three different values of the dark count rate of the detectors, $p_d\in\{10^{-6}, 10^{-7}, 10^{-8}\}$. The plots are obtained by numerically optimizing\footnote{The optimization is carried out by using the built-in function ``NMaximize'' of the software Wolfram Mathematica 10.0.} the secret key rate --for every value of the loss-- over the signal intensity ($\alpha^2$) and over one decoy intensity, while for simplicity the other decoy intensities are fixed to near-to-optimal values for all values of the overall loss. More specifically, we preliminarily performed an optimization of the key rate over the whole set of intensity settings and noticed that most of the decoy intensities are roughly constant with the loss and tend to be as low as possible. For instance, if we consider the case with two decoy intensity settings ($\mu_0$ and $\mu_1$, with $\mu_0>\mu_1$), we observe that the optimal value for the weakest decoy $\mu_1$ is basically the lowest possible for any value of the loss. In practice, however, it might be difficult to generate very weak signals due to the finite extinction ratio of a practical intensity modulator~\cite{ros}, so we fix $\mu_1$ to a reasonable small value from an experimental point of view, say $\mu_1=10^{-5}$~\cite{experiment-Toshiba,experiment-Toronto}, while keeping the optimization over the remaining intensities. Similarly, if we consider the case with three decoy intensity settings ($\mu_0$, $\mu_1$ and $\mu_2$, with $\mu_0>\mu_1>\mu_2$), we find that the optimal values for the weakest decoys $\mu_1$ and $\mu_2$ are also the lowest possible for any value of the loss.~Moreover, in this last case, we show in \ref{stronger_decoys} that the system performance remains basically unchanged if one increases the value of the weakest intensity to say $\mu_2=10^{-3}$, which might be even easier to implement experimentally than $10^{-5}$. Thus, we fix $\mu_2=10^{-3}$ and we differentiate it from $\mu_1$ by, for example, one order of magnitude (i.e. we take $\mu_1=10^{-2}$). The same argument holds in the case with four decoy intensity settings (see \ref{stronger_decoys}), where we fix $\mu_2=10^{-3}$, $\mu_1=10^{-2}$, and $\mu_0=10^{-1}$. We remark, however, that our method is general in the sense that the analytical upper bounds on the yields can be evaluated with any desired combination of intensity settings, while we select these particular decoy intensity values only for illustration purposes. Also, let us emphasize that the optimal decoy intensity values in the finite-key regime might be different from the values mentioned above. The analysis of the finite-key regime is, however, beyond the scope of this paper. Importantly, it turns out that the resulting asymptotic secret key rates in these scenarios are almost indistinguishable from those obtained by optimizing the value of all the intensity settings. 

\noindent The optimal values of the signal and decoy intensities which are optimized as a function of the loss are also plotted in this Section. In this regard, we also study how the key rate is affected when the intensities are subjected to fluctuations around their optimal values in \autoref{intensity-fluct}.

\subsection{Two decoy intensity settings}  \label{2decoys-plots}
\begin{figure}[!htb]
	\centering
	\includegraphics[width=0.8\linewidth,keepaspectratio]{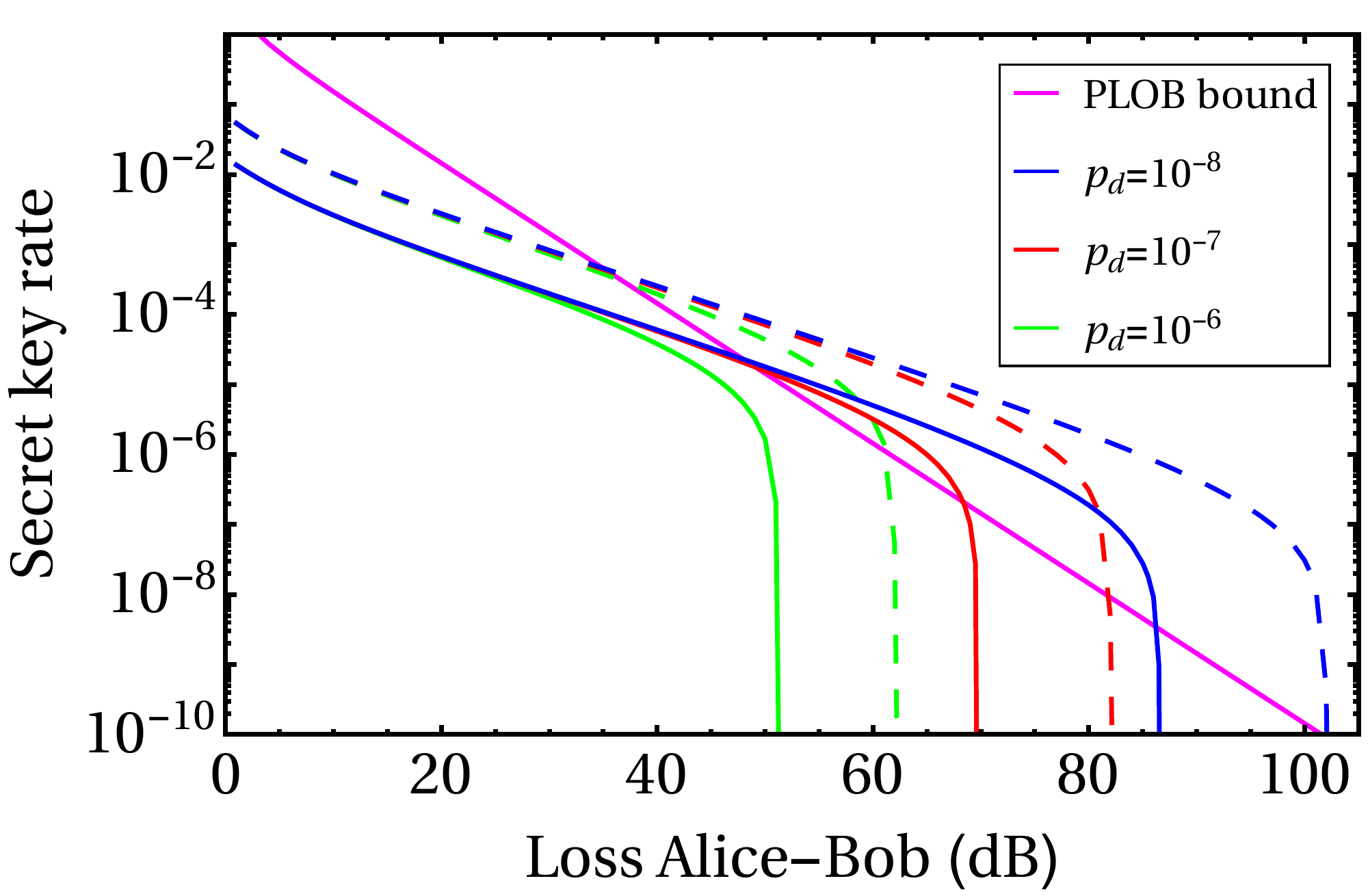}
	\caption{Secret key rate in logarithmic scale as a function of the overall loss in the channels Alice-$C$ and Bob-$C$ for three different dark count rates ($10^{-6}$ green, $10^{-7}$ red, $10^{-8}$ blue). The solid lines correspond to the case where the yields $Y_{00},Y_{02},Y_{20}$ and $Y_{11}$ are estimated by means of two decoy intensity settings through the bounds presented in \autoref{yields-bounds-2decoys} and the key rate is optimized over the signal intensity $\alpha^2$ (see \autoref{optimal-alpha-2decoys}) and the decoy intensity $\mu_0$ (see \autoref{optimal-mu0-2decoys}). The other decoy intensity, $\mu_1$, is fixed to $\mu_1=10^{-5}$. The dashed lines assume that all the yields are known from the channel model and the secret key rate is optimized over $\alpha^2$. That is, these lines show the maximum value of the secret key rate which could be achieved with an infinite number of decoy intensity settings and the security analysis reported in \cite{Curty-security-TF}. The solid magenta line illustrates the PLOB bound \cite{PLOB}. The plot shows that in the presence of a dark count rate of at most about $p_d=10^{-7}$ the protocol can beat the PLOB bound even with just two decoy intensity settings.}
	\label{key-rate-2decoys}
\end{figure}
\begin{figure}[!htb]
	\centering
	\begin{subfigure}[t]{.5\textwidth}
		\centering
		\includegraphics[width=1\linewidth,keepaspectratio]{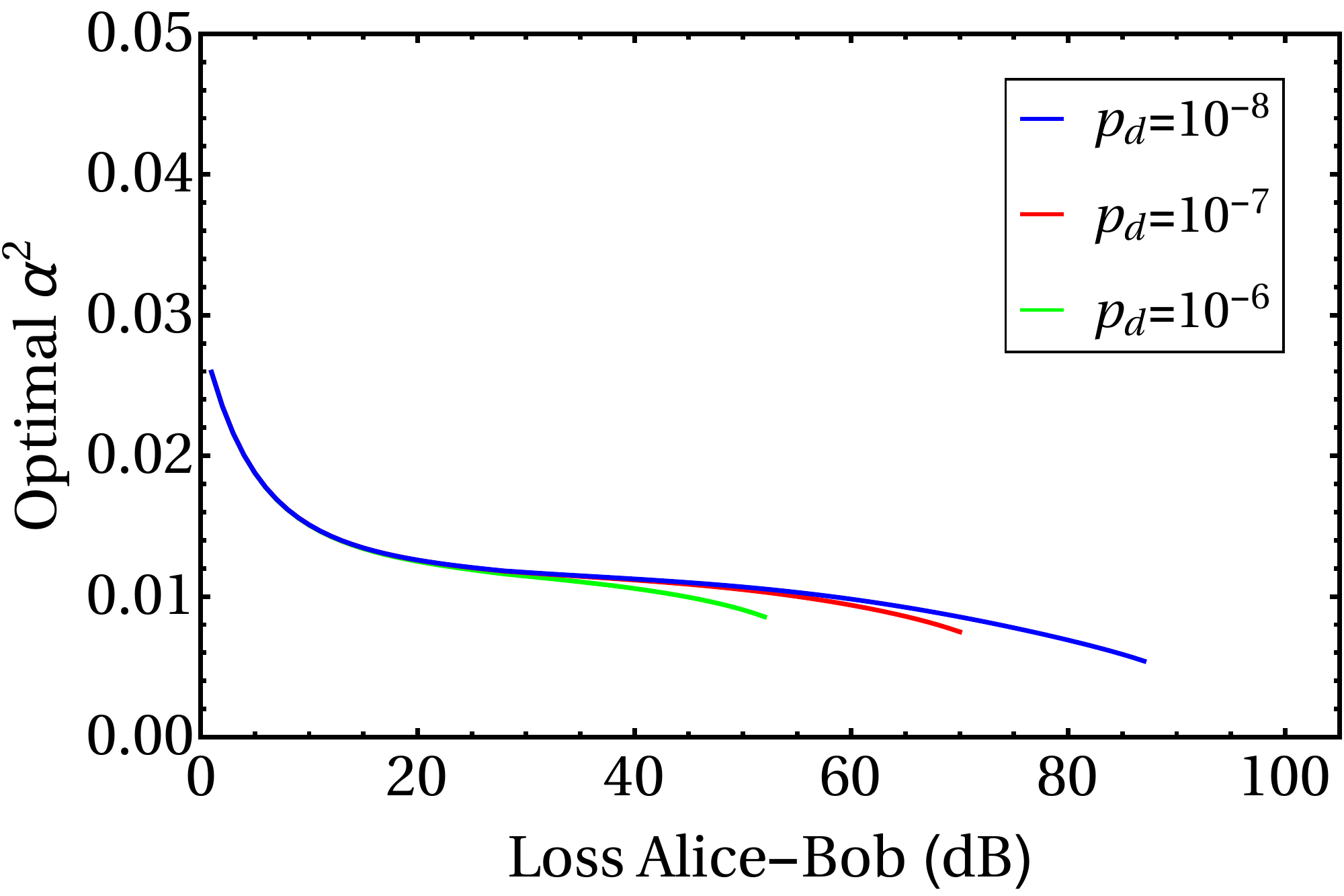}
		\caption{Optimal values of the signal intensity $\alpha^2$ as a function of the loss between Alice and Bob for three different dark count rates. These values are obtained from the optimization of the secret key rate (solid lines) of \autoref{key-rate-2decoys}.}
		\label{optimal-alpha-2decoys}
	\end{subfigure}%
	\begin{subfigure}[t]{.5\textwidth}
		\centering
		\includegraphics[width=1\linewidth,keepaspectratio]{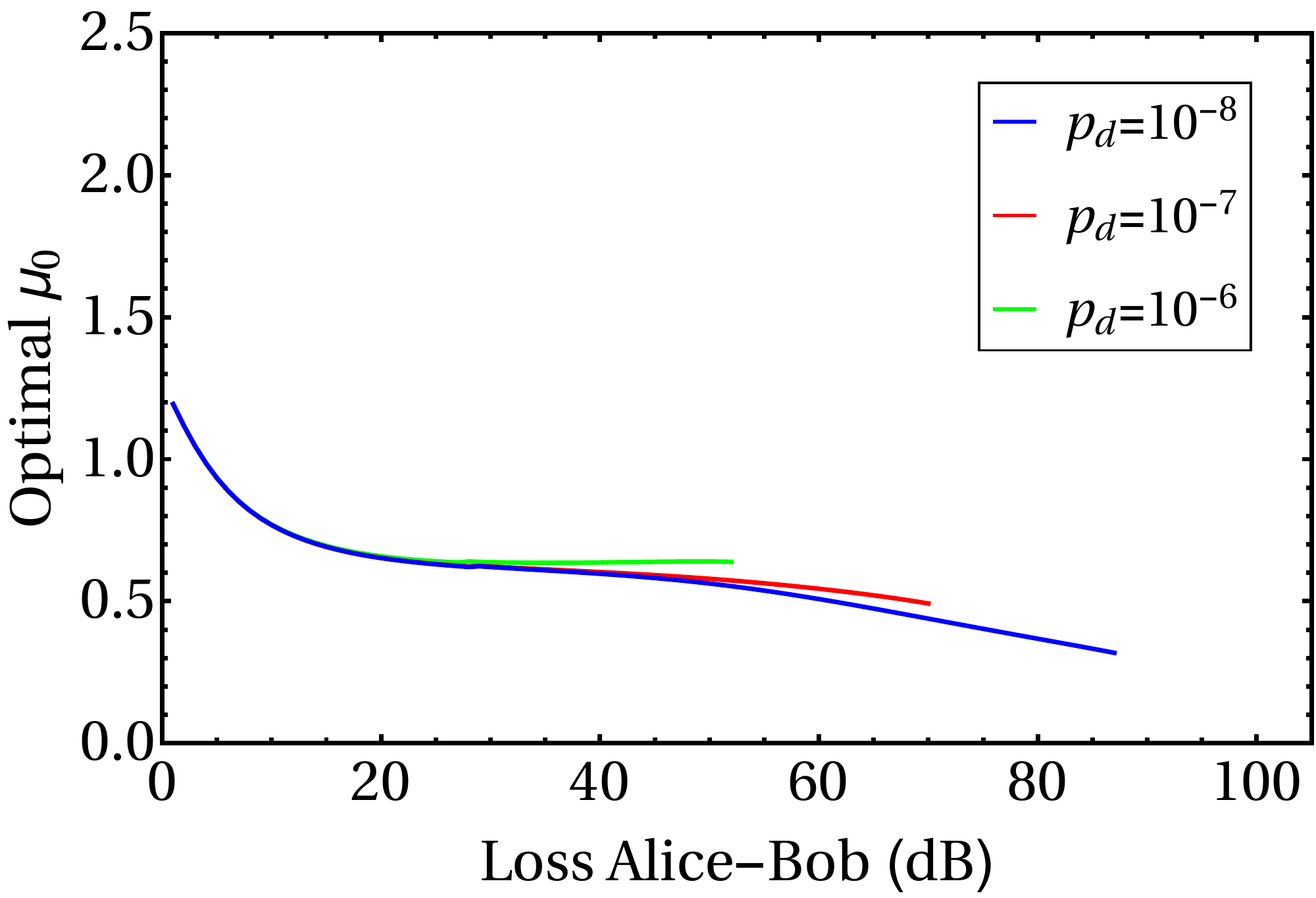}
		\caption{Optimal values of the decoy intensity $\mu_0$ as a function of the loss between Alice and Bob for three different dark count rates. These values are obtained from the optimization of the secret key rate (solid lines) of \autoref{key-rate-2decoys}. The other decoy intensity is set to: $\mu_1=10^{-5}$.}
		\label{optimal-mu0-2decoys}
	\end{subfigure}
	\caption{Optimal values of the signal and decoy intensities $\alpha^2$ and $\mu_0$ for the TF-QKD protocol \cite{Curty-security-TF} when the parties have at their disposal two decoy intensity settings to estimate the yields. }
	\label{optimal-intensities-2decoys}
\end{figure}
In \autoref{key-rate-2decoys} we plot the secret key rate against the overall loss for the case where Alice and Bob use two decoy intensity settings each. The solid lines are obtained by bounding from above the yields $Y_{00},Y_{02},Y_{20}$ and $Y_{11}$ by means of the expressions derived in \autoref{yields-bounds-2decoys} and by optimizing the rate over the signal intensity $\alpha^2$ and the decoy intensity $\mu_0$, while the other decoy intensity is fixed to $\mu_1=10^{-5}$ as explained above. The optimal values for $\alpha^2$ and $\mu_0$ are shown in \autoref{optimal-alpha-2decoys} and \autoref{optimal-mu0-2decoys}, respectively. The dashed lines are instead obtained by employing the exact expression of the yields\footnote{By ``exact expression'' we mean that if the experimental apparatus were accurately described by the channel model in \ref{channel-model}, then the yields associated to that experimental setup would be precisely predicted by (\ref{yields-model}).} which is given by (\ref{yields-model}) for the channel model considered. This represents the ideal scenario in which the parties have an infinite number of decoys through which they can estimate all the yields precisely. Note that in order to obtain the dashed lines in \autoref{key-rate-2decoys} we use the exact expression of the yields $Y_{nm}$ only for $n,m \leq 12$ while we set the other yields to 1. This is enough to basically reproduce the behavior of the secret key rate when all the infinite number of yields are computed via the channel model's formula given by (\ref{yields-model}), as argued in \cite{Curty-security-TF}. The dashed lines are only optimized over the signal intensity, since the yields are directly given by the channel model. Finally, we also insert in \autoref{key-rate-2decoys} the PLOB bound on the secret key capacity \cite{PLOB}, which reads as follows in terms of the transmittance $\eta$ :
\begin{equation}
K(\eta)=-\log_2 (1-\eta)  \label{PLOB-bound} \,\,.
\end{equation}
In \autoref{key-rate-2decoys} we observe that even by means of just two decoy intensity settings the key rate can beat the PLOB bound, provided that the dark count rate is $p_d \lesssim 10^{-7}$ . This happens because with two decoys the parties can already non-trivially estimate the yields $Y_{nm}$ with $n+m \leq 2$ as we showed in \autoref{yields-bounds-2decoys}, and these yields are the most relevant terms in the phase-error rate formula given by (\ref{phase-error-rate}) \cite{Curty-security-TF}. Note that we did not estimate the yields $Y_{01}$ and $Y_{10}$ since only the yields $Y_{nm}$ with $n+m$ an even number contribute to the phase-error rate (\ref{phase-error-rate}).\\
However, \autoref{key-rate-2decoys} also shows that there is a sensible gap between the rates where the yields are estimated with two decoys (solid lines) and the best possible rates one could achieve (dashed lines) if all the yields were known. This clearly indicates that, although two decoys allow to estimate the yields of largest contribution in the phase-error rate, such estimations are not sufficiently tight and the ability to estimate a larger number of yields would increase the performance of the protocol.\\
By considering \autoref{optimal-intensities-2decoys} and the fixed value of the decoy intensity $\mu_1$, one notices that the optimal intensities are rather small and thus, in a real experimental implementation, intensity fluctuations might be an issue. In \autoref{intensity-fluct} we address this problem by studying how the key rate is affected under intensity fluctuations and show that for fluctuations up to about 40\% the change in the key rate performance is minimal.\\
Also, we notice that the optimal values of the signal intensity $\alpha^2$ (see \autoref{optimal-alpha-2decoys}) and the decoy intensity $\mu_0$ (see \autoref{optimal-mu0-2decoys}) are almost constant with the loss, for losses $\gtrsim 20 \mbox{ dB}$. This means that in a scenario where the loss in the quantum channels varies dynamically with time within a reasonable interval, one could still fix the signal intensity and both decoy intensities to constant values which happen to be close to the optimal ones. This argument also holds in the case of three (see \autoref{3decoys-plots}) and four decoy intensity settings (see \autoref{4decoys-plots}).

\subsection{Three decoy intensity settings} \label{3decoys-plots}
\begin{figure}[!htb]
	\centering
	\includegraphics[width=0.8\linewidth,keepaspectratio]{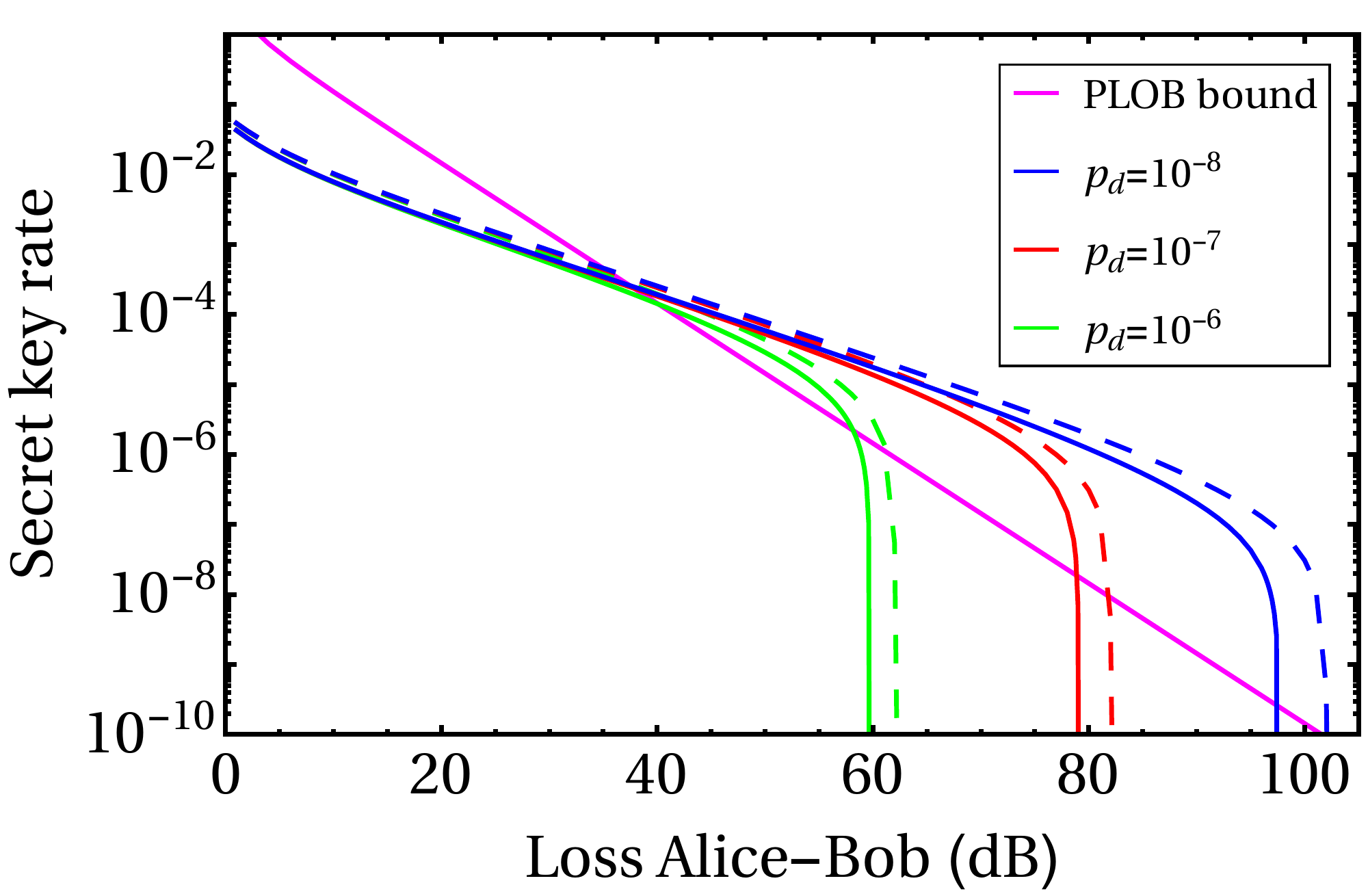}
	\caption{Secret key rate in logarithmic scale as a function of the overall loss in the channels Alice-$C$ and Bob-$C$ for three different dark count rates ($10^{-6}$ green, $10^{-7}$ red, $10^{-8}$ blue). The solid lines correspond to the case where the yields $Y_{00},Y_{02},Y_{20},Y_{11},Y_{13},Y_{31},Y_{04},Y_{40}$ and $Y_{22}$ are estimated by means of three decoy intensity settings through the bounds presented in \ref{yields-bounds-3decoys} and the key rate is optimized over the signal intensity $\alpha^2$ (see \autoref{optimal-alpha-3decoys}) and the decoy intensity $\mu_0$ (see \autoref{optimal-mu0-3decoys}). The other decoy intensities are fixed to $\mu_1=10^{-2}$ and $\mu_2=10^{-3}$. The dashed lines assume that all the yields are known from the channel model and the secret key rate is optimized over $\alpha^2$. That is, these lines show the maximum value of the secret key rate which could be achieved with an infinite number of decoy intensity settings and the security analysis reported in \cite{Curty-security-TF}. The solid magenta line illustrates the PLOB bound \cite{PLOB}. The plot shows that already with three decoy intensity settings the key rate (solid lines) is sensibly close to the ideal one in which all the yields are known (dashed lines), meaning that the contribution of the other yields trivially bounded by 1 in the phase error rate is minimal.}
	\label{key-rate-3decoys}
\end{figure}
\begin{figure}[!thb]
	\centering
	\begin{subfigure}[t]{.5\textwidth}
		\centering
		\includegraphics[width=1\linewidth,keepaspectratio]{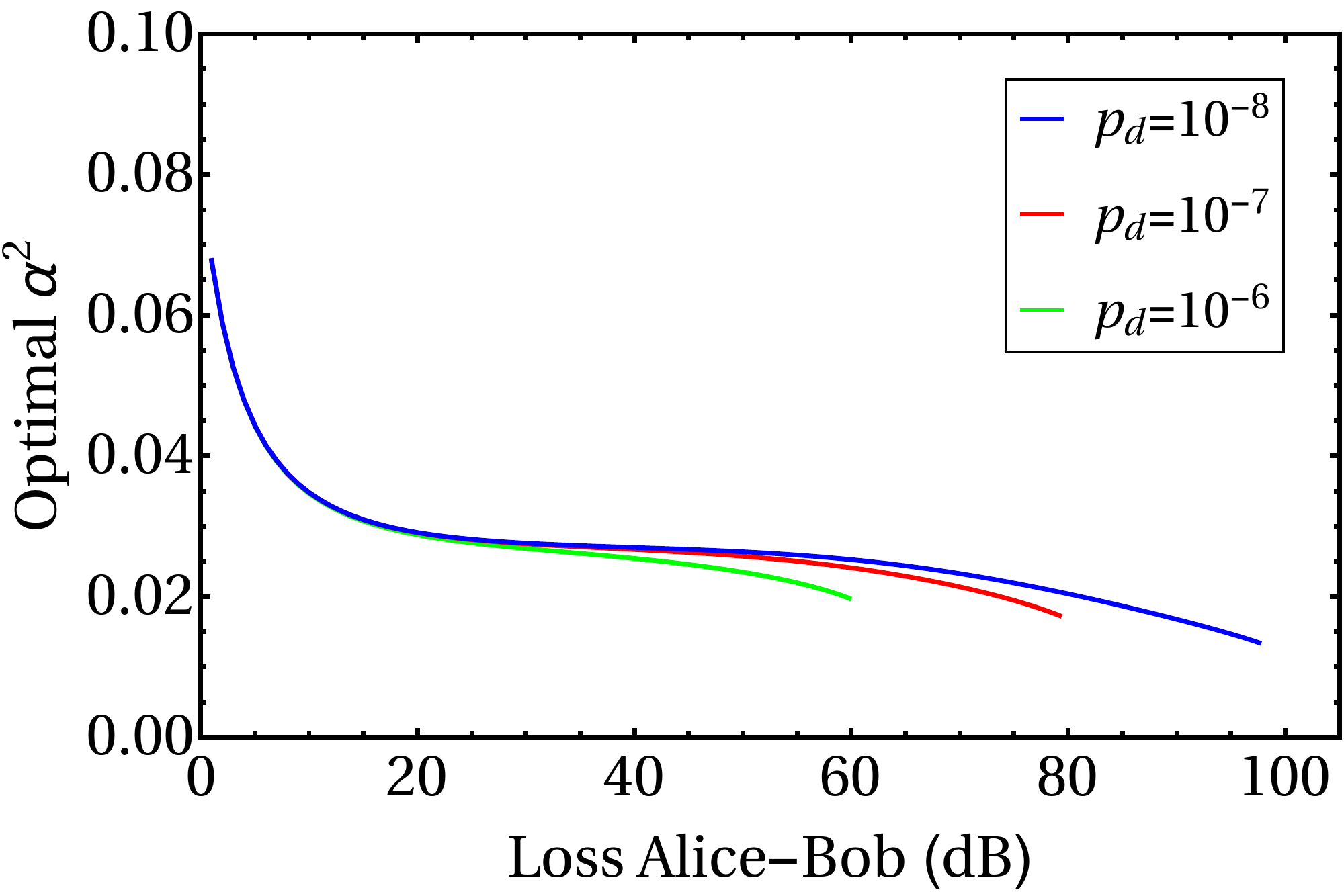}
		\caption{Optimal values of the signal intensity $\alpha^2$ as a function of the loss between Alice and Bob for three different dark count rates. These values are obtained from the optimization of the secret key rate (solid lines) of \autoref{key-rate-3decoys}. We observe that the optimal signal intensity is roughly doubled with respect to the two-decoys case (\autoref{optimal-alpha-2decoys}).}
		\label{optimal-alpha-3decoys}
	\end{subfigure}%
	\begin{subfigure}[t]{.5\textwidth}
		\centering
		\includegraphics[width=1\linewidth,keepaspectratio]{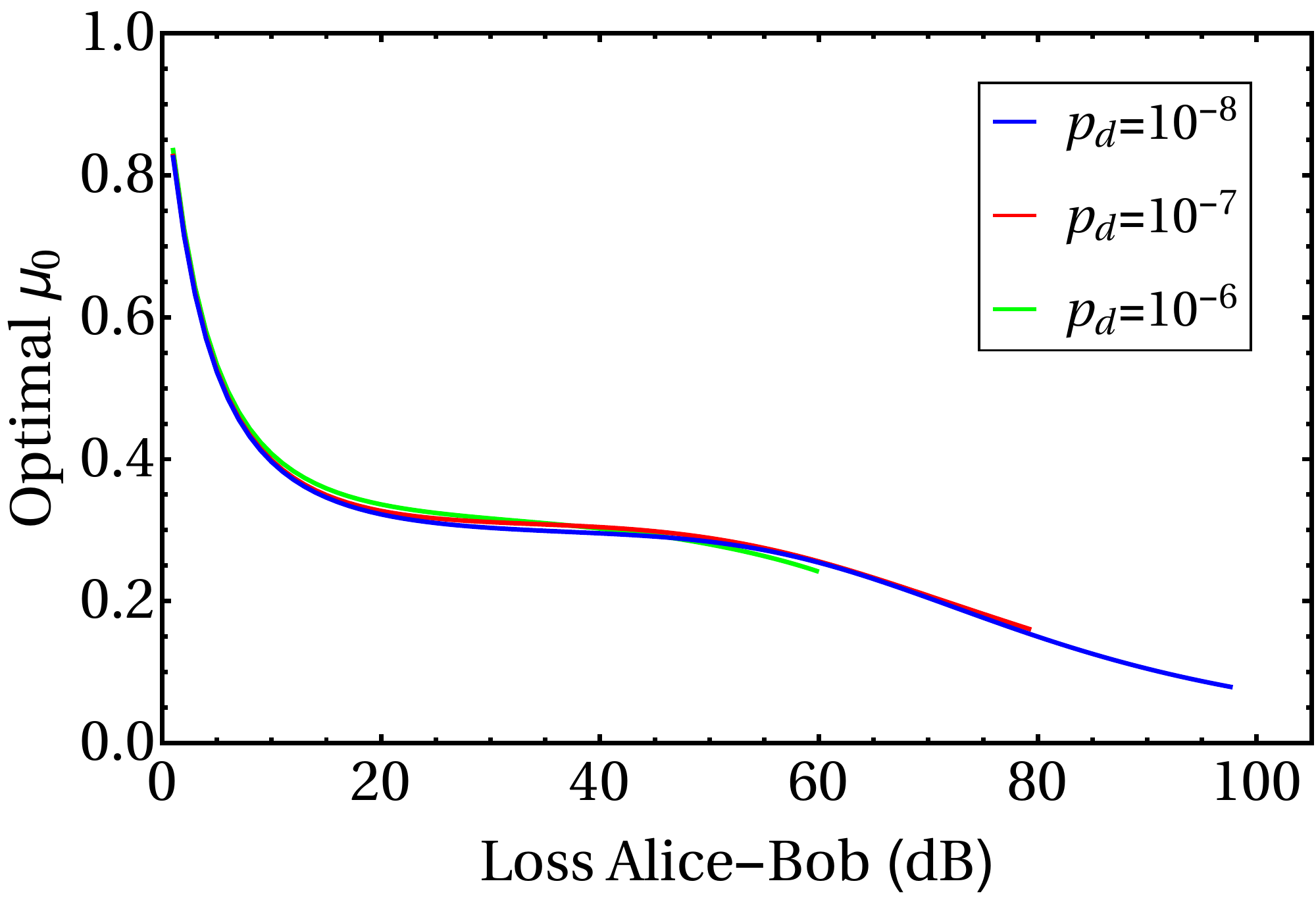}
		\caption{Optimal values of the decoy intensity $\mu_0$ as a function of the loss between Alice and Bob for three different dark count rates. These values are obtained from the optimization of the secret key rate (solid lines) of \autoref{key-rate-3decoys}. The other decoy intensities are set to: $\mu_1=10^{-2}$ and $\mu_2=10^{-3}$.}
		\label{optimal-mu0-3decoys}
	\end{subfigure}
	\caption{Optimal values of the signal and decoy intensities $\alpha^2$ and $\mu_0$ for the TF-QKD protocol \cite{Curty-security-TF} when the parties have at their disposal three decoy intensity settings to estimate the yields. }
	\label{optimal-intensities-3decoys}
\end{figure}
In \autoref{key-rate-3decoys} we plot the secret key rate against the overall loss for the case where Alice and Bob use three decoy intensity settings each. The solid lines are obtained by bounding from above the relevant yields $Y_{nm}\,\mbox{such that}\,n+m \leq 4$ (i.e. we upper bound the yields $Y_{00},Y_{02},Y_{20},Y_{11},Y_{13},Y_{31},Y_{04},Y_{40}$ and $Y_{22}$). The exact expressions for the different upper bounds on the yields can be found in \ref{yields-bounds-3decoys}, and we omit them here for simplicity. The solid lines are optimized over the signal intensity $\alpha^2$ and the decoy intensity $\mu_0$, while the weakest decoy intensities are fixed for simplicity to $\mu_1=10^{-2}$ and $\mu_2=10^{-3}$. As explained above, the resulting secret key rate in this scenario is almost indistinguishable from that obtained by optimizing over all the intensity settings. The optimal values for $\alpha^2$ and $\mu_0$ are shown in \autoref{optimal-alpha-3decoys} and \autoref{optimal-mu0-3decoys}, respectively. The dashed lines are again obtained by employing the exact expression of the yields given by the channel model (\ref{yields-model}) and coincide with those plotted in \autoref{key-rate-2decoys}.\\
We observe in \autoref{key-rate-3decoys} that the use of three decoys yields a significant improvement in the protocol's performance with respect to the two-decoys case (see \autoref{key-rate-2decoys}). As a matter of fact, in \autoref{key-rate-3decoys} the solid lines are almost overlapping the dashed lines for most values of the channel loss. This is due to the fact that with three decoys the parties constrain the yields with nine independent equations (instead of four equations as in the two-decoys case), which enable a tighter estimation of $Y_{00},Y_{02},Y_{20}$ and $Y_{11}$ and the non-trivial estimation of five additional yields.\\
Moreover, in the case of three decoys the optimal signal intensity $\alpha^2$ (see \autoref{optimal-alpha-3decoys}) is roughly double the value of the correspondent intensity when using two decoys (see \autoref{optimal-alpha-2decoys}). The reason for this is connected to the role of $\alpha^2$ in the protocol's key rate. In fact, the prefactor $p(k_c,k_d)$ with $k_c+k_d=1$ of the key rate formula given by (\ref{R_kckd}) increases for increasing $\alpha^2$: the higher the mean number of photons sent by the parties (within certain limits) the higher the probability of having a click in one of the two detectors. On the other hand, increasing $\alpha^2$ excessively also affects the phase-error rate. Note that by setting some yields to 1 in the phase error rate formula given by (\ref{phase-error-rate}) we give rise to addends like $c_{2n} c_{2m}$ and $c_{2n+1} c_{2m+1}$ which increase for increasing $\alpha^2$, leading to an overall increase of the phase-error rate and thus decrease of the key rate. The optimal value of $\alpha^2$ is thus given by the trade-off between the effect of the prefactor $p(k_c,k_d)$ and that of the terms $c_{2n} c_{2m}$ and $c_{2n+1} c_{2m+1}$. Now, by noting that the contribution of the therms $c_{2n} c_{2m}$ and $c_{2n+1} c_{2m+1}$ decreases for increasing $n,m$, we understand that their negative effect on the key rate is diminished in the case of three decoys since we non-trivially estimate more yields, i.e. a lower number of yields is set to 1. This allows $\alpha^2$ to acquire higher values with respect to the two-decoys case, as we observed in \autoref{optimal-alpha-3decoys}.\\
Finally we point out that such an argument does not apply to the discussion about the optimal value of the decoy intensity $\mu_0$ in the case of two and three decoys. As a matter of fact, the key rate does not depend on the decoy intensities in the same way as on the signal intensity: the decoy intensities only appear in the yields' bounds inserted in the phase-error rate. Additionally, the analytical bounds on the yields when using two or three decoys cannot be compared in a straightforward way. Nonetheless we observe a similar behavior of the optimal $\mu_0$ for two (see \autoref{optimal-mu0-2decoys}) and three decoys (see \autoref{optimal-mu0-3decoys}). 

\subsection{Four decoy intensity settings} \label{4decoys-plots}
\begin{figure}[!htb]
	\centering
	\includegraphics[width=0.8\linewidth,keepaspectratio]{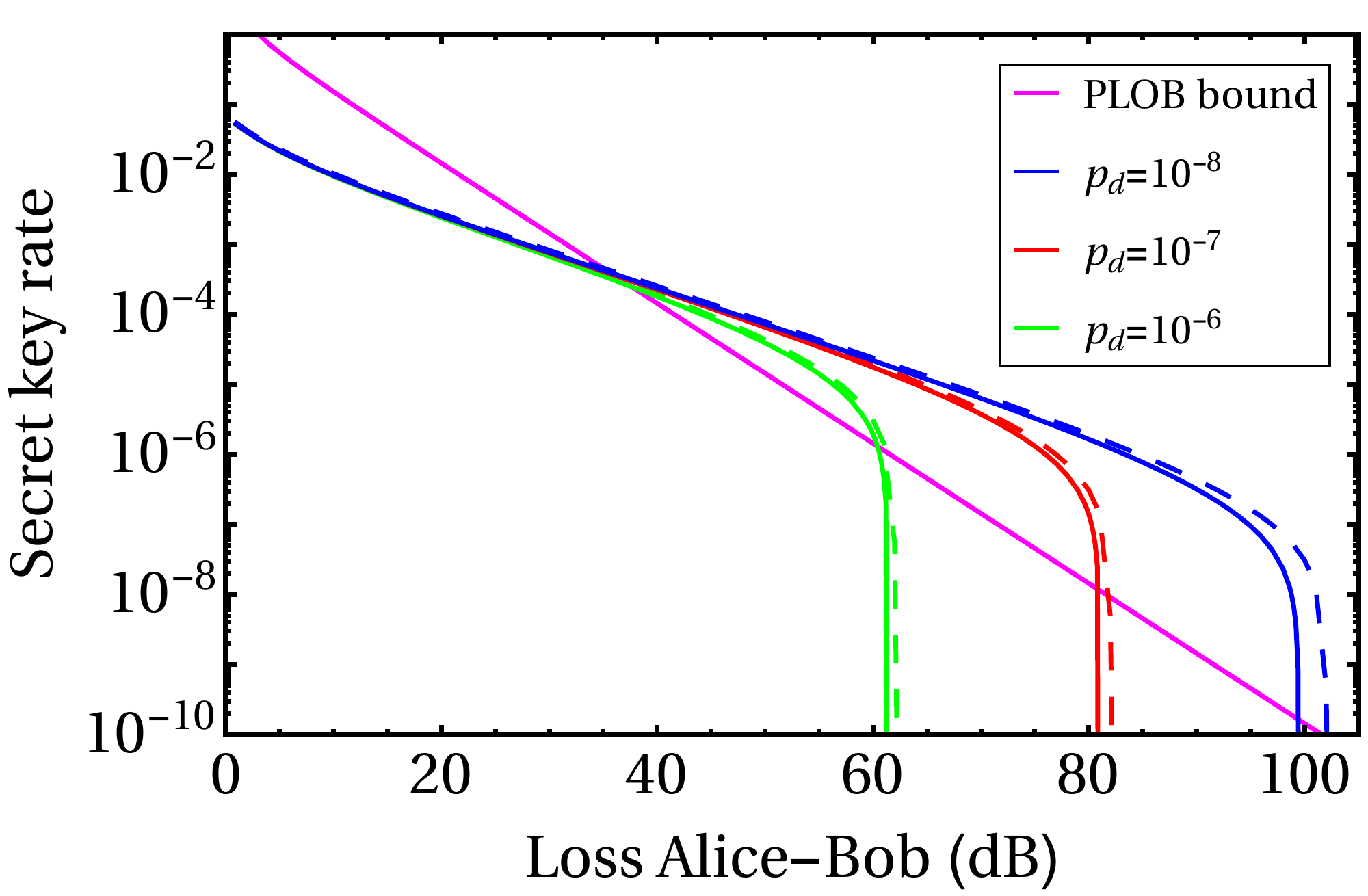}
	\caption{Secret key rate in logarithmic scale as a function of the overall loss in the channels Alice-$C$ and Bob-$C$ for three different dark count rates ($10^{-6}$ green, $10^{-7}$ red, $10^{-8}$ blue). The solid lines correspond to the case where the yields $Y_{00},Y_{02},Y_{20},Y_{11}$ and $Y_{22}$ are estimated by means of three decoys through the bounds presented in \ref{yields-bounds-3decoys} (i.e., for simplicity here we disregard the information provided by the additional fourth decoy intensity setting) and $Y_{13},Y_{31},Y_{04}$ and $Y_{40}$ are estimated with four decoys via the bounds in \ref{yields-bounds-4decoys}. The key rate is optimized over the signal intensity $\alpha^2$ (see \autoref{optimal-alpha-4decoys}) and the decoy intensity $\mu_3$ (see \autoref{optimal-mu3-4decoys}), while the other decoy intensities are fixed to $\mu_0=10^{-1},\mu_1=10^{-2}$ and $\mu_2=10^{-3}$. The dashed lines are optimized over $\alpha^2$ and assume that all the yields are known from the channel model. They correspond to the maximum value of the secret key rate which could be achieved with an infinite number of decoy intensity settings. The solid magenta line illustrates the PLOB bound \cite{PLOB}. The plot indicates that the tighter estimation of the yields $Y_{13},Y_{31},Y_{04}$ and $Y_{40}$ with respect to the case of three decoy intensity settings is enough to basically reproduce the ideal scenario in which all the yields are known (dashed lines).}
	\label{key-rate-4decoys}
\end{figure}
\begin{figure}[!thb]
	\centering
	\begin{subfigure}[t]{.5\textwidth}
		\centering
		\includegraphics[width=1\linewidth,keepaspectratio]{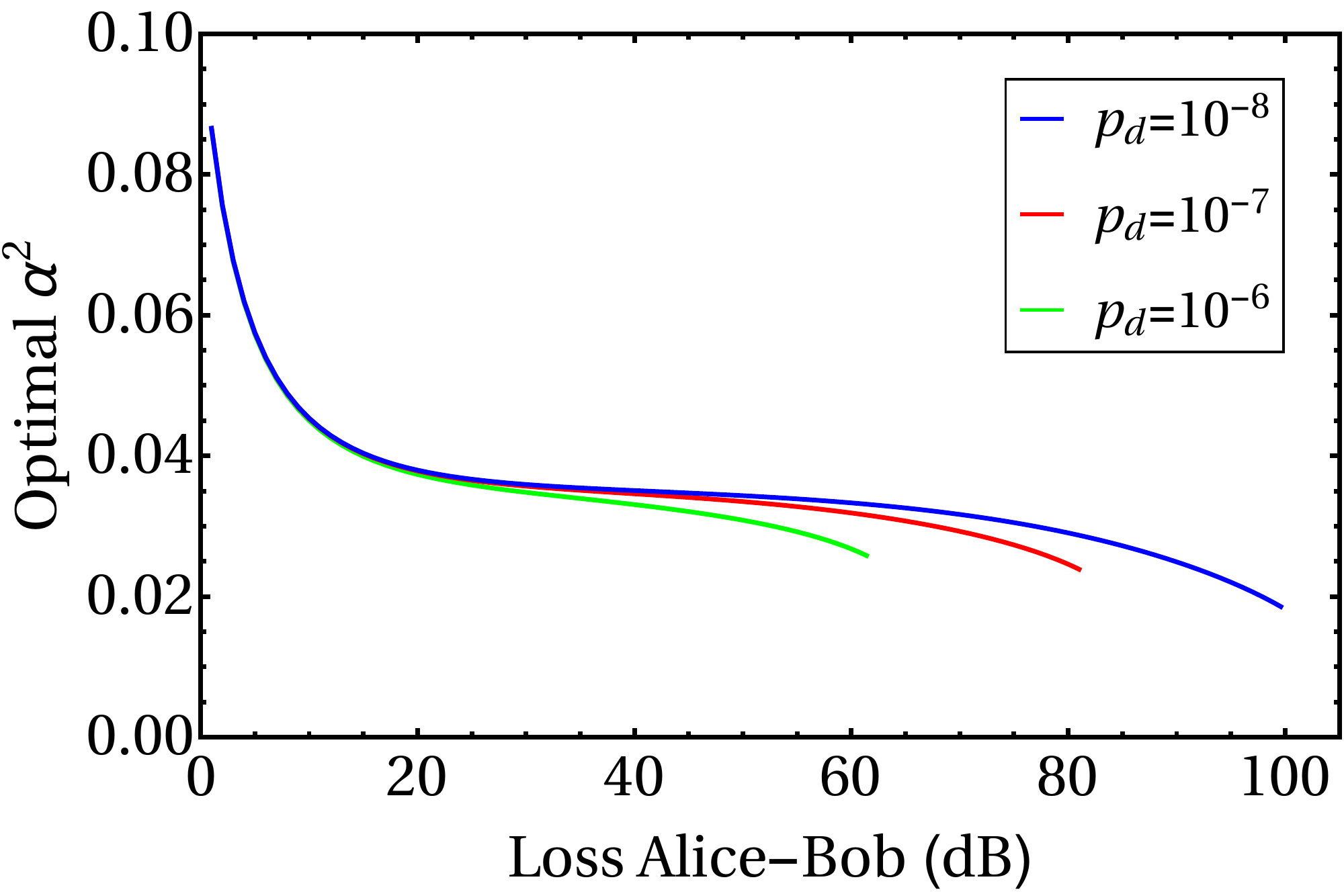}
		\caption{Optimal values of the signal intensity $\alpha^2$ as a function of the loss between Alice and Bob for three different dark count rates. These values are obtained from the optimization of the secret key rate (solid lines) of \autoref{key-rate-4decoys}.}
		\label{optimal-alpha-4decoys}
	\end{subfigure}%
	\begin{subfigure}[t]{.5\textwidth}
		\centering
		\includegraphics[width=1\linewidth,keepaspectratio]{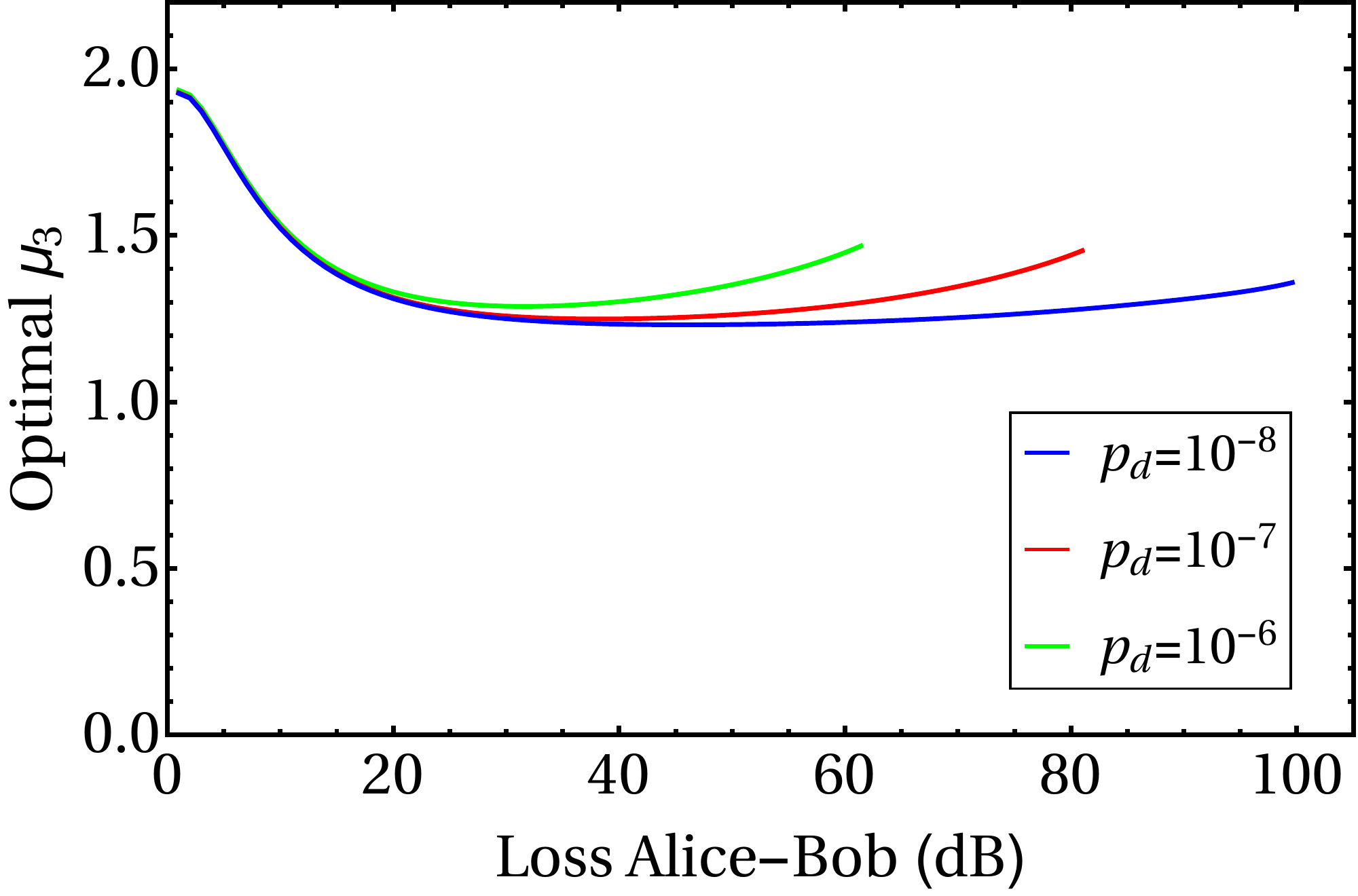}
		\caption{Optimal values of the decoy intensity $\mu_3$ as a function of the loss between Alice and Bob for three different dark count rates. These values are obtained from the optimization of the secret key rate (solid lines) of \autoref{key-rate-4decoys}. The other decoy intensities are set to: $\mu_0=10^{-1},\mu_1=10^{-2}$ and $\mu_2=10^{-3}$. The difference between this plot and the optimal $\mu_0$ plots (see \autoref{optimal-mu0-2decoys} and \autoref{optimal-mu0-3decoys}) in the case of two and three decoys is due to the fact that, unlike $\mu_0$, the intensity $\mu_3$ does not appear in all the yields' bounds since we used the fourth decoy just for bounding $Y_{13},Y_{31},Y_{04}$ and $Y_{40}$.}
		\label{optimal-mu3-4decoys}
	\end{subfigure}
	\caption{Optimal values of the signal and decoy intensities $\alpha^2$ and $\mu_3$ for the TF-QKD protocol \cite{Curty-security-TF} when the parties have at their disposal four decoy intensity settings to estimate the yields. }
	\label{optimal-intensities-4decoys}
\end{figure}
In \autoref{key-rate-4decoys} we plot the secret key rate against the overall loss for the case where Alice and Bob use four decoy intensity settings each. Like in the three-decoys case, the solid lines are obtained by bounding from above the yields $Y_{00},Y_{02},Y_{20},Y_{11},Y_{13},Y_{31},Y_{04},Y_{40}$ and $Y_{22}$ by means of four decoys. In particular, for the yields $Y_{00},Y_{02},Y_{20},Y_{11}$ and $Y_{22}$ we use the exact same analytical bounds derived with three decoys since they are tight enough, and the use of a fourth decoy intensity would just make them more cumbersome without providing a significant improvement of the resulting secret key rate. For the remaining four yields we instead derived tighter bounds with the help of the fourth intensity $\mu_3$ (see \ref{yields-bounds-4decoys}). The solid lines are obtained by optimizing the rate over the signal intensity $\alpha^2$ and the fourth decoy intensity $\mu_3$. It turns out that the optimal values for the other decoy intensities are basically the lowest possible for any value of the loss, so, as explained above, for simplicity we fix the smallest one to an experimentally reasonable small value (say  $\mu_2=10^{-3}$), and then we differentiate it from the other two decoys, $\mu_1$ and $\mu_0$, by one order of magnitude, i.e., we take $\mu_1=10^{-2}$ and $\mu_0=10^{-1}$. Importantly, this decision has a neglectable effect on the resulting secret key rate, when compared to that obtained by optimizing over all intensity settings. The optimal values for $\alpha^2$ and $\mu_3$ are shown in \autoref{optimal-alpha-4decoys} and \autoref{optimal-mu3-4decoys}, respectively. The dashed lines are the same as in \autoref{key-rate-2decoys} and \autoref{key-rate-3decoys}.\\
With four decoys (see \autoref{key-rate-4decoys}) the key rates basically reproduces the ideal ones (dashed lines) in which all the yields are known, with the gap being at maximum of 1 dB at the very end of the plot lines (i.e. in the very high loss regime). This demonstrates that there is no need to bound further yields than the nine yields we bounded in the cases of three and four decoys. Of course, the tighter estimation of the yields $Y_{13},Y_{31},Y_{04}$ and $Y_{40}$ achieved with four decoys results in an improvement of the key rate with respect to the case of three decoys (see \autoref{key-rate-3decoys}), especially in the region of high losses.\\
Concerning the optimal signal intensity (see \autoref{optimal-alpha-4decoys}), we notice a slight increase with respect to the three-decoys case (see \autoref{optimal-alpha-3decoys}) due to the tighter estimation of some yields in the phase-error rate formula, which allows their correspondent coefficients to acquire a slightly higher value under an increase of $\alpha^2$.\\
Finally, the reason why the optimal $\mu_3$ plot (see \autoref{optimal-mu3-4decoys}) looks quite different (with values above $1$) from the optimal $\mu_0$ plots for the cases of two and three decoys (see \autoref{optimal-mu0-2decoys} and \autoref{optimal-mu0-3decoys}) is the following. In the TF-QKD protocol considered, the most important yields (i.e., those with a bigger impact on the resulting phase error rate) are those associated to pairs of pulses with zero or with a very low number of photons. It is therefore very important to be able to estimate these yields as tightly as possible. For this, we have that the optimal intensities $\mu_0$ and $\mu_1$ ($\mu_0$, $\mu_1$ and $\mu_2$) for the case with two (three) decoys are well below $1$, just like in standard decoy-state QKD protocols~\cite{decoy-state-method-Lo,decoy-state-method-Wang}. However, as explained above, here we use the intensity $\mu_3$ to improve the upper bounds for the yields $Y_{13}$, $Y_{31}$, $Y_{04}$ and $Y_{40}$. That is, the intensity $\mu_3$ is only used to estimate yields associated to pairs of pulses with a total number of photons equal to four. Thus, it is natural that the optimal value of $\mu_3$ is not too low and greater than $1$. 

\subsection{Intensity fluctuations}  \label{intensity-fluct}
\begin{figure}[!b]
	\centering
	\fbox{\includegraphics[width=0.65\linewidth,keepaspectratio]{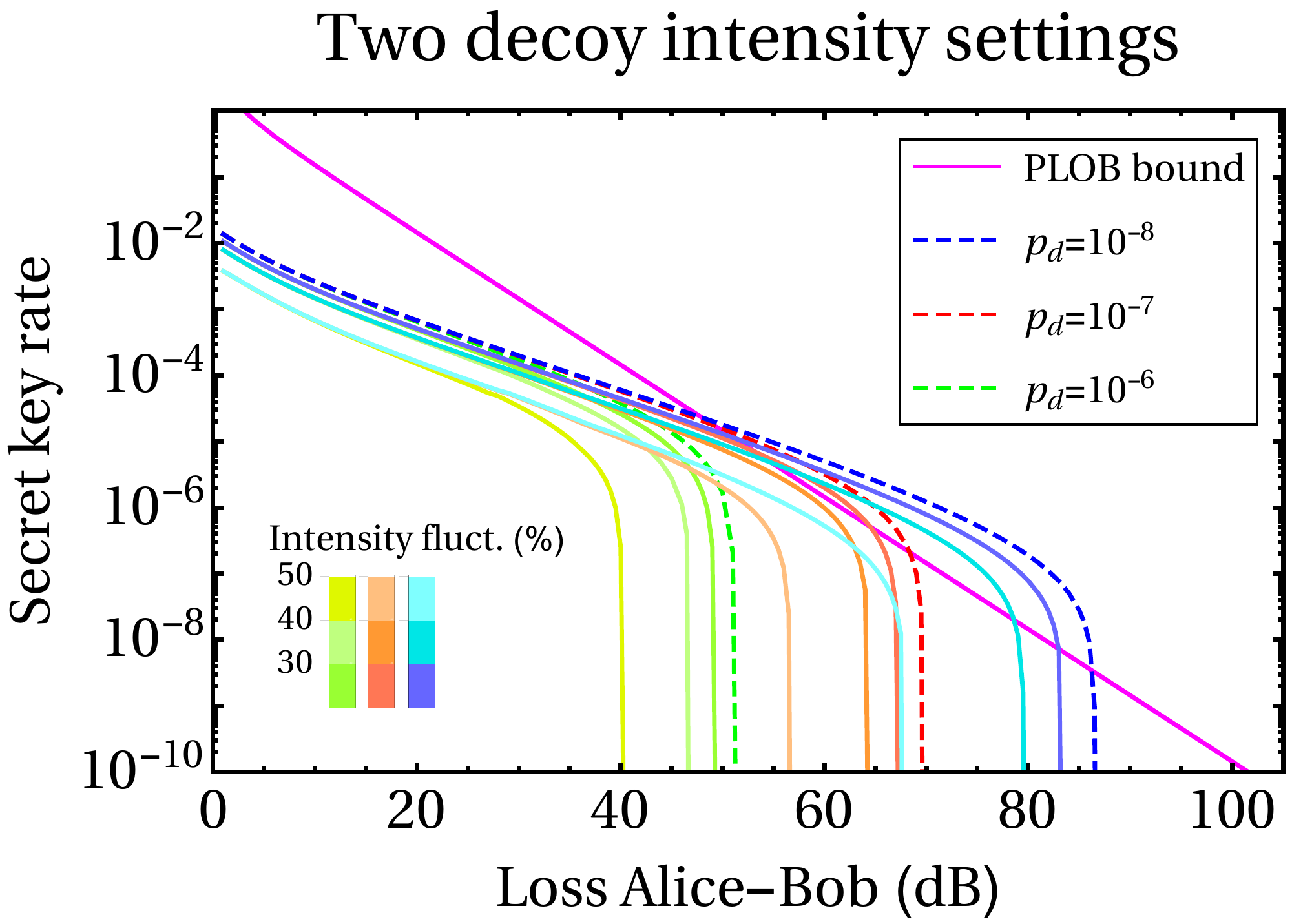}}
\end{figure}
\begin{figure}[!htb]
	\centering
	\fbox{\includegraphics[width=0.65\linewidth,keepaspectratio]{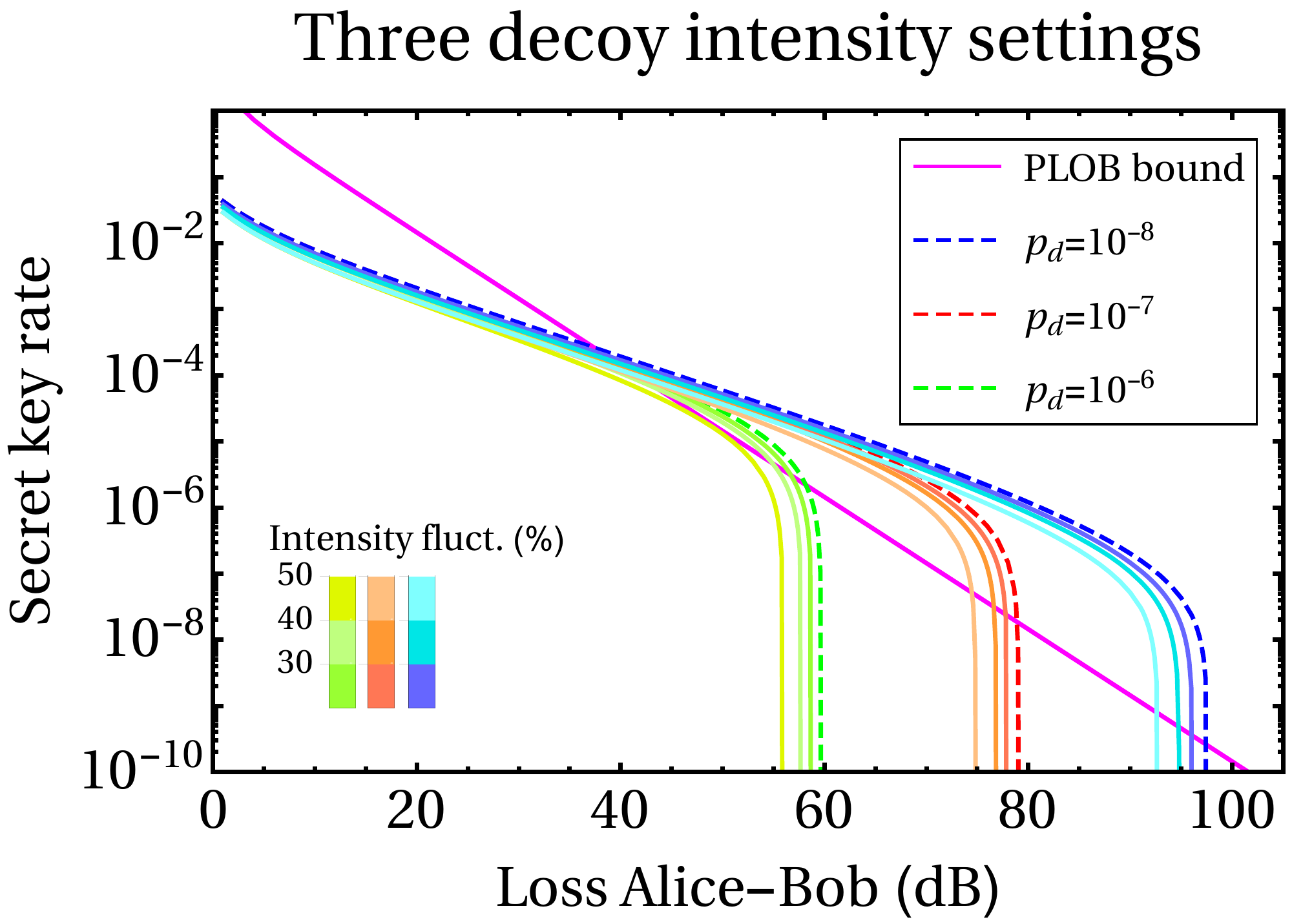}}
	\fbox{\includegraphics[width=0.65\linewidth,keepaspectratio]{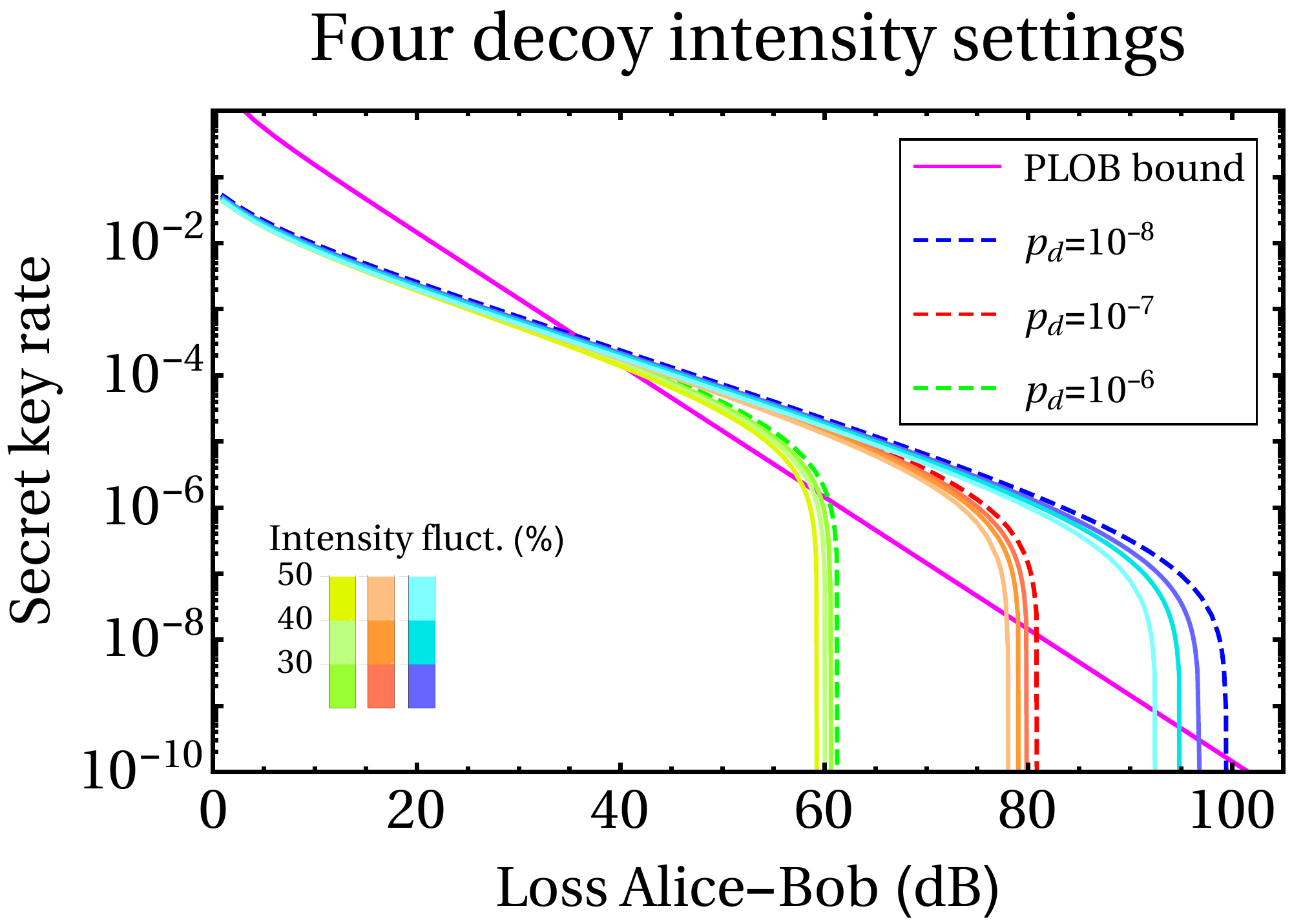}}
	\caption{Comparison of the secret key rate with optimal signal and decoys intensities (dashed lines, computed in \autoref{simulations}) with the secret key rates affected by increasing intensity fluctuations (solid lines): 30\%,40\% and 50\% (brighter colors; right to left). We assume that the fluctuations affect each decoy intensity and the signal intensity as well. The plots show that the TF-QKD protocol is quite robust against intensity fluctuations, and that its robustness increases with the number of decoys.}
	\label{int-fluct}
\end{figure}
Here we investigate the robustness of the TF-QKD protocol against intensity fluctuations that may occur in the preparation of the pulses sent by Alice and Bob. This is motivated by the fact that the optimal signal and decoy intensities that the parties should adopt in order to maximize the key rate for a given loss are quite small, thus the effect of intensity fluctuations might be an issue in practice. On the other hand, we also note that the optimal value of a given decoy or signal intensity is either constant or varies very moderately with the loss.\\
Here we consider the simple scenario in which the intensity fluctuations are symmetric, i.e., we assume that the intensity of Alice's signal matches perfectly with the intensity of Bob's signal. Or, to put it in other words, we consider that Alice's and Bob's signals suffer from the same intensity fluctuations and thus their intensities are equal. This means that such analysis is only valid to evaluate auto-compensating TF-QKD set-ups like, for instance, the one introduced in~\cite{experiment-Toronto}. It cannot be used however to analyze set-ups where more than one laser source is used~\cite{experiment-Toshiba,experiment-chinese}. Although we do not expect a dramatic change of our results when asymmetric intensity fluctuations are considered in the latter case, specially if they are not too large.
\\
Also, we assume that the signal and all the decoy intensities suffer from a fluctuation of magnitude 30\%, 40\% or 50\% around their optimal value. This means for example that, for a fluctuation say of 30\%, the signal intensity $\alpha^2$ and all the decoy intensities $\mu_k$ fluctuate in the intervals: $0.7\, \alpha_{\mathrm{opt}}^2 \leq \alpha^2 \leq 1.3\, \alpha_{\mathrm{opt}}^2$ and $0.7\, \mu^{\mathrm{opt}}_{k} \leq \mu_k \leq 1.3\, \mu^{\mathrm{opt}}_{k}$, respectively, where $\alpha^2_{\mathrm{opt}}$ and $\mu^{\mathrm{opt}}_{k}$ represent the optimal values. We then account for the worst-case scenario by numerically minimizing the key rate over all the intensities constrained in their respective fluctuation interval. Only in this way we can still guarantee that the resulting key rate is associated to a secure protocol.\\
The results of this study are given in \autoref{int-fluct}. Here we plot the original key rates --i.e. without fluctuations of the signal and decoy intensities-- as dashed lines\footnote{The dashed lines of the key rates without fluctuations correspond to the solid lines in \autoref{key-rate-2decoys}, \autoref{key-rate-3decoys} and \autoref{key-rate-4decoys}.} and the key rates affected by intensity fluctuations as solid lines. The plots are given for the same dark count rates and misalignments used in \autoref{simulations}, in the case of two, three and four decoy intensity settings. The color of the solid lines becomes brighter for increasing fluctuation magnitude.\\
We observe that the performance of the protocol is considerably affected by intensity fluctuations in the case of two decoys, while the effect becomes almost negligible for three and four decoys. The reason for this lies in the fact that the tightness of the yields' bounds has a stronger dependence on the value of the decoy intensities when the number of decoys --and thus constraints on the yields-- is low. In other words, if the parties have at their disposal a larger number of decoys, they can properly combine the numerous constraints on the yields and obtain inherently tight bounds, i.e. bounds that are tight regardless of the actual values of the intensities involved. If, instead, the parties have few decoys, say two, then the bounds they derive on the yields can be tight or loose depending on the values assigned to the decoy intensities, since the constraints on the yields are fewer.\\
In conclusion, in the case of two decoys the parties can tolerate intensity fluctuations up to 40\%, which correspond to a decrease in the protocol's key rate especially in the high-loss region, quantified by a reduction of about 5 to 6 dB of the maximum tolerated loss\footnote{By ``maximum tolerated loss'' we mean the loss threshold above which the protocol's key rate becomes roughly zero.}. Remarkably, with three decoys the decrease of the maximum tolerated loss would be under 5 dB for fluctuations up to 50\%. Finally, for four decoys the protocol's performance remains almost the same for fluctuations up to about 50\% around the optimal values (except when the dark count probability is the smallest considered: $p_d=10^{-8}$). We deduce that the TF-QKD protocol introduced in \cite{Curty-security-TF} seems to be quite robust against intensity fluctuations.

\section{Conclusions} \label{conclusions}

In this paper we have investigated in detail the performance of the Twin-Field quantum key distribution (TF-QKD) protocol presented in \cite{Curty-security-TF} in the realistic scenario of a finite number of decoy intensity settings at the parties' disposal. Indeed, the protocol requires that Alice and Bob use the decoy-state method \cite{decoy-state-method-Hwang,decoy-state-method-Lo,decoy-state-method-Wang} to estimate the phase-error rate by upper bounding certain yields. Unlike most QKD protocols which employ such method, in this case the protocol's key rate depends --in principle-- on infinitely many yields and it is essential to upper bound (rather than lower bound) their values. Clearly, the more yields the parties tightly upper bound, the better the protocol's performance is. We have introduced an analytical method to perform such estimation when Alice and Bob use two, three or four decoy intensity settings each. The yields' analytical bounds provided in this work imply a fully-analytical expression for the protocol's secret key rate, which is very convenient for performance optimization (e.g. in the finite-key scenario). Also, we remark that the secret key rates obtained with our analytical bounds basically overlap those achievable with numerical tools like linear programming for most values of the overall loss, which confirms that the analytical approach is actually quite tight.
\\ 
In so doing, we have shown that the TF-QKD protocol can beat the PLOB bound \cite{PLOB} even with just two decoys for reasonable values of the setup parameters, which include: the loss, the dark count rate, the polarization misalignment and the phase mismatch. Furthermore the plots assuming four decoys demonstrate that one can approximately achieve the best possible performance by tightly estimating only nine yields.
The optimization of the key rate over the signal and decoy intensities indicates that their optimal values are all either constant or weakly-dependent on the loss of the channel. This means that the protocol is particularly suitable for contexts where the channel loss varies in time, for instance in the scalable MDI-QKD networks conceived in \cite{asymmetric-MDI-QKD}. Finally we have investigated the scenario where the intensities of the optical states prepared by Alice and Bob are affected by fluctuations and observed that the protocol seems to be very robust against such phenomena.\\
A natural continuation of this work would take into account the finite-key effects due to the finite number of pulses sent by the parties to the central relay. This could be done by combining the results presented in this paper with the finite-keys estimation techniques used in \cite{Curty-finitekey}.

\ack
We thank Dagmar Bru\ss\ and Hermann Kampermann for helpful discussions, and an anonymous referee for very useful comments.
This project has received funding from the European Union's Horizon 2020 research and innovation programme under the Marie Skłodowska-Curie grant agreement No 675662, and from the Spanish Ministry of Economy and Competitiveness (MINECO), the Fondo Europeo de Desarrollo Regional (FEDER) through grant TEC2017-88243-R.

	\appendix
	\section{Channel model}  \label{channel-model}
	The channel model that we employ to simulate the gains that would be observed experimentally in the $X$-basis (i.e. the probabilities $p(k_c,k_d|b_A,b_B)$) and $Z$-basis (i.e. the probabilities $Q^{k,l}_{k_c,k_d}$)  is taken from \cite{Curty-security-TF}. In all the expressions of this Section we assume $k_c+k_d=1$.\\
	In particular, a beam splitter of transmittance $\sqrt{\eta}$ accounts for the loss in the quantum channel linking Alice (Bob) to node $C$ and for the non-unity detection efficiency of detectors $D_c$ and $D_d$. The polarization misalignment introduced by the channel Alice-$C$ (Bob-$C$) is modeled with a unitary operation mapping the polarization input modes $a^\dagger_{\mathrm{in}}$ ($b^\dagger_{\mathrm{in}}$) to the orthogonal polarization output modes $a^\dagger_{\mathrm{out}}$ and $a^\dagger_{\mathrm{out}\perp}$ ($b^\dagger_{\mathrm{out}}$ and $b^\dagger_{\mathrm{out}\perp}$) according to: $a^\dagger_{\mathrm{in}} \rightarrow \cos\theta_A a^\dagger_{\mathrm{out}}- \sin\theta_A a^\dagger_{\mathrm{out}\perp}$ ($b^\dagger_{\mathrm{in}} \rightarrow \cos\theta_B b^\dagger_{\mathrm{out}}- \sin\theta_B b^\dagger_{\mathrm{out}\perp}$), for an angle $\theta_A$ ($\theta_B$). Moreover, the phase mismatch between Alice and Bob's signals arriving at node $C$ is modeled by shifting the phase of Bob's signals by an angle $\phi=\delta \pi$, for a certain parameter $\delta$.
	Finally the model considers that both detectors are affected by a dark count probability $p_d$, which is independent of the signals received and has the same value for both detectors.\\
	With this setup, the gains in the $X$-basis can be written as:
	\begin{equation}
		p(k_c,k_d|b_A,b_B)= (1-p_d)\left[p_d e^{-2\gamma} + q(k_c,k_d|b_A,b_B)\right] \,\,,  \label{gainsX}
	\end{equation}
	where $\gamma=\sqrt{\eta}\alpha^2$ (with $\alpha$ being the amplitude of the signal states) and
	\begin{equation}
		q(k_c,k_d|b_A,b_B)= \left\{
		{\begin{array}{lcl}
		e^{-\gamma(1-\cos\phi \cos\theta)}-e^{-2\gamma}	 & \mbox{if} & k_c \oplus b_A \oplus b_B =1 \\
		e^{-\gamma(1+\cos\phi \cos\theta)}-e^{-2\gamma}	 & \mbox{if} & k_c \oplus b_A \oplus b_B =0  
			\end{array}}
		\right.   \label{gainsX-nodarkcount}
	\end{equation}
	with $\theta=\theta_A -\theta_B$. Starting from (\ref{gainsX}), one can readily compute the probability $p(k_c,k_d)$ and the bit-error rate $e_{k_c,k_d}$ by means of equations (\ref{prob-kckd}) and (\ref{e10}),(\ref{e01}), respectively:
	\begin{eqnarray}
		p(k_c,k_d) &= \frac{1}{2}(1-p_d)\left(e^{-\gamma\cos\phi \cos\theta}+e^{\gamma\cos\phi \cos\theta}\right)e^{-\gamma} -(1-p_d)^2 e^{-2\gamma} \label{prob-kckd-model}\,\,, \\
		e_{k_c,k_d} &= \frac{e^{-\gamma\cos\phi \cos\theta}-(1-p_d)e^{-\gamma}}{e^{-\gamma\cos\phi \cos\theta}+e^{\gamma\cos\phi \cos\theta} -2(1-p_d)e^{-\gamma}}
		\label{bit-error-rate-model} \,\,.
	\end{eqnarray}
	The gains in the $Z$-basis instead read:
	\begin{equation}
		Q^{k,l}_{k_c,k_d}= (1-p_d)\left[(p_d -1) e^{-\sqrt{\eta}(\mu_k+\mu_l)} + e^{-\sqrt{\eta}(\mu_k+\mu_l)/2} I_0 (\sqrt{\eta \mu_k \mu_l}\cos\theta)\right] \label{gainsZ} \,\,,
	\end{equation}
	where the function $I(z)=\frac{1}{2\pi i} \oint e^{(z/2)(t+1/t)}t^{-1}dt$ is the modified Bessel function of first kind.\\
	In the simulations shown in \autoref{simulations} we compare the key rate computed with our analytical bounds on the yields with the key rate evaluated with the exact expressions of the yields, i.e. the expressions obtained directly from the channel model. According to the above channel model, the yields read:
	\begin{equation}
		Y^{k_c,k_d}_{nm}= (1-p_d)\left[(p_d -1)(1-\sqrt{\eta})^{n+m} + y^{k_c,k_d}_{nm} \right] \label{yields-model} \,\,,
	\end{equation}
	where
	\begin{eqnarray}
		&\fl y^{k_c,k_d}_{nm}=\sum_{k=0}^{n} {n\choose k} \sum_{l=0}^{m} {m\choose l} \frac{\sqrt{\eta}^{k+l}(1-\sqrt{\eta})^{n+m-k-l}}{2^{k+l} k!l!}\sum_{r=0}^{k} {k\choose r} \sum_{p=0}^{l} {l \choose p} \sum_{q=\max(0,r+p-l)}^{\min(k,r+p)} {k \choose q} \nonumber\\
		&\fl {l \choose r+p-q} (r+p)!(k+l-r-p)! \cos^{r+q}(\theta_A)\cos^{r+2p-q}(\theta_B) \sin^{2k-r-q} (\theta_A) \sin^{2l-r-2p+q} (\theta_B) 
		\,\,. \nonumber\\
		\fl \label{yields2-model} 
	\end{eqnarray}
	To conclude, we remark that all the quantities entering the key rate formula (\ref{key-rate}) --i.e. (\ref{prob-kckd-model}),(\ref{bit-error-rate-model}) and the gains (\ref{gainsZ}) indirectly through the yields' bounds-- are symmetric under the swap $k_c \leftrightarrow k_d$ due to the symmetries of the channel model.\\
	In all the simulations shown in \autoref{simulations} we fix both polarization and phase misalignments to 2\%, which means that: $\theta_A=-\theta_B=\arcsin\sqrt{0.02}$ and $\delta=0.02$.
	
	\section{Stronger and weaker decoy intensities} \label{stronger_decoys}
	\begin{figure}[!htb]
		\centering
		\begin{subfigure}[t]{.5\textwidth}
			\centering
			\includegraphics[width=1\linewidth,keepaspectratio]{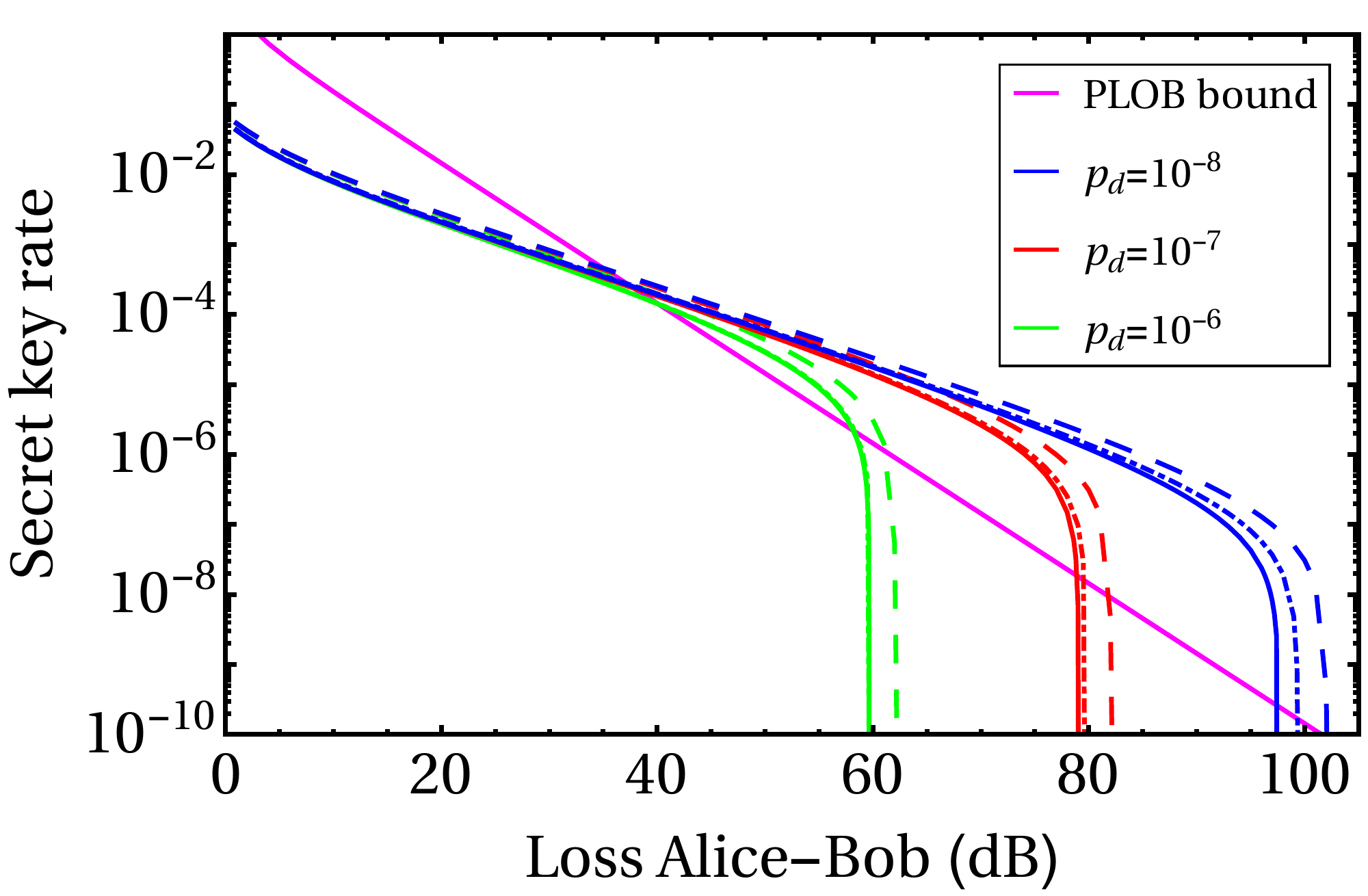}
			\caption{Optimal key rate as a function of the overall loss when the parties use three decoy intensity settings, for three different values of the dark count rate ($p_d$). The solid lines are obtained by fixing the weaker decoy intensities to $\mu_2=10^{-3}$ and $\mu_1=10^{-2}$, while the dot-dashed lines are obtained by fixing the same intensities to $\mu_2=10^{-5}$ and $\mu_1=10^{-4}$. The dashed lines assume that all the yields are known from the channel model and the magenta line is the PLOB bound \cite{PLOB}. Note that the green dot-dashed lines and green solid lines ($p_d=10^{-6}$) are almost perfectly overlapping.}
			\label{3strongerdecoys-vs-3decoys}
		\end{subfigure}%
		\begin{subfigure}[t]{.5\textwidth}
			\centering
			\includegraphics[width=1\linewidth,keepaspectratio]{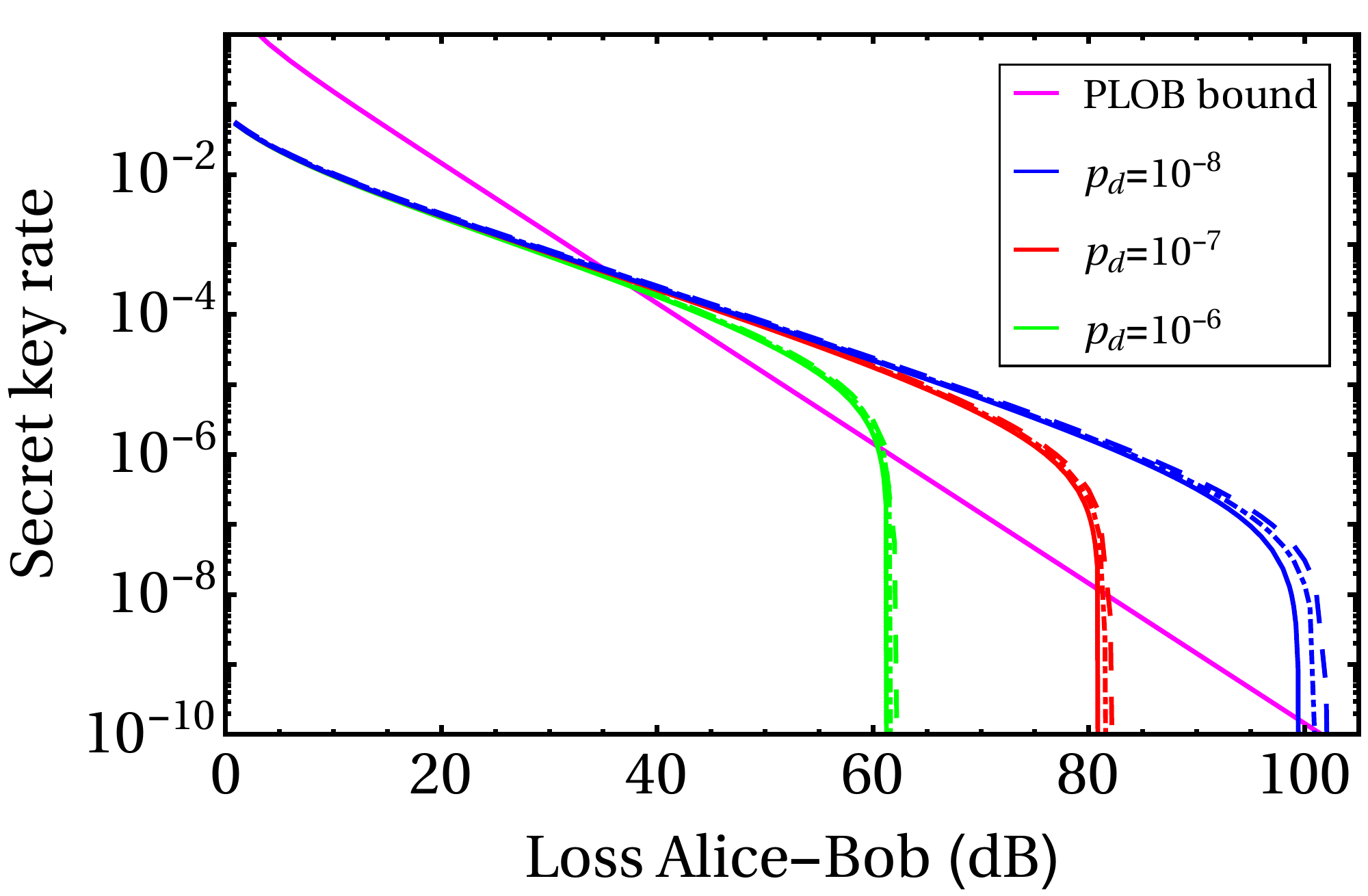}
			\caption{Optimal key rate as a function of the overall loss when the parties use four decoy intensity settings, for three different values of the dark count rate ($p_d$). The solid lines are obtained by fixing the weaker decoy intensities to $\mu_2=10^{-3}$, $\mu_1=10^{-2}$ and $\mu_0=10^{-1}$, while the dot-dashed lines are obtained by fixing the same intensities to $\mu_2=10^{-5}$, $\mu_1=10^{-4}$ and $\mu_0=10^{-3}$. The dashed lines assume that all the yields are known from the channel model and the magenta line is the PLOB bound \cite{PLOB}.}
			\label{4strongerdecoys-vs-4decoys}
		\end{subfigure}
		\caption{Comparison of the optimal key rates achievable with different fixed values of the weaker decoy intensities. The two cases analyzed (solid and dot-dashed lines) are almost indistinguishable.}
		\label{stronger-vs-weaker_decoys}
	\end{figure}
	As explained in \autoref{simulations}, the optimal key rates are basically not affected if their optimization is only performed over the signal intensity ($\alpha$) and over one decoy intensity, while having the remaining weaker decoy intensities fixed to near-to-optimal values for all losses. In \autoref{stronger-vs-weaker_decoys}, we compare the optimal key rate that the parties can achieve when fixing their weaker decoy intensities to substantially different values, in the case of three (left) and four (right) decoy intensity settings. In particular, the solid lines are the same plotted in \autoref{key-rate-3decoys} and \autoref{key-rate-4decoys} for the three- and four-decoys case, respectively, i.e. they are obtained by fixing the weaker decoy intensities to $\mu_2=10^{-3}$ and $\mu_1=10^{-2}$ (three decoy intensity settings) and to $\mu_2=10^{-3}$, $\mu_1=10^{-2}$ and $\mu_0=10^{-1}$ (four decoy intensity settings). The dot-dashed lines, instead, are obtained by fixing the weaker intensities to values which are two orders of magnitude lower, that is $\mu_2=10^{-5}$ and $\mu_1=10^{-4}$ in the case of three decoy intensity settings and $\mu_2=10^{-5}$, $\mu_1=10^{-4}$ and $\mu_0=10^{-3}$ in the case of four decoy intensity settings. Clearly, the optimal key rates are basically not affected by employing relatively stronger pulses (those with $\mu_2=10^{-3}$ as the weakest intensity) for the weaker decoy intensity settings. Such stronger pulses could be more easily implemented experimentally and, for this, have been chosen in our simulations.
	
	\section{Yields' bounds with three decoys} \label{yields-bounds-3decoys}
	Here we derive analytical upper bounds on the yields appearing in (\ref{phase-error-rate}), following the same lines of \autoref{yields-bounds-2decoys}. In this case we assume that Alice and Bob can prepare their phase-randomized coherent pulses with three different intensity settings: $\{\mu_0,\mu_1,\mu_2\}$, which are the same for both parties. This choice is optimal since we assumed that the two optical channels linking the parties to the central node $C$ have equal transmittance $\sqrt{\eta}$ \cite{asymmetric-MDI-QKD}.\\
	The whole set of infinite yields is subjected to the following nine equality constraints:
	\begin{equation}
	\tilde{Q}^{k,l} \equiv e^{\mu_k + \mu_l} Q^{k,l} =\sum_{n,m=0}^{\infty} \frac{Y_{nm}}{n!m!} {\mu_k}^n {\mu_l}^m \quad k,l \in \{0,1,2\}\,\,, \label{constr-3decoys}
	\end{equation}
	and to the inequality constraints given by (\ref{ineq-constr}).\\
	We derive bounds on the yields $Y_{00},Y_{11},Y_{02},Y_{20},Y_{22},Y_{13},Y_{31},Y_{04}$ and $Y_{40}$.
	
	\subsection{Upper bound on $Y_{22}$} \label{bound_on_Y22-3decoys}
	Consider the following combinations of gains in which all the terms $Y_{1m}$ and $Y_{n1}$ are removed (i.e. their coefficients are equal to zero):
	\begin{eqnarray}
		G_{22}^{0,1} &= \mu^2_1 \tilde{Q}^{0,0}+\mu^2_0 \tilde{Q}^{1,1} -\mu_0 \mu_1 (\tilde{Q}^{0,1} +\tilde{Q}^{1,0}) \nonumber\\
		&= \sum_{n,m=0}^{\infty} \frac{Y_{nm}}{n!m!} \left(\mu^n_0 \mu_1 -\mu_0 \mu^n_1\right) \left(\mu^m_0 \mu_1 -\mu_0 \mu^m_1\right) \,;\nonumber\\
		G_{22}^{0,2} &= \mu^2_2 \tilde{Q}^{0,0}+\mu^2_0 \tilde{Q}^{2,2} -\mu_0 \mu_2 (\tilde{Q}^{0,2} +\tilde{Q}^{2,0}) \nonumber\\
		&= \sum_{n,m=0}^{\infty} \frac{Y_{nm}}{n!m!} \left(\mu^n_0 \mu_2 -\mu_0 \mu^n_2\right) \left(\mu^m_0 \mu_2 -\mu_0 \mu^m_2\right) \,;\nonumber\\
		G_{22}^{1,2} &= \mu^2_2 \tilde{Q}^{1,1}+\mu^2_1 \tilde{Q}^{2,2} -\mu_1 \mu_2 (\tilde{Q}^{1,2} +\tilde{Q}^{2,1}) \nonumber\\
		&= \sum_{n,m=0}^{\infty} \frac{Y_{nm}}{n!m!} \left(\mu^n_1 \mu_2 -\mu_1 \mu^n_2\right) \left(\mu^m_1 \mu_2 -\mu_1 \mu^m_2\right)  \label{G-22}\,\,,
	\end{eqnarray}
	where the superscripts in $G_{22}^{k,l}$ indicate which intensities are involved, while the subscripts indicate the yield that is going to be bounded.\\
	We now combine $G_{22}^{0,1},G_{22}^{0,2}$ and $G_{22}^{1,2}$ with arbitrary real coefficients $c_0$ and $c_1$ and impose that the resulting expression has the yields $Y_{0m}$ and $Y_{n0}$ removed as well:
	\begin{eqnarray}
	\fl	G_{22}^{0,1}+c_0 \,G_{22}^{0,2}+c_1 \,G_{22}^{1,2} &=  \sum_{n,m=0}^{\infty} \frac{Y_{nm}}{n!m!} \left[\left(\mu^n_0 \mu_1 -\mu_0 \mu^n_1\right) 
		\left(\mu^m_0 \mu_1 -\mu_0 \mu^m_1\right) \right. \nonumber\\
	\fl	&\left. +c_0 \left(\mu^n_0 \mu_2 -\mu_0 \mu^n_2\right) \left(\mu^m_0 \mu_2 -\mu_0 \mu^m_2\right) 
		+c_1 \left(\mu^n_1 \mu_2 -\mu_1 \mu^n_2\right) \left(\mu^m_1 \mu_2 -\mu_1 \mu^m_2\right)\right] \,. \label{combG-22}
	\end{eqnarray}
	Note that the linear combination above is already the most general for our needs. As a matter of fact, for every linear combination of $G_{22}^{0,1},G_{22}^{0,2}$ and $G_{22}^{1,2}$ one can always factor out the coefficient in front of $G_{22}^{0,1}$, as far as it is not zero. However, if the particular combination of gains which removes the terms $Y_{0m}$ and $Y_{n0}$ has a null coefficient in front of $G_{22}^{0,1}$, for symmetry reasons there would also exist another combination --that also removes the yields $Y_{0m}$ and $Y_{n0}$-- with a null coefficient in front of say $G_{22}^{0,2}$, and this one could be found in our case given by (\ref{combG-22}).\\
	For $Y_{0m}$ and $Y_{n0}$ to be removed in (\ref{combG-22}) it suffices that:
	\begin{eqnarray}
	\fl	(\mu_1-\mu_0)(\mu^m_0 \mu_1 -\mu_0 \mu^m_1)+c_0 (\mu_2-\mu_0)(\mu^m_0 \mu_2 -\mu_0 \mu^m_2)+c_1(\mu_2-\mu_1)(\mu^m_1 \mu_2 -\mu_1 \mu^m_2) = 0 \quad \forall \, m\,,
	\end{eqnarray}
	which implies:
	\begin{eqnarray}
		&\mu_0^m\left[\mu_1(\mu_1-\mu_0)+ c_0 \mu_2(\mu_2 -\mu_0)\right] + \mu_1^m\left[-\mu_0(\mu_1-\mu_0)+ c_1 \mu_2(\mu_2 -\mu_1)\right] \nonumber\\
		&+ \mu_2^m\left[-c_0 \mu_0(\mu_2-\mu_0) - c_1 \mu_1 (\mu_2 -\mu_1)\right] = 0 	\quad  \forall\, m \,\,.
	\end{eqnarray}
	A sufficient condition for this is that every coefficient of $\mu_i^m$ is identically zero, which happens for:
	\begin{eqnarray}
		c_0 &= -\frac{\mu_1(\mu_0-\mu_1)}{\mu_2(\mu_0-\mu_2)} \label{c0-22}\,\,, \\
		c_1 &=\frac{ \mu_0(\mu_0-\mu_1)}{\mu_2(\mu_1-\mu_2)}  \label{c1-22} \,\,.
	\end{eqnarray}
	Substituting (\ref{c0-22}) and (\ref{c1-22}) back into (\ref{combG-22}) and multiplying both sides by $\mu_2$, we get an expression where all the terms $Y_{0m},Y_{1m},Y_{n0}$ and $Y_{n1}$ are removed and where the term $Y_{22}$ gives the largest contribution:
	\begin{eqnarray}
	\fl	&\mu_2 G_{22}^{0,1} -\mu_1\frac{(\mu_0-\mu_1)}{(\mu_0-\mu_2)}\,G_{22}^{0,2} + \mu_0\frac{(\mu_0-\mu_1)}{(\mu_1-\mu_2)}\,G_{22}^{1,2} = 
		  \sum_{n,m=2}^{\infty} \frac{Y_{nm}}{n!m!} \bigg[\mu_2 \left(\mu^n_0 \mu_1 -\mu_0 \mu^n_1\right)\left(\mu^m_0 \mu_1 -\mu_0 \mu^m_1\right) \nonumber\\
	\fl	  &-\mu_1\frac{(\mu_0-\mu_1)}{(\mu_0-\mu_2)}\left(\mu^n_0 \mu_2 -\mu_0 \mu^n_2\right) \left(\mu^m_0 \mu_2 -\mu_0 \mu^m_2\right)+
		   \mu_0\frac{(\mu_0-\mu_1)}{(\mu_1-\mu_2)}\left(\mu^n_1 \mu_2 -\mu_1 \mu^n_2\right) \left(\mu^m_1 \mu_2 -\mu_1 \mu^m_2\right)\bigg] \,\,. \label{comb2G-22}
	\end{eqnarray}
	In order to extract a bound for $Y_{22}$ we need to recast the yields' coefficients in such a way that their sign becomes manifest. Each term of the sum in (\ref{comb2G-22}) may be recast as follows:
	\begin{eqnarray}
	\fl	\frac{Y_{nm}}{n!m!}\mu_0 \mu_1 \mu_2\bigg[\left(\mu^{n-1}_0 -\mu^{n-1}_1\right)\left(\mu^m_0 \mu_1 -\mu_0 \mu^m_1\right) 
		&-\frac{(\mu_0-\mu_1)}{(\mu_0-\mu_2)}\left(\mu^{n-1}_0 -\mu^{n-1}_2\right) \left(\mu^m_0 \mu_2 -\mu_0 \mu^m_2\right) \nonumber\\
	\fl	&+\frac{(\mu_0-\mu_1)}{(\mu_1-\mu_2)}\left(\mu^{n-1}_1 - \mu^{n-1}_2\right) \left(\mu^m_1 \mu_2 -\mu_1 \mu^m_2\right) \bigg] \,\,,
	\end{eqnarray}
	or equivalently as:
	\begin{eqnarray}
		\frac{Y_{nm}}{n!m!}\frac{\mu_0 \mu_1 \mu_2}{(\mu_0-\mu_2)(\mu_1-\mu_2)} \,\,A_{22}(\mu_0,\mu_1,\mu_2,m)\cdot A_{22}(\mu_0,\mu_1,\mu_2,n) \,\,, \label{Ynm-22}
	\end{eqnarray}
	where
	\begin{eqnarray}
		A_{22}(\mu_0,\mu_1,\mu_2,m) &\equiv \mu_1^m (\mu_0 - \mu_2) + \mu_2^m (\mu_1 - \mu_0) + \mu_0^m (\mu_2 - \mu_1) \label{A-22}  \,\,.
	\end{eqnarray}
	We can now rewrite factor $A_{22}$ as:
	\begin{eqnarray}
	\fl	&A_{22}(\mu_0,\mu_1,\mu_2,m) = \mu_1 \left[\mu_1^{m-1}(\mu_0 - \mu_2)-(\mu_0^m - \mu_2^m)\right]+ \mu_0 \mu_2 (\mu_0^{m-1}-\mu_2^{m-1}) \nonumber\\
	\fl	&= \mu_1 \left[\mu_1^{m-1}(\mu_0 - \mu_2)-(\mu_0 - \mu_2)\left(\sum_{k=0}^{m-1}\mu_0^{m-1-k}\mu_2^k\right)\right]+ 
		\mu_0 \mu_2 (\mu_0-\mu_2)\left(\sum_{j=0}^{m-2}\mu_0^{m-2-j}\mu_2^j\right) \nonumber\\
	\fl	&= (\mu_0 - \mu_2) \left[\mu_1^m -\mu_1 \sum_{k=0}^{m-1}\mu_0^{m-1-k}\mu_2^k +\mu_0 \mu_2 \sum_{j=0}^{m-2}\mu_0^{m-2-j}\mu_2^j \right] \nonumber\\
	\fl	&= (\mu_0 - \mu_2) \left[\mu_1^m + \sum_{k=0}^{m-1} \mu_2^k \left(-\mu_1 \mu_0 ^{m-1-k} + \mu_0\mu_2\mu_0^{m-2-k}\right) -\mu_0\mu_2\frac{\mu_2^{m-1}}{\mu_0} \right]
		   \nonumber\\
	\fl	&= (\mu_0 - \mu_2) \left[-(\mu_2^m - \mu_1^m) + \sum_{k=0}^{m-1} \mu_2^k \mu_0^{m-1-k} (\mu_2 - \mu_1)\right] \nonumber\\
	\fl	&= (\mu_0 - \mu_2)(\mu_2 - \mu_1)\left[\sum_{k=0}^{m-1} \mu_2^k \mu_0^{m-1-k} - \sum_{j=0}^{m-1} \mu_2^j \mu_1^{m-1-j}\right]  \nonumber\\
	\fl	&= (\mu_0 - \mu_2)(\mu_2 - \mu_1)\sum_{k=0}^{m-1} \mu_2^k (\mu_0^{m-1-k}-\mu_1^{m-1-k})  \label{A-22_1} \,\,.
	\end{eqnarray}
	Of course we can employ this expression also for $A_{22}(\mu_0,\mu_1,\mu_2,n)$, under the substitution $m\rightarrow n$. We will apply this consideration from now on to similar scenarios. 
	By substituting (\ref{A-22_1}) into (\ref{Ynm-22}), we get the final expression for each term of the sum in (\ref{comb2G-22}):
	\begin{eqnarray}
	\fl	&\frac{Y_{nm}}{n!m!}\frac{\mu_0 \mu_1 \mu_2}{(\mu_0-\mu_2)(\mu_1-\mu_2)} (\mu_0 - \mu_2)^2(\mu_2 - \mu_1)^2 \nonumber\\
		 &\times \,  \left[\sum_{k=0}^{m-1} \mu_2^k (\mu_0^{m-1-k}-\mu_1^{m-1-k})\right] \left[\sum_{j=0}^{n-1} \mu_2^j (\mu_0^{n-1-j}-\mu_1^{n-1-j})\right] \label{Ynmcoeff-22} \,\,.
	\end{eqnarray}
	That is, the sign of $Y_{nm}$'s coefficient is independent of $n$ and $m$ and it is the same for all terms in (\ref{comb2G-22}) (note that the product of the two sums in (\ref{Ynmcoeff-22}) is always positive). Thus a valid upper bound for $Y_{22}$ is obtained by setting all the other yields to zero in (\ref{comb2G-22}), except for $Y_{22}$. We obtain:
	\begin{eqnarray}
	\fl	\mu_2 G_{22}^{0,1} -\mu_1\frac{(\mu_0-\mu_1)}{(\mu_0-\mu_2)}\,G_{22}^{0,2} + \mu_0\frac{(\mu_0-\mu_1)}{(\mu_1-\mu_2)}\,G_{22}^{1,2} =
		\frac{Y^U_{22}\mu_0 \mu_1 \mu_2}{4}(\mu_0-\mu_2)(\mu_1-\mu_2)(\mu_0 - \mu_1)^2 \,\,,
	\end{eqnarray}
	which implies the following expression for the upper bound on $Y_{22}$:
	\begin{eqnarray}
		Y^U_{22}=4\frac{\frac{G_{22}^{0,1}}{\mu_0 \mu_1 (\mu_0 - \mu_1)} -\frac{G_{22}^{0,2}}{\mu_0 \mu_2(\mu_0-\mu_2)} +
			 \frac{G_{22}^{1,2}}{\mu_1 \mu_2(\mu_1-\mu_2)}}{(\mu_0-\mu_1)(\mu_0-\mu_2)(\mu_1-\mu_2)}   \,\,. \label{Y22-upperbound-3decoys}
	\end{eqnarray}
	We remark that the bound given by (\ref{Y22-upperbound-3decoys}) is not valid when any of the intensities $\mu_0$, $\mu_1$ or $\mu_2$ is equal to zero. As a matter of fact, in any of these cases the starting expression given by (\ref{comb2G-22}) becomes trivial. However, in most practical situations, due to the finite extinction ratio of amplitude modulators, none of the decoy intensities is actually equal to zero.    
	
	\subsection{Upper bound on $Y_{11}$}
	Consider the following combinations of gains in which all the terms $Y_{0m}$ and $Y_{n0}$ are removed:
	\begin{eqnarray}
	G_{11}^{0,1} &= \tilde{Q}^{0,0}+ \tilde{Q}^{1,1} - (\tilde{Q}^{0,1} +\tilde{Q}^{1,0}) \nonumber\\
	&= \sum_{n,m=0}^{\infty} \frac{Y_{nm}}{n!m!} \left(\mu^n_0 - \mu^n_1\right) \left(\mu^m_0 -\mu^m_1\right) \,;\nonumber\\
	G_{11}^{0,2} &= \tilde{Q}^{0,0}+ \tilde{Q}^{2,2} - (\tilde{Q}^{0,2} +\tilde{Q}^{2,0}) \nonumber\\
	&= \sum_{n,m=0}^{\infty} \frac{Y_{nm}}{n!m!} \left(\mu^n_0 - \mu^n_2\right) \left(\mu^m_0 - \mu^m_2\right) \,; \nonumber\\
	G_{11}^{1,2} &= \tilde{Q}^{1,1}+ \tilde{Q}^{2,2} - (\tilde{Q}^{1,2} +\tilde{Q}^{2,1}) \nonumber\\
	&= \sum_{n,m=0}^{\infty} \frac{Y_{nm}}{n!m!} \left(\mu^n_1 - \mu^n_2\right) \left(\mu^m_1 - \mu^m_2\right)  \label{G-11-3decoys} \,\,.
	\end{eqnarray}
	We now combine $G_{11}^{0,1},G_{11}^{0,2}$ and $G_{11}^{1,2}$ with arbitrary real coefficients $c_0$ and $c_1$ and impose that the resulting expression has the yields $Y_{2m}$ and $Y_{n2}$ also removed:
	\begin{eqnarray}
	\fl G_{11}^{0,1} &+ c_0 \,G_{11}^{0,2}+c_1 \,G_{11}^{1,2} = \nonumber\\
	 \fl &\sum_{n,m=0}^{\infty} \frac{Y_{nm}}{n!m!} \left[(\mu^n_0 - \mu^n_1)(\mu^m_0 - \mu^m_1) +c_0 (\mu^n_0 - \mu^n_2)(\mu^m_0 - \mu^m_2)+ 
	 c_1 (\mu^n_1 - \mu^n_2)(\mu^m_1 - \mu^m_2)\right]  \label{combG-11} \,\,.
	\end{eqnarray}
	For $Y_{2m}$ and $Y_{n2}$ to be removed it suffices:
	\begin{eqnarray}
		(\mu^n_0 - \mu^n_1)(\mu^2_0 - \mu^2_1) +c_0 (\mu^n_0 - \mu^n_2)(\mu^2_0 - \mu^2_2)+	c_1 (\mu^n_1 - \mu^n_2)(\mu^2_1 - \mu^2_2) =0 \quad\forall\, n\,, 
	\end{eqnarray}
	which is fulfilled by:
	\begin{eqnarray}
	c_0 &= -\frac{(\mu_0^2-\mu_1^2)}{(\mu^2_0-\mu^2_2)} \label{c0-11} \,\,,\\
	c_1 &=\frac{(\mu^2_0-\mu^2_1)}{(\mu^2_1-\mu^2_2)}  \label{c1-11} \,\,.
	\end{eqnarray}
	Substituting these terms back into (\ref{combG-11}) yields a combination of gains in which the terms $Y_{0m},Y_{n0},Y_{2m}$ and $Y_{n2}$ are removed:
	\begin{eqnarray}
	\fl	&G_{11}^{0,1} -\frac{(\mu_0^2-\mu_1^2)}{(\mu^2_0-\mu^2_2)} \,G_{11}^{0,2}+\frac{(\mu^2_0-\mu^2_1)}{(\mu^2_1-\mu^2_2)} \,G_{11}^{1,2} = \nonumber\\
	\fl	&Y_{11} (\mu_0 - \mu_1)\left[(\mu_0 -\mu_1) -\frac{(\mu_0 +\mu_1)}{(\mu_0+\mu_2)}(\mu_0-\mu_2) +\frac{(\mu_0 +\mu_1)}{(\mu_1+\mu_2)}(\mu_1-\mu_2) \right] \nonumber\\
	\fl	&+\sum_{m=3}^{\infty} \frac{Y_{1m}}{m!}(\mu_0 -\mu_1)\left[(\mu^m_0 -\mu^m_1) -\frac{(\mu_0 +\mu_1)}{(\mu_0+\mu_2)}(\mu^m_0-\mu^m_2) 
		                                                           +\frac{(\mu_0 +\mu_1)}{(\mu_1+\mu_2)}(\mu^m_1-\mu^m_2) \right] \nonumber\\
	\fl	&+\sum_{n=3}^{\infty} \frac{Y_{n1}}{n!}(\mu_0 -\mu_1)\left[(\mu^n_0 -\mu^n_1) -\frac{(\mu_0 +\mu_1)}{(\mu_0+\mu_2)}(\mu^n_0-\mu^n_2) 
	                                                              	+\frac{(\mu_0+\mu_1)}{(\mu_1+\mu_2)}(\mu^n_1-\mu^n_2)\right] \nonumber\\
	 \fl   &+\sum_{n,m=3}^{\infty} \frac{Y_{nm}}{n!m!} \left[(\mu^n_0 - \mu^n_1)(\mu^m_0 - \mu^m_1) 
	    -\frac{(\mu_0^2-\mu_1^2)}{(\mu^2_0-\mu^2_2)} (\mu^n_0 - \mu^n_2)(\mu^m_0 - \mu^m_2)+ 
	    \frac{(\mu^2_0-\mu^2_1)}{(\mu^2_1-\mu^2_2)} (\mu^n_1 - \mu^n_2)(\mu^m_1 - \mu^m_2)\right]  \,\,. \nonumber\\
	  \fl   \label{comb2G-11}
	\end{eqnarray}
	In order to get a valid upper bound for $Y_{11}$ we need to determine the signs of the coefficients of the remaining yields. We start by recasting each term of the sum in (\ref{comb2G-11}) corresponding to the $Y_{nm}$, with $n,m\geq 3$, as follows:
	\begin{eqnarray}
		\frac{Y_{nm}}{n!m!} \frac{1}{(\mu^2_0-\mu^2_2)(\mu^2_1-\mu^2_2)} A_{11}(\mu_0,\mu_1,\mu_2,m)\cdot  A_{11}(\mu_0,\mu_1,\mu_2,n) \,\,, \label{Ynm-11}
	\end{eqnarray}
	where
	\begin{eqnarray}
		A_{11}(\mu_0,\mu_1,\mu_2,m) &\equiv \mu_1^m (\mu^2_0 - \mu^2_2) + \mu_2^m (\mu^2_1 - \mu^2_0) + \mu_0^m (\mu^2_2 - \mu^2_1) \label{A-11} \,\,.
	\end{eqnarray}
	The factor $A_{11}$ can be rewritten as:
	\begin{eqnarray}
\fl	&A_{11}(\mu_0,\mu_1,\mu_2,m) = \mu^2_1 \left[\mu_1^{m-2}(\mu^2_0 - \mu^2_2)-(\mu_0^m - \mu_2^m)\right]+ \mu^2_0 \mu^2_2 (\mu_0^{m-2}-\mu_2^{m-2}) \nonumber\\
\fl	&= (\mu_0 - \mu_2) \left[\mu_1^m(\mu_0+\mu_2)  -\mu^2_1 \sum_{k=0}^{m-1}\mu_0^{m-1-k}\mu_2^k +\mu^2_0 \mu^2_2 \sum_{j=0}^{m-3}\mu_0^{m-3-j}\mu_2^j \right] \nonumber\\
\fl	&= (\mu_0 - \mu_2) \left[\mu_1^m(\mu_0+\mu_2) + \sum_{k=0}^{m-1} \mu_2^k \left(-\mu^2_1 \mu_0 ^{m-1-k} + \mu^2_0\mu^2_2\mu_0^{m-3-k}\right)
	      -\mu^2_0\mu^2_2\left(\frac{\mu_2^{m-2}}{\mu_0}+\frac{\mu_2^{m-1}}{\mu^2_0}\right) \right]  \nonumber\\
\fl	&= (\mu_0 - \mu_2) \left[(\mu_1^m-\mu_2^m)(\mu_0+\mu_2) + \sum_{k=0}^{m-1} \mu_2^k \mu_0 ^{m-1-k} \left(\mu_2^2 - \mu_1^2\right)\right] \nonumber\\
\fl	&= (\mu_0 - \mu_2) \left[(\mu_0+\mu_2)(\mu_1-\mu_2)\left(\sum_{j=0}^{m-1}\mu_1^{m-1-j} \mu_2^j\right) -(\mu_1+\mu_2)(\mu_1-\mu_2)
	                      \sum_{k=0}^{m-1} \mu_2^k \mu_0 ^{m-1-k} \right] \nonumber\\
\fl	&= (\mu_0 - \mu_2)(\mu_1-\mu_2) \sum_{k=0}^{m-1} \mu_2^k \left[(\mu_0+\mu_2)\mu_1^{m-1-k} - (\mu_1+\mu_2)\mu_0 ^{m-1-k} \right] \nonumber\\
\fl	&= (\mu_0 - \mu_2)(\mu_1-\mu_2)\bigg\{\sum_{k=0}^{m-3} \mu_2^k \left[\mu_2(\mu_1^{m-1-k}-\mu_0 ^{m-1-k}) + \mu_0\mu_1(\mu_1 ^{m-2-k}-\mu_0 ^{m-2-k}) \right] 
		\nonumber\\	\fl &\hspace{3.5cm} +\mu_2^{m-1}(\mu_0 -\mu_1) + \mu_2^{m-1}(\mu_1-\mu_0) \bigg\} \nonumber\\
\fl	&= (\mu_0 - \mu_2)(\mu_1-\mu_2)(\mu_1-\mu_0)\sum_{k=0}^{m-3} \mu_2^k \left[\mu_2 \sum_{j=0}^{m-2-k}\mu_1^{m-2-k-j}\mu_0^j +\mu_0\mu_1 
	      \sum_{j=0}^{m-3-k}\mu_1^{m-3-k-j}\mu_0^j\right]  \nonumber\\
\fl	&= (\mu_0 - \mu_2)(\mu_1-\mu_2)(\mu_1-\mu_0)\sum_{k=0}^{m-3} \mu_2^k \left[(\mu_2+\mu_0)\sum_{j=0}^{m-3-k}\mu_1^{m-2-k-j}\mu_0^j +\mu_2\mu_0^{m-2-k}\right] \nonumber\\
\fl	&\equiv (\mu_0 - \mu_2)(\mu_1-\mu_2)(\mu_1-\mu_0) F(m)    \label{A-11_1} \,\,,
	\end{eqnarray}
	where the factor $F(m)\geq 0 \,,\,\, \forall\, m\geq 3$. Substituting (\ref{A-11_1}) back into (\ref{Ynm-11}), we recast each term of the sum in (\ref{comb2G-11}) corresponding to the $Y_{nm}$, with $n,m\geq 3$, as:
	\begin{eqnarray}
	\frac{Y_{nm}}{n!m!} \frac{(\mu_0 - \mu_2)^2(\mu_1-\mu_2)^2(\mu_1-\mu_0)^2}{(\mu^2_0-\mu^2_2)(\mu^2_1-\mu^2_2)} F(n)F(m)  \label{Ynmcoeff-11} \,\,,
	\end{eqnarray}
	so that its sign is manifestly dependent on the factor $(\mu_0 - \mu_2)(\mu_1-\mu_2)$.\\
	In a similar fashion, one can rewrite each term of the sum in (\ref{comb2G-11}) corresponding to the $Y_{1m}$, with $m\geq 3$, as:
	\begin{eqnarray}
	-\frac{Y_{1m}}{m!} \frac{(\mu_0 - \mu_1)^2(\mu_1-\mu_2)(\mu_0-\mu_2)}{(\mu_0+\mu_2)(\mu_1+\mu_2)} F(m)  \label{Y1mcoeff-11} \,\,,
	\end{eqnarray}
	thus deducing that this expression has opposite sign with respect to that given by (\ref{Ynmcoeff-11}). Same holds for $Y_{n1}$, since it can be shown that its coefficient is exactly (\ref{Y1mcoeff-11}) with the	substitution $m\rightarrow n$.\\
	Finally, by showing that the term corresponding to $Y_{11}$ in (\ref{comb2G-11}) can be factorized as:
	\begin{eqnarray}
	Y_{11} \frac{(\mu_0 - \mu_1)^2(\mu_1-\mu_2)(\mu_0-\mu_2)}{(\mu_0+\mu_2)(\mu_1+\mu_2)}   \label{Y11coeff-11} \,\,,
	\end{eqnarray}
	one concludes that this expression has the same sign as that given by (\ref{Ynmcoeff-11}).\\
	Putting together these considerations into (\ref{comb2G-11}), a valid upper bound on $Y_{11}$ is obtained when the yields $Y_{nm}$, with $n,m \geq 3$, are set to zero and the yields $Y_{1m}$ and $Y_{n1}$ are set to their maximum allowed value. Since in \ref{bound_on_Y13-3decoys} and \ref{bound_on_Y31-3decoys} we derive upper bounds on $Y_{13}$ and $Y_{31}$ (see \ref{Y13-upperbound-3decoys} and \ref{Y31-upperbound-3decoys}), we can employ them in (\ref{comb2G-11}) instead of trivially bounding these yields with 1. In this way we obtain:
	\begin{eqnarray}
	\fl	G_{11}^{0,1} &-\frac{(\mu_0^2-\mu_1^2)}{(\mu^2_0-\mu^2_2)} \,G_{11}^{0,2}+\frac{(\mu^2_0-\mu^2_1)}{(\mu^2_1-\mu^2_2)} \,G_{11}^{1,2} =
		    Y^U_{11} \frac{(\mu_0 - \mu_1)^2(\mu_1-\mu_2)(\mu_0-\mu_2)}{(\mu_0+\mu_2)(\mu_1+\mu_2)}  \nonumber\\
	\fl	    &+\frac{(\mu_0-\mu_1)}{6} (Y^U_{13}+Y^U_{31})  \left[\mu^3_0 -\mu^3_1 -\frac{(\mu_0 +\mu_1)}{(\mu_0+\mu_2)}(\mu^3_0-\mu^3_2)+\frac{(\mu_0+\mu_1)}{(\mu_1+\mu_2)}(\mu^3_1-\mu^3_2)\right] \nonumber\\
	\fl	    &+ 2(\mu_0-\mu_1) \sum_{n=4}^{\infty} \left[\frac{(\mu^n_0 -\mu^n_1)}{n!} -\frac{(\mu_0 +\mu_1)}{(\mu_0+\mu_2)}\frac{(\mu^n_0-\mu^n_2)}{n!} 
		    +\frac{(\mu_0+\mu_1)}{(\mu_1+\mu_2)}\frac{(\mu^n_1-\mu^n_2)}{n!}\right] \,\,,
	\end{eqnarray}
	which leads to the following upper bound on $Y_{11}$:
	\begin{eqnarray}
	\fl	Y^U_{11}=  &\frac{(\mu_0+\mu_2)(\mu_1+\mu_2)}{(\mu_0 - \mu_1)^2(\mu_1-\mu_2)(\mu_0-\mu_2)}
		\left[G_{11}^{0,1} -\frac{(\mu_0^2-\mu_1^2)}{(\mu^2_0-\mu^2_2)} \,G_{11}^{0,2}+\frac{(\mu^2_0-\mu^2_1)}{(\mu^2_1-\mu^2_2)} \,G_{11}^{1,2} - 2(\mu_0-\mu_1)E_{11}\right]  \nonumber\\
	\fl	&+\frac{(\mu_1\mu_2+\mu_0\mu_1+\mu_0\mu_2)}{6}(Y^U_{13}+Y^U_{31}) \,\,,	 \label{Y11-upperbound-3decoys}
	\end{eqnarray} 
	where the term $E_{11}$ is defined as:
	\begin{eqnarray}
		E_{11} = \,\,&e^{\mu_0}-e^{\mu_1}-(\mu_0 -\mu_1)\left(1+\frac{\mu_0}{2}+\frac{\mu_1}{2}+\frac{\mu^2_0}{6}+\frac{\mu^2_1}{6}+\frac{\mu_0\mu_1}{6}\right)  \nonumber\\
		&+\frac{\mu_0+\mu_1}{\mu_1+\mu_2}\left[e^{\mu_1}-e^{\mu_2}-(\mu_1 -\mu_2)\left(1+\frac{\mu_1}{2}+\frac{\mu_2}{2}+\frac{\mu^2_1}{6}+\frac{\mu^2_2}{6}+\frac{\mu_1\mu_2}{6}\right)\right] \nonumber\\
		&-\frac{\mu_0+\mu_1}{\mu_0+\mu_2}\left[e^{\mu_0}-e^{\mu_2}-(\mu_0 -\mu_2)\left(1+\frac{\mu_0}{2}+\frac{\mu_2}{2}+\frac{\mu^2_0}{6}+\frac{\mu^2_2}{6}+\frac{\mu_0\mu_2}{6}\right)\right] \,\,.  \label{E-11}
	\end{eqnarray}
	
	\subsection{Upper bound on $Y_{02}$ and $Y_{04}$}  \label{bound_on_Y02-3decoys}
	Consider the following combinations of gains in which all the terms $Y_{1m}$ and $,Y_{n0}$ are removed:
	\begin{eqnarray}
	G_{02}^{0,1} &= \mu_1\tilde{Q}^{0,0}+ \mu_0\tilde{Q}^{1,1} - \mu_1\tilde{Q}^{0,1} -\mu_0\tilde{Q}^{1,0} \nonumber\\
	&= \sum_{n,m=0}^{\infty} \frac{Y_{nm}}{n!m!} \left(\mu^n_0\mu_1 - \mu_0\mu^n_1\right) \left(\mu^m_0 -\mu^m_1\right) \,;\nonumber\\
	G_{02}^{0,2} &= \mu_2\tilde{Q}^{0,0}+ \mu_0\tilde{Q}^{2,2} - \mu_2\tilde{Q}^{0,2} -\mu_0\tilde{Q}^{2,0} \nonumber\\
	&= \sum_{n,m=0}^{\infty} \frac{Y_{nm}}{n!m!} \left(\mu^n_0\mu_2 - \mu_0\mu^n_2\right) \left(\mu^m_0 -\mu^m_2\right) \,;\nonumber\\
	G_{02}^{1,2} &= \mu_2\tilde{Q}^{1,1}+ \mu_1\tilde{Q}^{2,2} - \mu_2\tilde{Q}^{1,2} -\mu_1\tilde{Q}^{2,1} \nonumber\\
	&= \sum_{n,m=0}^{\infty} \frac{Y_{nm}}{n!m!} \left(\mu^n_1\mu_2 - \mu_1\mu^n_2\right) \left(\mu^m_1 -\mu^m_2\right)  \label{G-02-3decoys} \,\,.
	\end{eqnarray}
	We now combine $G_{02}^{0,1},G_{02}^{0,2}$ and $G_{02}^{1,2}$ with arbitrary real coefficients $c_0$ and $c_1$ and impose that the resulting expression has the yields $Y_{2m}$ and $Y_{n1}$ also removed:
	\begin{eqnarray}
	\fl &G_{02}^{0,1} + c_0 \,G_{02}^{0,2}+c_1 \,G_{02}^{1,2} = \nonumber\\
	\fl &\sum_{n,m=0}^{\infty} \frac{Y_{nm}}{n!m!} \left[(\mu^n_0\mu_1 - \mu_0\mu^n_1)(\mu^m_0 - \mu^m_1) +c_0 (\mu^n_0\mu_2 - \mu_0\mu^n_2)(\mu^m_0 - \mu^m_2)+ 
	c_1 (\mu^n_1\mu_2 - \mu_1\mu^n_2)(\mu^m_1 - \mu^m_2)\right] \,\,. \nonumber\\
	\fl  \label{combG-02} 
	\end{eqnarray}
	For $Y_{2m}$ and $Y_{n1}$ to be removed the coefficients $c_0$ and $c_1$ must satisfy:
	\begin{eqnarray}
	\fl \left\{
	{\begin{array}{lcl}
	 (\mu^n_0\mu_1 - \mu_0\mu^n_1)(\mu_0 - \mu_1) +c_0 (\mu^n_0\mu_2 - \mu_0\mu^n_2)(\mu_0 - \mu_2)+ c_1 (\mu^n_1\mu_2 - \mu_1\mu^n_2)(\mu_1 - \mu_2) & = &0  \quad\forall\,n\\
	(\mu^2_0\mu_1 - \mu_0\mu^2_1)(\mu^m_0 - \mu^m_1) +c_0 (\mu^2_0\mu_2 - \mu_0\mu^2_2)(\mu^m_0 - \mu^m_2)+c_1 (\mu^2_1\mu_2 - \mu_1\mu^2_2)(\mu^m_1 - \mu^m_2) & = & 0
	   \quad\forall\, m  
	\end{array}}
	 \right.  \nonumber\\
	 \fl \label{system-02}
	\end{eqnarray}
	or equivalently:
	\begin{eqnarray}
	 \def\arraystretch{2.2}
	\fl \left\{
	{\begin{array}{l}
		\mu_0^n\left[\mu_1(\mu_0-\mu_1)+c_0\mu_2(\mu_0-\mu_2)\right]+\mu_1^n\left[-\mu_0(\mu_0-\mu_1)+c_1\mu_2(\mu_1-\mu_2)\right]\\
		 -\mu_2^n\left[\mu_0 c_0(\mu_0-\mu_2)+\mu_1 c_1(\mu_1-\mu_2)\right] =0 \quad\forall\,n \\
		\mu_0^m\left[\mu_1\mu^2_0-\mu_0\mu^2_1+c_0(\mu_2\mu^2_0-\mu_0\mu^2_2)\right]+\mu_1^m\left[-(\mu_1\mu^2_0-\mu_0\mu^2_1)+c_1(\mu_2\mu^2_1-\mu_1\mu^2_2)\right]\\
		-\mu_2^m\left[c_0(\mu_2\mu^2_0-\mu_0\mu^2_2)+c_1(\mu_2\mu^2_1-\mu_1\mu^2_2)\right] =0 \quad\forall\, m\,.
		\end{array}}
	\right.
	\end{eqnarray}
	A sufficient condition for this is that the coefficient of every $\mu_i^n$ and every $\mu_i^m$ is identically zero. This imposes six conditions on $c_0$ and $c_1$, however thanks to the inherent symmetries of the system a solution exists, and reads:
	\begin{eqnarray}
	c_0 &= -\frac{\mu_1(\mu_0-\mu_1)}{\mu_2(\mu_0-\mu_2)} \label{c0-02} \,\,, \\
	c_1 &= \frac{\mu_0(\mu_0-\mu_1)}{\mu_2(\mu_1-\mu_2)}  \label{c1-02} \,\,.
	\end{eqnarray}
	Substituting these expressions back into (\ref{combG-02}) and multiplying both sides by $\mu_2$, yields a combination of gains in which the terms $Y_{n0},Y_{n1},Y_{1m}$ and $Y_{2m}$ are removed. In particular, we obtain:
	\begin{eqnarray}
	\fl &\mu_2 G_{02}^{0,1} -\frac{\mu_1(\mu_0-\mu_1)}{(\mu_0-\mu_2)} \,G_{02}^{0,2}+\frac{\mu_0(\mu_0-\mu_1)}{(\mu_1-\mu_2)} \,G_{02}^{1,2} = \nonumber\\
	\fl &\sum_{m=2}^{\infty} \frac{Y_{0m}}{m!}(\mu_0 -\mu_1)\left[-\mu_2(\mu^m_0 -\mu^m_1) +\mu_1(\mu^m_0-\mu^m_2) -\mu_0 (\mu^m_1-\mu^m_2) \right] \nonumber\\
	\fl &+\sum_{\stackrel[m=2]{n=3}{}}^{\infty} \frac{Y_{nm}}{n!m!} \left[\mu_2 (\mu^n_0\mu_1 - \mu_0\mu^n_1)(\mu^m_0 - \mu^m_1) -\frac{\mu_1(\mu_0-\mu_1)}{(\mu_0-\mu_2)}
	 (\mu^n_0\mu_2 - \mu_0\mu^n_2)(\mu^m_0 - \mu^m_2) \right. \nonumber\\
	\fl &\left.\hspace{2cm}+ \frac{\mu_0(\mu_0-\mu_1)}{(\mu_1-\mu_2)}(\mu^n_1\mu_2 - \mu_1\mu^n_2)(\mu^m_1 - \mu^m_2) \right]    \label{comb2G-02} \,\,.
	\end{eqnarray}
	In order to get a valid upper bound for $Y_{02}$ and $Y_{04}$ we need to study the sign of the coefficients of the remaining yields. We start by recasting each term of the sum corresponding to the $Y_{nm}$, with $n\geq 3$ and $m\geq 2$, in (\ref{comb2G-02}) as follows:
	\begin{eqnarray}
	\frac{Y_{nm}}{n!m!} \frac{1}{(\mu_0-\mu_2)(\mu_2-\mu_1)} A_{22}(\mu_0,\mu_1,\mu_2,m)\cdot B_{02}(\mu_0,\mu_1,\mu_2,n) \label{Ynmcoeff-02} \,\,,
	\end{eqnarray}
	where
	\begin{eqnarray}
	B_{02}(\mu_0,\mu_1,\mu_2,n) &\equiv \mu_1\mu_2\mu_0^n(\mu_1-\mu_2) + \mu_0^2(\mu_1\mu_2^n-\mu_2\mu_1^n)+\mu_0(\mu_2^2\mu_1^n-\mu_1^2\mu_2^n)  \label{B-02} 
	\end{eqnarray}
	and $A_{22}$ is the one found when bounding $Y_{22}$, thus we know from (\ref{A-22_1}) it can be recast as:
	\begin{eqnarray}
		A_{22}(\mu_0,\mu_1,\mu_2,m)=(\mu_0 - \mu_2)(\mu_2 - \mu_1)\sum_{k=0}^{m-1} \mu_2^k (\mu_0^{m-1-k}-\mu_1^{m-1-k}) \,\,. \label{A-02_1}
	\end{eqnarray}
	We can rewrite $B_{02}$ as:
	\begin{eqnarray}
	&B_{02}(\mu_0,\mu_1,\mu_2,n) =\mu_0\mu_1\mu_2 (\mu_1-\mu_2)\left[\mu_0^{n-1} -\mu_0 \sum_{k=0}^{n-2}\mu_1^{n-2-k}\mu_2^k +\mu_1\mu_2\sum_{j=0}^{n-3}\mu_1^{n-3-j}\mu_2^j\right] \nonumber\\
	&= \mu_0\mu_1\mu_2 (\mu_1-\mu_2)\left[\mu_0^{n-1} + \sum_{k=0}^{n-2}\mu_1^{n-2-k}\mu_2^k(\mu_2-\mu_0) -\mu_2^{n-1}\right] \nonumber\\
	&= \mu_0\mu_1\mu_2 (\mu_1-\mu_2)\left[\mu_0^{n-1}-\mu_2^{n-1} - \sum_{k=0}^{n-2}\mu_1^{n-2-k}\mu_2^k(\mu_0-\mu_2)\right] \nonumber\\
	&= \mu_0\mu_1\mu_2 (\mu_1-\mu_2)(\mu_0-\mu_2)\left[\sum_{k=0}^{n-2}\mu_2^k \mu_0^{n-2-k} - \sum_{k=0}^{n-2}\mu_2^k \mu_1^{n-2-k}\right] \nonumber\\
	&= \mu_0\mu_1\mu_2 (\mu_1-\mu_2)(\mu_0-\mu_2)\sum_{k=0}^{n-2}\mu_2^k (\mu_0^{n-2-k} - \mu_1^{n-2-k}) \,\,. \label{B-02_1}
	\end{eqnarray}
	Employing (\ref{A-02_1}) and (\ref{B-02_1}) into (\ref{Ynmcoeff-02}) we get:
	\begin{eqnarray}
	\fl \frac{Y_{nm}}{n!m!} \mu_0\mu_1\mu_2 (\mu_1-\mu_2)(\mu_0-\mu_2)\left[\sum_{k=0}^{n-2}\mu_2^k (\mu_0^{n-2-k} - \mu_1^{n-2-k})\right]
	  \left[\sum_{k=0}^{m-1} \mu_2^k (\mu_0^{m-1-k}-\mu_1^{m-1-k})\right]  \label{Ynmcoeff-02_1} \,\,,
	\end{eqnarray}
	which means that the sign of this expression is fully determined by the factor $(\mu_1-\mu_2)(\mu_0-\mu_2)$ (note that the product of the two sums in (\ref{Ynmcoeff-02_1}) is always positive).\\
	Concerning the terms that appear in the sum in (\ref{comb2G-02}) corresponding to the $Y_{0m}$, with $m\geq 2$, we have:
	\begin{eqnarray}
		&\frac{Y_{0m}}{m!}(\mu_1 -\mu_0)\left[\mu_2(\mu^m_0 -\mu^m_1) -\mu_1(\mu^m_0-\mu^m_2) +\mu_0 (\mu^m_1-\mu^m_2) \right] \nonumber\\
		&= \frac{Y_{0m}}{m!}(\mu_1 -\mu_0) A_{22}(\mu_0,\mu_1,\mu_2,m)  \nonumber\\
		&= \frac{Y_{0m}}{m!}(\mu_0 - \mu_2)(\mu_1 - \mu_2)(\mu_0 -\mu_1)  \sum_{k=0}^{m-1} \mu_2^k (\mu_0^{m-1-k}-\mu_1^{m-1-k}) \label{Y0mcoeff-02}\,\,,
	\end{eqnarray}
	where we used (\ref{A-22}) in the first equality and (\ref{A-02_1}) in the second equality. Expression (\ref{Y0mcoeff-02}) implies that its sign is always equal to the sign of the terms given by (\ref{Ynmcoeff-02_1}), since it is determined by the same factor $(\mu_1-\mu_2)(\mu_0-\mu_2)$ (note that the product of the last two factors in (\ref{Y0mcoeff-02}) is always positive).\\
	A valid upper bound on $Y_{02}$ is thus obtained by setting all the other yields to zero in (\ref{comb2G-02}). By doing so, we obtain:
	\begin{eqnarray}
		Y^U_{02}=2\frac{\frac{\mu_2 G_{02}^{0,1}}{\mu_0-\mu_1} -\frac{\mu_1 G_{02}^{0,2}}{\mu_0-\mu_2} +\frac{\mu_0 G_{02}^{1,2}}{\mu_1-\mu_2}}
		{(\mu_0 - \mu_2)(\mu_1 - \mu_2)(\mu_0 -\mu_1)}  \label{Y02-upperbound-3decoys} \,\,.
	\end{eqnarray}
	One can do the same when bounding $Y_{04}$, i.e. setting all the other yields to zero except for $Y_{04}$, in (\ref{comb2G-02}). We find that:
	\begin{eqnarray}
	Y^U_{04}=4!\frac{\frac{\mu_2 G_{02}^{0,1}}{\mu_0-\mu_1} -\frac{\mu_1 G_{02}^{0,2}}{\mu_0-\mu_2} +\frac{\mu_0 G_{02}^{1,2}}{\mu_1-\mu_2}}
	{\mu_1(\mu^4_0 - \mu^4_2)-\mu_0 (\mu^4_1 - \mu^4_2) -\mu_2(\mu^4_0 -\mu^4_1)}  \label{Y04-upperbound-3decoys} \,\,.
	\end{eqnarray}

	\subsection{Upper bound on $Y_{20}$ and $Y_{40}$}
	Consider the following combinations of gains in which all the terms $Y_{0m}$ and $Y_{n1}$ are removed:
	\begin{eqnarray}
	G_{20}^{0,1} &= \mu_1\tilde{Q}^{0,0}+ \mu_0\tilde{Q}^{1,1} - \mu_0\tilde{Q}^{0,1} -\mu_1\tilde{Q}^{1,0} \nonumber\\
	&= \sum_{n,m=0}^{\infty} \frac{Y_{nm}}{n!m!} (\mu^n_0 -\mu^n_1)(\mu^m_0\mu_1 - \mu_0\mu^m_1)  \,;\nonumber\\
	G_{20}^{0,2} &= \mu_2\tilde{Q}^{0,0}+ \mu_0\tilde{Q}^{2,2} - \mu_0\tilde{Q}^{0,2} -\mu_2\tilde{Q}^{2,0} \nonumber\\
	&= \sum_{n,m=0}^{\infty} \frac{Y_{nm}}{n!m!} (\mu^n_0 -\mu^n_2)(\mu^m_0\mu_2 - \mu_0\mu^m_2) \,;\nonumber\\
	G_{20}^{1,2} &= \mu_2\tilde{Q}^{1,1}+ \mu_1\tilde{Q}^{2,2} - \mu_1\tilde{Q}^{1,2} -\mu_2\tilde{Q}^{2,1} \nonumber\\
	&= \sum_{n,m=0}^{\infty} \frac{Y_{nm}}{n!m!} (\mu^n_1 -\mu^n_2)(\mu^m_1\mu_2 - \mu_1\mu^m_2)  \label{G-20-3decoys} \,\,.
	\end{eqnarray}
	We now combine $G_{20}^{0,1},G_{20}^{0,2}$ and $G_{20}^{1,2}$ with arbitrary real coefficients $c_0$ and $c_1$ and impose that the resulting expression has the yields $Y_{1m}$ and $Y_{n2}$ also removed:
	\begin{eqnarray}
	\fl &G_{20}^{0,1} + c_0 \,G_{20}^{0,2}+c_1 \,G_{20}^{1,2} = \nonumber\\
	\fl &\sum_{n,m=0}^{\infty} \frac{Y_{nm}}{n!m!} \left[(\mu^n_0 -\mu^n_1)(\mu^m_0\mu_1 - \mu_0\mu^m_1) +c_0 (\mu^n_0 -\mu^n_2)(\mu^m_0\mu_2 - \mu_0\mu^m_2)+ 
	c_1 (\mu^n_1 -\mu^n_2)(\mu^m_1\mu_2 - \mu_1\mu^m_2)\right]   \,\,. \nonumber \\
	\fl \label{combG-20}
	\end{eqnarray}
	For $Y_{1m}$ and $Y_{n2}$ to be removed the coefficients $c_0$ and $c_1$ must satisfy:
	\begin{eqnarray}
	\fl	\left\{
	{\begin{array}{lcl}
		(\mu^m_0\mu_1 - \mu_0\mu^m_1)(\mu_0 - \mu_1) +c_0 (\mu^m_0\mu_2 - \mu_0\mu^m_2)(\mu_0 - \mu_2)+ c_1 (\mu^m_1\mu_2 - \mu_1\mu^m_2)(\mu_1 - \mu_2) & = &0  \quad\forall\,m\\
		(\mu^2_0\mu_1 - \mu_0\mu^2_1)(\mu^n_0 - \mu^n_1) +c_0 (\mu^2_0\mu_2 - \mu_0\mu^2_2)(\mu^n_0 - \mu^n_2)+c_1 (\mu^2_1\mu_2 - \mu_1\mu^2_2)(\mu^n_1 - \mu^n_2) & = & 0
		\quad\forall\, n  \,.
		\end{array}}
	\right.  \nonumber\\
	\fl \mbox{ }
	\end{eqnarray}
	This system of linear equations coincides with the one given by (\ref{system-02}) that we found when bounding $Y_{02}$, thus the solution is given by (\ref{c0-02}) for $c_0$ and by (\ref{c1-02}) for $c_1$.
	Substituting these expressions back into (\ref{combG-20}) and multiplying both sides by $\mu_2$, yields a combination of gains in which the terms $Y_{n1},Y_{n2},Y_{0m}$ and $Y_{1m}$ are removed:
	\begin{eqnarray}
\fl	&\mu_2 G_{20}^{0,1} -\frac{\mu_1(\mu_0-\mu_1)}{(\mu_0-\mu_2)} \,G_{20}^{0,2}+\frac{\mu_0(\mu_0-\mu_1)}{(\mu_1-\mu_2)} \,G_{20}^{1,2} = \nonumber\\
\fl	&\sum_{n=2}^{\infty} \frac{Y_{n0}}{n!}(\mu_0 -\mu_1)\left[-\mu_2(\mu^n_0 -\mu^n_1) +\mu_1(\mu^n_0-\mu^n_2) -\mu_0 (\mu^n_1-\mu^n_2) \right] \nonumber\\
\fl	&+\sum_{\stackrel[m=3]{n=2}{}}^{\infty} \frac{Y_{nm}}{n!m!} \left[\mu_2 (\mu^n_0 -\mu^n_1)(\mu^m_0\mu_1 - \mu_0\mu^m_1) -\frac{\mu_1(\mu_0-\mu_1)}{(\mu_0-\mu_2)}
	(\mu^n_0 -\mu^n_2)(\mu^m_0\mu_2 - \mu_0\mu^m_2) \right. \nonumber\\
\fl	&\left.\hspace{2cm}+ \frac{\mu_0(\mu_0-\mu_1)}{(\mu_1-\mu_2)}(\mu^n_1 -\mu^n_2)(\mu^m_1\mu_2 - \mu_1\mu^m_2) \right]    \label{comb2G-20} \,\,.
	\end{eqnarray}
	Since the coefficients of $Y_{n0}$ and $Y_{nm}$ coincide with those found when bounding $Y_{02}$ if one exchanges $m\longleftrightarrow n$, we can directly use the results obtained in \ref{bound_on_Y02-3decoys} to recast the terms that contain the $Y_{nm}$ with $n\geq 2$ and $m\geq 3$. In particular, according to (\ref{Ynmcoeff-02_1}), we obtain:
	\begin{eqnarray}
	\fl \frac{Y_{nm}}{n!m!} \mu_0\mu_1\mu_2 (\mu_1-\mu_2)(\mu_0-\mu_2)\left[\sum_{k=0}^{m-2}\mu_2^k (\mu_0^{m-2-k} - \mu_1^{m-2-k})\right]
	\left[\sum_{k=0}^{n-1} \mu_2^k (\mu_0^{n-1-k}-\mu_1^{n-1-k})\right]  \label{Ynmcoeff-20_1} \,\,,
	\end{eqnarray}
	and according to (\ref{Y0mcoeff-02}) the terms that contain the yields $Y_{n0}$ can be written as:
	\begin{eqnarray}
	\fl \frac{Y_{n0}}{n!}(\mu_0 - \mu_2)(\mu_1 - \mu_2)(\mu_0 -\mu_1)  \sum_{k=0}^{n-1} \mu_2^k (\mu_0^{n-1-k}-\mu_1^{n-1-k})  \,\,. 
	\end{eqnarray}
	Like in the case of $Y_{02}$ (see \ref{bound_on_Y02-3decoys}), a valid upper bound on $Y_{20}$ is thus obtained setting all the other yields to zero in (\ref{comb2G-20}). We obtain:
	\begin{eqnarray}
	Y^U_{20}=2\frac{\frac{\mu_2 G_{20}^{0,1}}{\mu_0-\mu_1} -\frac{\mu_1 G_{20}^{0,2}}{\mu_0-\mu_2} +\frac{\mu_0 G_{20}^{1,2}}{\mu_1-\mu_2}}
	{(\mu_0 - \mu_2)(\mu_1 - \mu_2)(\mu_0 -\mu_1)}  \label{Y20-upperbound-3decoys} \,\,.
	\end{eqnarray}
	One can do the same to bound $Y_{40}$, i.e. to set all the other yields to zero, except for $Y_{40}$. In this case we obtain:
	\begin{eqnarray}
	Y^U_{04}=4!\frac{\frac{\mu_2 G_{20}^{0,1}}{\mu_0-\mu_1} -\frac{\mu_1 G_{20}^{0,2}}{\mu_0-\mu_2} +\frac{\mu_0 G_{20}^{1,2}}{\mu_1-\mu_2}}
	{\mu_1(\mu^4_0 - \mu^4_2)-\mu_0 (\mu^4_1 - \mu^4_2) -\mu_2(\mu^4_0 -\mu^4_1)}  \label{Y40-upperbound-3decoys} \,\,.
	\end{eqnarray}

	\subsection{Upper bound on $Y_{13}$} \label{bound_on_Y13-3decoys}
	We look for that combination of gains in which all the terms proportional to $Y_{n0},Y_{n1},Y_{0m}$ and $Y_{2m}$ are removed. In order to find it, we consider the most general combination of all gains:
	\begin{eqnarray}
		G_{13}=\sum_{i,j=0}^{2} c_{i,j} \tilde{Q}^{i,j} = \sum_{n,m=0}^{\infty} \frac{Y_{nm}}{n!m!}\left[\sum_{i,j=0}^{2} c_{i,j} \mu_i^n \mu_j^m \right] \,\,, \label{G-13}
	\end{eqnarray}
	and impose proper conditions on the real coefficients $c_{i,j}$:
	\begin{eqnarray}
	\fl &Y_{n0}\mbox{ removed:}\,\, \sum_{i=0}^{2} \mu_i^n \left(\sum_{j=0}^{2} c_{i,j}\right) =0 \quad\forall\, n \quad\Leftarrow\quad c_{i,0}+c_{i,1}+c_{i,2}=0 
	     \quad \mbox{for} \,\, i=0,1,2  \label{Yn0removed-13} \\
	\fl &Y_{n1}\mbox{ removed:}\,\, \sum_{i=0}^{2} \mu_i^n \left(\sum_{j=0}^{2} \mu_j c_{i,j}\right) =0 \quad\forall\, n \quad\Leftarrow\quad 
	\mu_0 c_{i,0}+ \mu_1 c_{i,1}+\mu_2 c_{i,2}=0 \quad \mbox{for} \,\, i=0,1,2  \label{Yn1removed-13} \\
	\fl &Y_{0m}\mbox{ removed:}\,\, \sum_{j=0}^{2} \mu_j^m \left(\sum_{i=0}^{2} c_{i,j}\right) =0 \quad\forall\, m \quad\Leftarrow\quad c_{0,j}+c_{1,j}+c_{2,j}=0 
	\quad \mbox{for} \,\, j=0,1,2  \label{Y0mremoved-13} \\
	\fl &Y_{2m}\mbox{ removed:}\,\, \sum_{j=0}^{2} \mu_j^m \left(\sum_{i=0}^{2} \mu_i^2 c_{i,j}\right) =0 \quad\forall\, m \quad\Leftarrow\quad 
	 \mu_0^2 c_{0,j}+\mu_1^2 c_{1,j}+ \mu_2^2 c_{2,j}=0 \quad \mbox{for} \,\, j=0,1,2  \label{Y2mremoved-13}  \,.
	\end{eqnarray}
	The conditions given by equations (\ref{Yn0removed-13}, \ref{Yn1removed-13}, \ref{Y0mremoved-13}, \ref{Y2mremoved-13}) form an overdetermined system of equations for the nine variables $c_{i,j}$. However, thanks to the symmetries of the problem, a unique solution for $c_{i,j}$ exists and reads (we rescale every coefficient by requiring $c_{0,0}=1$):
	\begin{eqnarray}
		&c_{0,0}=1 \nonumber\,\,,\\
		&c_{0,1}=-\frac{(\mu_0-\mu_2)}{\mu_1-\mu_2}  \nonumber\,\,,\\
		&c_{0,2}=-1-c_{0,1}=\frac{\mu_0-\mu_1}{\mu_1 -\mu_2} \nonumber\,\,,\\
		&c_{1,0}=-\frac{(\mu_0^2-\mu_2^2)}{\mu_1^2-\mu_2^2} \nonumber\,\,,\\
		&c_{1,1}= c_{1,0}c_{0,1} =\frac{(\mu_0^2-\mu_2^2)(\mu_0-\mu_2)}{(\mu_1^2-\mu_2^2)(\mu_1 -\mu_2)}  \nonumber\,\,,\\
		&c_{1,2}= -c_{1,0}-c_{1,1}=c_{1,0}c_{0,2}=-\frac{(\mu_0^2-\mu_2^2)(\mu_0-\mu_1)}{(\mu_1^2-\mu_2^2)(\mu_1 -\mu_2)} \nonumber\,\,,\\
		&c_{2,0}=-1-c_{1,0}=\frac{\mu_0^2-\mu_1^2}{\mu_1^2-\mu_2^2} \nonumber\,\,,\\
		&c_{2,1}=-c_{0,1}-c_{1,1}=c_{0,1}c_{2,0}= \frac{(\mu^2_1-\mu^2_0)(\mu_0-\mu_2)}{(\mu_1^2-\mu_2^2)(\mu_1 -\mu_2)} \nonumber\,\,,\\
		&c_{2,2}=-c_{2,0}-c_{2,1}=(1+c_{1,0})(1+c_{0,1})=\frac{(\mu_0^2-\mu_1^2)(\mu_0-\mu_1)}{(\mu_1^2-\mu_2^2)(\mu_1-\mu_2)}  \label{cij-13}\,\,.
	\end{eqnarray}
	By substituting (\ref{cij-13}) back into (\ref{G-13}) we get an expression in which the terms $Y_{n0},Y_{n1},Y_{0m}$ and $Y_{2m}$ are removed:
	\begin{eqnarray}
	 \fl G_{13}= &\sum_{m=2}^{\infty}\frac{Y_{1m}}{m!} \frac{(\mu_0-\mu_1)(\mu_0-\mu_2)}{(\mu_1-\mu_2)(\mu_1+\mu_2)}\cdot A_{22}(\mu_0,\mu_1,\mu_2,m) \nonumber\\
	 \fl &+ \sum_{\stackrel[m=2]{n=3}{}}^{\infty} \frac{Y_{nm}}{n!m!}\frac{A_{22}(\mu_0,\mu_1,\mu_2,m)\cdot A_{11}(\mu_0,\mu_1,\mu_2,n)}{(\mu_1-\mu_2)^2(\mu_1+\mu_2)}  \label{combG-13} \,\,,
	\end{eqnarray}
	where $A_{22}$ is the factor given by (\ref{A-22}) also present in the bounds for $Y_{02}$ and $Y_{22}$, whereas $A_{11}$ is the factor given by (\ref{A-11}) which appears in the bound on $Y_{11}$. Note that this is somehow expected: when bounding $Y_{02}$ and $Y_{22}$ we removed the terms $Y_{n0}$ and $Y_{n1}$ as we just did for $Y_{13}$, and in bounding $Y_{11}$ we removed the terms $Y_{0m}$ and $Y_{2m}$ as we did here. Therefore, by exploiting the result given by (\ref{A-22_1}) we can recast each term of the sum corresponding to the $Y_{1m}$, with $m\geq 2$, in (\ref{combG-13}) as:
	\begin{eqnarray}
		-\frac{Y_{1m}}{m!}\frac{(\mu_0-\mu_2)^2}{(\mu_1+\mu_2)}(\mu_0-\mu_1)\sum_{k=0}^{m-1} \mu_2^k (\mu_0^{m-1-k}-\mu_1^{m-1-k})\,\,,
	\end{eqnarray}
	and realize that it is always negative, regardless of the value of the intensities.\\
	By employing the results (\ref{A-22_1}, \ref{A-11_1}) we can recast each term of the sum corresponding to the $Y_{nm}$ with $n\geq 3$ and $m\geq 2$, in (\ref{combG-13}) as:
	\begin{eqnarray}
	\frac{Y_{nm}}{n!m!}\frac{(\mu_0 - \mu_2)^2(\mu_1-\mu_2)^2}{(\mu_1-\mu_2)^2(\mu_1+\mu_2)}(\mu_0-\mu_1)\sum_{k=0}^{m-1} \mu_2^k (\mu_0^{m-1-k}-\mu_1^{m-1-k})F(n)\,\,,
	\end{eqnarray} 
	and realize that it is always positive\footnote{$F(n)$ is defined in (\ref{A-11_1}).}, regardless of the intensities.\\
	A valid upper bound on $Y_{13}$ is then obtained by setting $Y_{1m}\rightarrow 0$ (except for $Y_{13}$) and $Y_{nm}\rightarrow 1$ for all $n\geq 3$ and $m\geq 2$ in (\ref{combG-13}). As a result we obtain:
	\begin{eqnarray}
	\fl	G_{13}= &-\frac{Y^U_{13}}{3!}\frac{(\mu_0-\mu_2)^2}{(\mu_1+\mu_2)}(\mu_0-\mu_1)\left[\mu_0^2-\mu_1^2+\mu_2(\mu_0-\mu_1)\right] \nonumber\\
	\fl	&+ \sum_{\stackrel[m=2]{n=3}{}}^{\infty}\frac{\left[\mu_1^m (\mu_0 - \mu_2) + \mu_2^m (\mu_1 - \mu_0) + \mu_0^m (\mu_2 - \mu_1)\right]\cdot
			\left[\mu_1^n (\mu^2_0 - \mu^2_2) + \mu_2^n (\mu^2_1 - \mu^2_0) + \mu_0^n (\mu^2_2 - \mu^2_1)\right]}{n!m!(\mu_1-\mu_2)^2(\mu_1+\mu_2)}\,,  \nonumber\\
	\fl	&\mbox{ }
	\end{eqnarray}
	which implies:
	\begin{eqnarray}
	\fl	&\frac{Y^U_{13}}{6}\frac{(\mu_0-\mu_2)^2(\mu_0-\mu_1)^2(\mu_0+\mu_1+\mu_2)}{\mu_1+\mu_2} =-G_{13} \nonumber\\
	\fl	& + \frac{(e^{\mu_1}-\mu_1-1)(\mu_0 - \mu_2) + (e^{\mu_2}-\mu_2-1)(\mu_1 - \mu_0) + (e^{\mu_0}-\mu_0-1)(\mu_2 - \mu_1)}{(\mu_1-\mu_2)^2(\mu_1+\mu_2)}\nonumber\\
	\fl	&\times \left[(e^{\mu_1}-\frac{\mu_1^2}{2}-\mu_1-1)(\mu^2_0 - \mu^2_2) + (e^{\mu_2}-\frac{\mu_2^2}{2}-\mu_2-1)(\mu^2_1 - \mu^2_0) + 
		(e^{\mu_0}-\frac{\mu_0^2}{2}-\mu_0-1)(\mu^2_2 - \mu^2_1)\right] \,\,. \nonumber\\
	\fl	&\mbox{ }
	\end{eqnarray}
	We thus obtain the following upper bound on $Y_{13}$:
	\begin{eqnarray}
	\fl	&Y^U_{13} = -\frac{6(\mu_1+\mu_2)G_{13}}{(\mu_0-\mu_2)^2(\mu_0-\mu_1)^2(\mu_0+\mu_1+\mu_2)}+ \frac{6}{(\mu_0-\mu_2)^2(\mu_1-\mu_2)^2(\mu_0-\mu_1)^2(\mu_0+\mu_1+\mu_2)}
		\nonumber\\
	\fl	&\times \left[e^{\mu_2}(\mu_1-\mu_0)+e^{\mu_1}(\mu_0-\mu_2)+e^{\mu_0}(\mu_2-\mu_1)\right]  \nonumber\\
	\fl	&\times \left[e^{\mu_2}(\mu^2_1-\mu^2_0)+e^{\mu_1}(\mu^2_0-\mu^2_2)+e^{\mu_0}(\mu^2_2-\mu^2_1)-(\mu_0-\mu_1)(\mu_1-\mu_2)(\mu_0-\mu_2)\right]
		\label{Y13-upperbound-3decoys} \,\,,
	\end{eqnarray}
	where $G_{13}$ is defined in (\ref{G-13}) and the coefficients of the combination of gains in (\ref{cij-13}).

	\subsection{Upper bound on $Y_{31}$} \label{bound_on_Y31-3decoys}
	We look for that combination of gains in which all the terms proportional to $Y_{n0},Y_{n2},Y_{0m}$ and $Y_{1m}$ are removed. In order to find it, we proceed like in the previous case. That is, we consider the most general combination of all gains:
	\begin{eqnarray}
	G_{31}=\sum_{i,j=0}^{2} c_{i,j} \tilde{Q}^{i,j} = \sum_{n,m=0}^{\infty} \frac{Y_{nm}}{n!m!}\left[\sum_{i,j=0}^{2} c_{i,j} \mu_i^n \mu_j^m \right] \,\,, \label{G-31}
	\end{eqnarray}
	and impose proper conditions on the real coefficients $c_{i,j}$:
	\begin{eqnarray}
	\fl &Y_{n0}\mbox{ removed:}\,\, \sum_{i=0}^{2} \mu_i^n \left(\sum_{j=0}^{2} c_{i,j}\right) =0 \quad\forall\, n \quad\Leftarrow\quad c_{i,0}+c_{i,1}+c_{i,2}=0 
	\quad \mbox{for} \,\, i=0,1,2  \label{Yn0removed-31} \\
	\fl &Y_{n2}\mbox{ removed:}\,\, \sum_{i=0}^{2} \mu_i^n \left(\sum_{j=0}^{2} \mu^2_j c_{i,j}\right) =0 \quad\forall\, n \quad\Leftarrow\quad 
	\mu^2_0 c_{i,0}+ \mu^2_1 c_{i,1}+\mu^2_2 c_{i,2}=0 \quad \mbox{for} \,\, i=0,1,2  \label{Yn2removed-31} \\
	\fl &Y_{0m}\mbox{ removed:}\,\, \sum_{j=0}^{2} \mu_j^m \left(\sum_{i=0}^{2} c_{i,j}\right) =0 \quad\forall\, m \quad\Leftarrow\quad c_{0,j}+c_{1,j}+c_{2,j}=0 
	\quad \mbox{for} \,\, j=0,1,2  \label{Y0mremoved-31} \\
	\fl &Y_{1m}\mbox{ removed:}\,\, \sum_{j=0}^{2} \mu_j^m \left(\sum_{i=0}^{2} \mu_i c_{i,j}\right) =0 \quad\forall\, m \quad\Leftarrow\quad 
	\mu_0 c_{0,j}+\mu_1 c_{1,j}+ \mu_2 c_{2,j}=0 \quad \mbox{for} \,\, j=0,1,2  \label{Y1mremoved-31}  \,.
	\end{eqnarray}
	The conditions (\ref{Yn0removed-31}, \ref{Yn2removed-31}, \ref{Y0mremoved-31}, \ref{Y1mremoved-31}) form an overdetermined system of equations for the nine variables $c_{i,j}$. However, thanks to the symmetries of the problem, a unique solution for $c_{i,j}$ exists and reads (we rescale every coefficient by requiring $c_{0,0}=1$):
	\begin{eqnarray}
	&c_{0,0}=1 \nonumber\,\,,\\
	&c_{0,1}=-\frac{(\mu_0^2-\mu_2^2)}{\mu_1^2-\mu_2^2}  \nonumber\,\,,\\
	&c_{0,2}=-1-c_{0,1}=\frac{\mu_0^2-\mu_1^2}{\mu_1^2-\mu_2^2} \nonumber\,\,,\\
	&c_{1,0}=-\frac{(\mu_0-\mu_2)}{\mu_1-\mu_2} \nonumber\,\,,\\
	&c_{1,1}= c_{1,0}c_{0,1} =\frac{(\mu_0^2-\mu_2^2)(\mu_0-\mu_2)}{(\mu_1^2-\mu_2^2)(\mu_1 -\mu_2)}  \nonumber\,\,,\\
	&c_{1,2}= -c_{1,0}-c_{1,1}=c_{1,0}c_{0,2}=\frac{(\mu^2_1-\mu^2_0)(\mu_0-\mu_2)}{(\mu_1^2-\mu_2^2)(\mu_1 -\mu_2)} \nonumber\,\,,\\
	&c_{2,0}=-1-c_{1,0}=\frac{\mu_0-\mu_1}{\mu_1 -\mu_2} \nonumber\,\,,\\
	&c_{2,1}=-c_{0,1}-c_{1,1}=c_{0,1}c_{2,0}= -\frac{(\mu_0^2-\mu_2^2)(\mu_0-\mu_1)}{(\mu_1^2-\mu_2^2)(\mu_1 -\mu_2)} \nonumber\,\,,\\
	&c_{2,2}=-c_{2,0}-c_{2,1}=(1+c_{1,0})(1+c_{0,1})=\frac{(\mu_0^2-\mu_1^2)(\mu_0-\mu_1)}{(\mu_1^2-\mu_2^2)(\mu_1-\mu_2)}  \,\,.\label{cij-31}
	\end{eqnarray}
	By substituting (\ref{cij-31}) back into (\ref{G-31}) we get an expression in which the terms $Y_{n0},Y_{n2},Y_{0m}$ and $Y_{1m}$ are removed:
	\begin{eqnarray}
	\fl G_{31}= &\sum_{n=2}^{\infty}\frac{Y_{n1}}{n!} \frac{(\mu_0-\mu_1)(\mu_0-\mu_2)}{(\mu_1-\mu_2)(\mu_1+\mu_2)}\cdot A_{22}(\mu_0,\mu_1,\mu_2,n) + \nonumber \\
	\fl &\sum_{\stackrel[m=3]{n=2}{}}^{\infty} \frac{Y_{nm}}{n!m!}\frac{A_{22}(\mu_0,\mu_1,\mu_2,n)\cdot A_{11}(\mu_0,\mu_1,\mu_2,m)}{(\mu_1-\mu_2)^2(\mu_1+\mu_2)}  \label{combG-31} \,\,,
	\end{eqnarray}
	where $A_{22}$ and $A_{11}$ are again the factors from $Y_{22}$ and $Y_{11}$ bounds given by equations (\ref{A-22},\ref{A-11}), similarly to what happens when bounding $Y_{13}$ (see \ref{bound_on_Y13-3decoys}). Therefore the analysis of the coefficients' sign is the same as in \ref{bound_on_Y13-3decoys}. Hence a valid upper bound on $Y_{31}$ is obtained by setting $Y_{n1}\rightarrow 0$ (except for $Y_{31}$) and $Y_{nm}\rightarrow 1$ in (\ref{combG-31}) for all $n\geq 2$ and $m\geq 3$ in (\ref{combG-31}).
	Analogous steps to those in \ref{bound_on_Y13-3decoys} lead to the following upper bound:
	\begin{eqnarray}
	\fl &Y^U_{31} = -\frac{6(\mu_1+\mu_2)G_{31}}{(\mu_0-\mu_2)^2(\mu_0-\mu_1)^2(\mu_0+\mu_1+\mu_2)}+ \frac{6}{(\mu_0-\mu_2)^2(\mu_1-\mu_2)^2(\mu_0-\mu_1)^2(\mu_0+\mu_1+\mu_2)}
	 \nonumber\\
	\fl &\times \left[e^{\mu_2}(\mu_1-\mu_0)+e^{\mu_1}(\mu_0-\mu_2)+e^{\mu_0}(\mu_2-\mu_1)\right]  \nonumber\\
	\fl &\times \left[e^{\mu_2}(\mu^2_1-\mu^2_0)+e^{\mu_1}(\mu^2_0-\mu^2_2)+e^{\mu_0}(\mu^2_2-\mu^2_1)-(\mu_0-\mu_1)(\mu_1-\mu_2)(\mu_0-\mu_2)\right]
	\label{Y31-upperbound-3decoys}\,\,,
	\end{eqnarray}
	where $G_{31}$ is defined in (\ref{G-31}) and the coefficients of the combination of gains in (\ref{cij-31}).

	\subsection{Upper bound on $Y_{00}$}
	Consider the following combinations of gains in which all the terms $Y_{1m}$ and $Y_{n1}$ are removed:
	\begin{eqnarray}
		G_{00}^{0,1} &= \mu^2_1 \tilde{Q}^{0,0}+\mu^2_0 \tilde{Q}^{1,1} -\mu_0 \mu_1 (\tilde{Q}^{0,1} +\tilde{Q}^{1,0}) \nonumber\\
		&= \sum_{n,m=0}^{\infty} \frac{Y_{nm}}{n!m!} \left(\mu^n_0 \mu_1 -\mu_0 \mu^n_1\right) \left(\mu^m_0 \mu_1 -\mu_0 \mu^m_1\right) \,;\nonumber\\
		G_{00}^{0,2} &= \mu^2_2 \tilde{Q}^{0,0}+\mu^2_0 \tilde{Q}^{2,2} -\mu_0 \mu_2 (\tilde{Q}^{0,2} +\tilde{Q}^{2,0}) \nonumber\\
		&= \sum_{n,m=0}^{\infty} \frac{Y_{nm}}{n!m!} \left(\mu^n_0 \mu_2 -\mu_0 \mu^n_2\right) \left(\mu^m_0 \mu_2 -\mu_0 \mu^m_2\right) \,;\nonumber\\
		G_{00}^{1,2} &= \mu^2_2 \tilde{Q}^{1,1}+\mu^2_1 \tilde{Q}^{2,2} -\mu_1 \mu_2 (\tilde{Q}^{1,2} +\tilde{Q}^{2,1}) \nonumber\\
		&= \sum_{n,m=0}^{\infty} \frac{Y_{nm}}{n!m!} \left(\mu^n_1 \mu_2 -\mu_1 \mu^n_2\right) \left(\mu^m_1 \mu_2 -\mu_1 \mu^m_2\right) \,\,. \label{G-00-3decoys}
	\end{eqnarray}
	We now combine $G_{00}^{0,1},G_{00}^{0,2}$ and $G_{00}^{1,2}$ with arbitrary real coefficients $c_0$ and $c_1$ and impose that the terms $Y_{2m}$ and $Y_{n2}$ are also removed in the resulting expression:
	\begin{eqnarray}
	\fl	&G_{00}^{0,1}+c_0 \,G_{00}^{0,2}+c_1 \,G_{00}^{1,2} =  \sum_{n,m=0}^{\infty} \frac{Y_{nm}}{n!m!} \left[\left(\mu^n_0 \mu_1 -\mu_0 \mu^n_1\right) 
		\left(\mu^m_0 \mu_1 -\mu_0 \mu^m_1\right) \right. \nonumber\\
	\fl	&\left. +c_0 \left(\mu^n_0 \mu_2 -\mu_0 \mu^n_2\right) \left(\mu^m_0 \mu_2 -\mu_0 \mu^m_2\right) 
		+c_1 \left(\mu^n_1 \mu_2 -\mu_1 \mu^n_2\right) \left(\mu^m_1 \mu_2 -\mu_1 \mu^m_2\right)\right]  \label{combG-00} \,\,.
	\end{eqnarray}
	For $Y_{2m}$ and $Y_{n2}$ to be removed it suffices that for every $m$ it holds:
	\begin{eqnarray}
	\fl	(\mu_0^2\mu_1-\mu_0\mu^2_1)(\mu^m_0 \mu_1 -\mu_0 \mu^m_1)+c_0 (\mu_0^2\mu_2-\mu_0\mu^2_2)(\mu^m_0 \mu_2 -\mu_0 \mu^m_2)+c_1(\mu_1^2\mu_2-\mu_1\mu^2_2)(\mu^m_1 \mu_2 -\mu_1 \mu^m_2) = 0 \,,	\nonumber\\
	\fl \mbox{ }
	\end{eqnarray}
	which is fulfilled by:
	\begin{eqnarray}
		c_0 &= -\frac{\mu^2_1(\mu_0-\mu_1)}{\mu^2_2(\mu_0-\mu_2)} \label{c0-00} \,\,, \\
		c_1 &=\frac{ \mu^2_0(\mu_0-\mu_1)}{\mu^2_2(\mu_1-\mu_2)}  \label{c1-00} \,\,.
	\end{eqnarray}
	Substituting (\ref{c0-00}) and (\ref{c1-00}) back into (\ref{combG-00}) and multiplying both sides by $\mu^2_2$, we get an expression where all the terms $Y_{0m},Y_{2m},Y_{n0}$ and $Y_{n2}$ are removed and where the term $Y_{00}$ gives the largest contribution. More precisely, we find that:
	\begin{eqnarray}
	\fl	&\mu^2_2 G_{00}^{0,1} -\mu^2_1\frac{(\mu_0-\mu_1)}{(\mu_0-\mu_2)}\,G_{00}^{0,2} + \mu^2_0\frac{(\mu_0-\mu_1)}{(\mu_1-\mu_2)}\,G_{00}^{1,2} = \nonumber\\ 
	\fl	&Y_{00}\left[\mu_2^2 (\mu_0 -\mu_1)^2 -\mu_1^2 (\mu_0 -\mu_1)(\mu_0 -\mu_2) +\mu_0^2 (\mu_0 -\mu_1)(\mu_1 -\mu_2)\right]  \nonumber\\
	\fl	&+\sum_{m=3}^{\infty} \frac{Y_{0m}}{m!} \left[\mu_2^2 (\mu_1 -\mu_0)(\mu^m_0 \mu_1 -\mu_0 \mu^m_1) + \mu_1^2(\mu_0 -\mu_1)(\mu^m_0 \mu_2 -\mu_0 \mu^m_2) -
		        \mu_0^2 (\mu_0-\mu_1)(\mu^m_1 \mu_2 -\mu_1 \mu^m_2)\right]  \nonumber\\
	\fl	&+\sum_{n=3}^{\infty} \frac{Y_{n0}}{n!} \left[\mu_2^2 (\mu_1 -\mu_0)(\mu^n_0 \mu_1 -\mu_0 \mu^n_1) + \mu_1^2(\mu_0 -\mu_1)(\mu^n_0 \mu_2 -\mu_0 \mu^n_2) -
		  	    \mu_0^2 (\mu_0-\mu_1)(\mu^n_1 \mu_2 -\mu_1 \mu^n_2)\right]  \nonumber\\
	\fl	&+\sum_{n,m=3}^{\infty} \frac{Y_{nm}}{n!m!} \mu^2_0\mu^2_1\mu^2_2 \bigg[(\mu^{n-1}_0 -\mu^{n-1}_1)(\mu^{m-1}_0 -\mu^{m-1}_1) \nonumber\\
	\fl	  &- \frac{(\mu_0-\mu_1)}{(\mu_0-\mu_2)}(\mu^{n-1}_0 -\mu^{n-1}_2)(\mu^{m-1}_0 -\mu^{m-1}_2) +\frac{(\mu_0-\mu_1)}{(\mu_1-\mu_2)}(\mu^{n-1}_1 -\mu^{n-1}_2)(\mu^{m-1}_1 -\mu^{m-1}_2)\bigg] \,\,.\label{comb2G-00}
	\end{eqnarray}
	In order to extract an upper bound on $Y_{00}$ we need to study the sign of the yields' coefficients. We start by recasting the term corresponding to $Y_{00}$ as:
	\begin{eqnarray}
		&  Y_{00} (\mu_0 -\mu_1)\left[\mu_2^2(\mu_0 -\mu_1)-\mu_1^2 (\mu_0 -\mu_2) +\mu_0^2 (\mu_1 -\mu_2)\right]  \nonumber\\
		&= Y_{00} (\mu_0 -\mu_1)^2 (\mu_1-\mu_2)(\mu_0-\mu_2) \,\,.  \label{Y00-00}
	\end{eqnarray}
	We observe that the sign of this expression is determined by the factors $(\mu_1-\mu_2)(\mu_0-\mu_2)$.\\
	We then proceed by recasting each term of the sum corresponding to the $Y_{nm}$, with $n,m\geq 3$ in (\ref{comb2G-00}) as:
	\begin{eqnarray}
		\frac{Y_{nm}}{n!m!}\frac{A_{00}(\mu_0,\mu_1,\mu_2,m)\cdot A_{00}(\mu_0,\mu_1,\mu_2,n)}{(\mu_0-\mu_2)(\mu_1-\mu_2)} \,\,,\label{Ynm-00}
	\end{eqnarray}
	where
	\begin{eqnarray}
	\fl	A_{00}(\mu_0,\mu_1,\mu_2,m) &\equiv \mu_1^m (\mu^2_2\mu_0 - \mu_2\mu^2_0) + \mu_2^m (\mu^2_0\mu_1 - \mu_0\mu^2_1) + \mu_0^m (\mu^2_1\mu_2 - \mu_1\mu_2^2) \label{A-00}  \,\,.
	\end{eqnarray}
	This factor can be rewritten as:
	\begin{eqnarray}
	\fl	A_{00}(\mu_0,\mu_1,\mu_2,m) &= \mu_0\mu_1 \mu_2\left[\mu_1^{m-1} (\mu_2 - \mu_0) + \mu_2^{m-1} (\mu_0 - \mu_1) + \mu_0^{m-1} (\mu_1 - \mu_2)\right] \nonumber\\
	\fl &= -\mu_0 \mu_1 \mu_2 \, A_{22}(\mu_0,\mu_1,\mu_2,m-1) \,\,, \label{A-00_1}
	\end{eqnarray}
	where $A_{22}$ is defined as (\ref{A-22}) in \ref{bound_on_Y22-3decoys}. Thus we can use the result (\ref{A-22_1}) obtained in \ref{bound_on_Y22-3decoys} to directly recast $A_{00}$ as:
	\begin{eqnarray}
		A_{00}(\mu_0,\mu_1,\mu_2,m)=  \mu_0\mu_1 \mu_2 (\mu_0 - \mu_2)(\mu_1 - \mu_2)\sum_{k=0}^{m-2} \mu_2^k (\mu_0^{m-2-k}-\mu_1^{m-2-k})  \,\,. \label{A-00_2}
	\end{eqnarray}
	By substituting (\ref{A-00_2}) back into (\ref{Ynm-00}), we get the final expression for each term of the sum corresponding to the $Y_{nm}$, with $n,m\geq 3$ in (\ref{comb2G-00}):
	\begin{eqnarray}
	\fl	\frac{Y_{nm}}{n!m!}\mu^2_0 \mu^2_1 \mu^2_2(\mu_0-\mu_2)(\mu_1-\mu_2)\left[\sum_{k=0}^{m-2} \mu_2^k (\mu_0^{m-2-k}-\mu_1^{m-2-k})\right]\left[\sum_{k=0}^{n-2} \mu_2^k (\mu_0^{n-2-k}-\mu_1^{n-2-k})\right]
		   \label{Ynmcoeff-00} \,\,.
	\end{eqnarray}
	which has manifestly the same sign as the expression given by (\ref{Y00-00}), for any value of the intensities (the product of the last two factors is always positive).\\
	Finally, we recast the $Y_{0m}$'s terms ($Y_{n0}$'s terms are identical under the replacement $m\rightarrow n$) as:
	\begin{eqnarray}
		 &\frac{Y_{0m}}{m!}\mu_0\mu_1 \mu_2(\mu_0-\mu_1) \left[\mu_2 (\mu^{m-1}_1  - \mu^{m-1}_0) + \mu_1(\mu^{m-1}_0 - \mu^{m-1}_2) - \mu_0(\mu^{m-1}_1 - \mu^{m-1}_2)\right]  \nonumber\\
		 &=\frac{Y_{0m}}{m!}(\mu_0-\mu_1) A_{00}(\mu_0,\mu_1,\mu_2,m) \nonumber\\
		 &= \frac{Y_{0m}}{m!}\mu_0\mu_1 \mu_2 (\mu_0-\mu_1) (\mu_0 - \mu_2)(\mu_1 - \mu_2)\sum_{k=0}^{m-2} \mu_2^k (\mu_0^{m-2-k}-\mu_1^{m-2-k}) \,\,,\label{Y0mcoeff-00}
	\end{eqnarray}
	where we employed (\ref{A-00_1}) in the first equality and (\ref{A-00_2}) in the second one. We observe that the sign of the $Y_{0m}$'s terms is again determined by the factors $(\mu_0 - \mu_2)(\mu_1 - \mu_2)$.\\
	We conclude that the coefficients of $Y_{0m}$, $Y_{n0}$ and $Y_{nm}$, with $n,m\geq 3$, carry the same sign as $Y_{00}$'s, which implies that a valid upper bound on $Y_{00}$ is obtained by setting all the other yields to zero in (\ref{comb2G-00}). In so doing, we find that:
	\begin{eqnarray}
		Y^U_{00}=\frac{\frac{\mu_2^2 G_{00}^{0,1}}{\mu_0 - \mu_1} -\frac{\mu_1^2 G_{00}^{0,2}}{\mu_0-\mu_2} + \frac{\mu_0^2 G_{00}^{1,2}}{\mu_1-\mu_2}}{(\mu_0-\mu_1)(\mu_0-\mu_2)(\mu_1-\mu_2)}  \,\,.
		    \label{Y00-upperbound-3decoys}
	\end{eqnarray}

	\section{Yields' bounds with four decoys} \label{yields-bounds-4decoys}
	Here we derive analytical upper bounds on the yields appearing in (\ref{phase-error-rate}), following the same lines of \autoref{yields-bounds-2decoys}. In this case we assume that Alice and Bob can prepare their phase-randomized coherent pulses with four different intensity settings: $\{\mu_0,\mu_1,\mu_2,\mu_3\}$, which are the same for both parties. This choice is optimal since we assumed that the two optical channels linking the parties to the central node $C$ have equal transmittance $\sqrt{\eta}$ \cite{asymmetric-MDI-QKD}.\\
	The whole set of infinite yields is subjected to the following sixteen equality constraints:
	\begin{equation}
	\tilde{Q}^{k,l} \equiv e^{\mu_k + \mu_l} Q^{k,l} =\sum_{n,m=0}^{\infty} \frac{Y_{nm}}{n!m!} {\mu_k}^n {\mu_l}^m \quad k,l \in \{0,1,2,3\} \,\,, \label{constr-4decoys}
	\end{equation}
	and to the same inequality constraints given by (\ref{ineq-constr}).\\
	In this appendix we only obtain bounds on the yields $Y_{13},Y_{31},Y_{04}$ and $Y_{40}$ since the bounds derived on the yields $Y_{00}, Y_{11}, Y_{02}, Y_{20}$ and $Y_{22}$ in \ref{yields-bounds-3decoys} are already good enough, i.e bounding them with one additional decoy intensity would not result in a significant improvement of the performance of the protocol.
	
	\subsection{Upper bound on $Y_{04}$} \label{bound_on_Y04-4decoys}
	Consider the following combinations of gains in which all the terms $Y_{1m}$ and $Y_{n0}$ are removed:
	\begin{eqnarray}
	\fl G_{04}^{i,j} &= \mu_j\tilde{Q}^{i,i}+ \mu_i\tilde{Q}^{j,j} - \mu_j\tilde{Q}^{i,j} -\mu_i\tilde{Q}^{j,i}= \sum_{n,m=0}^{\infty} \frac{Y_{nm}}{n!m!}(\mu_j \mu_i^n -\mu_i \mu_j^n)(\mu_i^m -\mu_j^m)  \,\,,
	  \label{G-04-4decoys} 
	\end{eqnarray}
	where $i,j \in\{0,1,2,3\} $. Since $G_{04}^{i,i}=0$ and $G_{04}^{i,j}=G_{04}^{j,i}$, we only have six distinct combinations that read (for $j>i$): $G_{04}^{0,1},G_{04}^{0,2},G_{04}^{0,3},G_{04}^{1,2},G_{04}^{1,3},G_{04}^{2,3}$.\\
	We now take the linear combination of the $G_{04}^{i,j}$ such that even the yields $Y_{2m},Y_{3m},Y_{n1}$ and $Y_{n2}$ are removed:
	\begin{eqnarray}
	\sum_{j>i} c_{i,j} G_{04}^{i,j} = \sum_{n,m=0}^{\infty} \frac{Y_{nm}}{n!m!} \sum_{j>i} c_{i,j} (\mu_j \mu_i^n -\mu_i \mu_j^n)(\mu_i^m -\mu_j^m)  \,\,, \label{combG-04} 
	\end{eqnarray}
	where we implicitly assume that both indexes $i,j$ run over the set $\{0,1,2,3\}$.
	For $Y_{2m},Y_{3m},Y_{n1}$ and $Y_{n2}$ to be removed, the real coefficients $c_{i,j}$ must satisfy:
	\begin{eqnarray}
	\left\{
	{\begin{array}{lcl}
	 \sum_{j>i} c_{i,j} (\mu_j \mu_i^2 -\mu_i \mu_j^2)(\mu_i^m -\mu_j^m) & = &0  \quad\forall\,m \\
	 \sum_{j>i} c_{i,j} (\mu_j \mu_i^3 -\mu_i \mu_j^3)(\mu_i^m -\mu_j^m) & = &0  \quad\forall\,m \\
	 \sum_{j>i} c_{i,j} (\mu_j \mu_i^n -\mu_i \mu_j^n)(\mu_i -\mu_j) & = &0  \quad\forall\,n \\
	 \sum_{j>i} c_{i,j} (\mu_j \mu_i^n -\mu_i \mu_j^n)(\mu_i^2 -\mu_j^2) & = &0  \quad\forall\,n \\
	\end{array}}
	 \right.  \label{system-04}
	\end{eqnarray}
	In order to solve system (\ref{system-04}), we look for those coefficients $c_{i,j}$ such that the multiplicative factors of $\mu_i^m$ and $\mu_i^n$ (for $i=0,1,2,3$) are all set to zero. This corresponds to imposing sixteen conditions on the six coefficients $c_{i,j}$. These conditions are not all independent, and a solution can be found even when we require (for simplicity) that $c_{0,1}=1$:
	\begin{eqnarray}
		&c_{0,1}=1 \,\,,\nonumber\\
		&c_{0,2}= -\frac{(\mu_0-\mu_1)\mu_1(\mu_1-\mu_3)}{(\mu_0-\mu_2)\mu_2 (\mu_2-\mu_3)} \,\,, \nonumber\\
		&c_{0,3}= \frac{(\mu_0-\mu_1)\mu_1(\mu_1-\mu_2)}{(\mu_0-\mu_3)\mu_3 (\mu_2-\mu_3)}  \,\,,\nonumber\\
		&c_{1,2}= \frac{(\mu_0-\mu_1)\mu_0(\mu_0-\mu_3)}{(\mu_1-\mu_2)\mu_2 (\mu_2-\mu_3)}  \,\,,\nonumber\\
		&c_{1,3}= -\frac{(\mu_0-\mu_1)\mu_0(\mu_0-\mu_2)}{(\mu_1-\mu_3)\mu_3 (\mu_2-\mu_3)}  \,\,,\nonumber\\
		&c_{2,3}= \frac{\mu_0\mu_1(\mu_0-\mu_1)^2}{\mu_2\mu_3 (\mu_2-\mu_3)^2}  \label{cij}\,\,.
	\end{eqnarray}
	By substituting the solution for the coefficients given by (\ref{cij}) back into (\ref{combG-04}), one gets:
	\begin{eqnarray}
	\fl \sum_{j>i} c_{i,j} G_{04}^{i,j} = \sum_{m=3}^{\infty}\frac{Y_{0m}}{m!} A_{04}(\mu_0,\mu_1,\mu_2,\mu_3,m) + \sum_{\stackrel[m=3]{n=4}{}}^{\infty} \frac{Y_{nm}}{n!m!} 
	B_{04}(\mu_0,\mu_1,\mu_2,\mu_3,n,m)\,\,,   \label{comb2G-04} 
	\end{eqnarray}
	where:
	\begin{eqnarray}
	 \fl &A_{04}(\mu_0,\mu_1,\mu_2,\mu_3,m) = -\frac{(\mu_0 -\mu_1)}{\mu_2 \mu_3 (\mu_2 - \mu_3)} \left[\mu_0^m (\mu_1 -\mu_2)(\mu_1 -\mu_3)(\mu_2 -\mu_3)-\mu_1^m (\mu_0 -\mu_2)(\mu_0 -\mu_3)(\mu_2 -\mu_3)  \right. \nonumber\\
	 \fl &\left.+\mu_2^m (\mu_0 -\mu_1)(\mu_0 -\mu_3)(\mu_1 -\mu_3)- \mu_3^m (\mu_0 -\mu_1)(\mu_0 -\mu_2)(\mu_1 -\mu_2) \right]   \nonumber \\ 
	 \fl &= -\frac{(\mu_0 -\mu_1)^2 (\mu_0 -\mu_2)(\mu_1 -\mu_2) (\mu_0 -\mu_3)(\mu_1 -\mu_3)}{\mu_2 \mu_3} \left(\sum_{i_1 \leq i_2 \leq \dots \leq i_{m-3}} \mu_{i_1}\mu_{i_2} \cdot \dots \cdot \mu_{i_{m-3}}\right) \label{A0m}\,\,,
	\end{eqnarray}
	and
	\begin{eqnarray}
	\fl &B_{04}(\mu_0,\mu_1,\mu_2,\mu_3,n,m) = \frac{-\mu_0 \mu_1}{(\mu_0 -\mu_2)(\mu_1 -\mu_2)(\mu_1 -\mu_3)(\mu_0 -\mu_3)(\mu_2 -\mu_3)^2} \left[ \mu_0^m (\mu_1 -\mu_2)(\mu_1 -\mu_3)(\mu_2 -\mu_3)\right. \nonumber\\
	 \fl &\left.-\mu_1^m (\mu_0 -\mu_2)(\mu_0 -\mu_3)(\mu_2 -\mu_3) +\mu_2^m (\mu_0 -\mu_1)(\mu_0 -\mu_3)(\mu_1 -\mu_3)- \mu_3^m (\mu_0 -\mu_1)(\mu_0 -\mu_2)(\mu_1 -\mu_2) \right] \nonumber\\
	 \fl &\times \left[- \mu_0^{n-1} (\mu_1 -\mu_2)(\mu_1 -\mu_3)(\mu_2 -\mu_3)+\mu_1^{n-1} (\mu_0 -\mu_2)(\mu_0 -\mu_3)(\mu_2 -\mu_3) \right. \nonumber\\
	 \fl & \left.-\mu_2^{n-1} (\mu_0 -\mu_1)(\mu_0 -\mu_3)(\mu_1 -\mu_3)+\mu_3^{n-1} (\mu_0 -\mu_1)(\mu_0 -\mu_2)(\mu_1 -\mu_2)\right] \nonumber \\
	 \fl &=  -\mu_0\mu_1 \mu_2 \mu_3 \,A_{04}(\mu_0,\mu_1,\mu_2,\mu_3,m)\cdot \left(\sum_{i_1 \leq i_2 \leq \dots \leq i_{n-4}} \mu_{i_1}\mu_{i_2} \cdot \dots \cdot \mu_{i_{n-4}}\right) \label{Bnm}  \,\,.
	\end{eqnarray}
	In (\ref{A0m},\ref{Bnm}) we again assume that the indexes in the sums run over the set $\{0,1,2,3\}$ and we define $\sum_{i_1 \leq i_2 \leq \dots \leq i_{m-3}} \mu_{i_1}\mu_{i_2} \cdot \dots \cdot \mu_{i_{m-3}}|_{m=3} =1$. From (\ref{A0m}) we deduce that the sign of $Y_{0m}$'s coefficient is independent of $m$, while from (\ref{Bnm}) we deduce that $Y_{nm}$'s coefficient has always opposite sign to that of $Y_{0m}$. Therefore a valid upper bound on $Y_{04}$ is obtained by setting to zero all the other yields $Y_{0m}$ and to 1 the yields $Y_{nm}$ with $n\geq 4$ and $m\geq 3$ in (\ref{comb2G-04}). We thus obtain:
	\begin{eqnarray}
	 \sum_{j>i} c_{i,j} G_{04}^{i,j} = \frac{Y^U_{04}}{4!} A_{04}(\mu_0,\mu_1,\mu_2,\mu_3,4) + \sum_{\stackrel[m=3]{n=4}{}}^{\infty} \frac{B_{04}(\mu_0,\mu_1,\mu_2,\mu_3,n,m)}{n!m!} \,\,, 
	\end{eqnarray}
	which implies the following upper bound on $Y_{04}$:
	\begin{eqnarray}
	Y^U_{04} = \frac{4!}{A_{04}(\mu_0,\mu_1,\mu_2,\mu_3,4)} \left[\sum_{j>i} c_{i,j} G_{04}^{i,j} - \sum_{\stackrel[m=3]{n=4}{}}^{\infty} \frac{B_{04}(\mu_0,\mu_1,\mu_2,\mu_3,n,m)}{n!m!} \right]  \label{Y04-upperbound-4decoys}\,\,,
	\end{eqnarray}
	where $c_{i,j}$ are given in (\ref{cij}), $G_{04}^{i,j}$ is defined in (\ref{G-04-4decoys}), the coefficient $A_{04}$ reads:
	\begin{eqnarray}
	\fl &A_{04}(\mu_0,\mu_1,\mu_2,\mu_3,4) = -\frac{(\mu_0 -\mu_1)^2 (\mu_0 -\mu_2)(\mu_1 -\mu_2) (\mu_0 -\mu_3)(\mu_1 -\mu_3)(\mu_0+\mu_1+\mu_2+\mu_3)}{\mu_2 \mu_3}  \label{A04}\,\,,
	\end{eqnarray}
	and the sum over the coefficient $B_{04}$ reads:
	\begin{eqnarray}
	\fl &\sum_{\stackrel[m=3]{n=4}{}}^{\infty} \frac{B_{04}(\mu_0,\mu_1,\mu_2,\mu_3,n,m)}{n!m!} = \frac{\mu_0 \mu_1}{(\mu_0 -\mu_2)(\mu_1 -\mu_2)(\mu_1 -\mu_3)(\mu_0 -\mu_3)(\mu_2 -\mu_3)^2}\,\,  \nonumber\\
	\fl &\times \left[ (e^{\mu_0}-1-\mu_0 -\frac{\mu_0^2}{2}) (\mu_1 -\mu_2)(\mu_1 -\mu_3)(\mu_2 -\mu_3)-(e^{\mu_1}-1-\mu_1 -\frac{\mu_1^2}{2}) (\mu_0 -\mu_2)(\mu_0 -\mu_3)(\mu_2 -\mu_3) \right. \nonumber\\
	\fl &\left. +(e^{\mu_2}-1-\mu_2 -\frac{\mu_2^2}{2}) (\mu_0 -\mu_1)(\mu_0 -\mu_3)(\mu_1 -\mu_3)- (e^{\mu_3}-1-\mu_3 -\frac{\mu_3^2}{2}) (\mu_0 -\mu_1)(\mu_0 -\mu_2)(\mu_1 -\mu_2) \right]^2  \,\,. \nonumber\\
	\fl \label{sum-Bnm}
	\end{eqnarray}

	\subsection{Upper bound on $Y_{40}$} \label{bound_on_Y40-4decoys}
	Consider the following combinations of gains in which all the terms $Y_{0m}$ and $Y_{n1}$ are removed:
	\begin{eqnarray}
	\fl G_{40}^{i,j} &= \mu_j\tilde{Q}^{i,i}+ \mu_i\tilde{Q}^{j,j} - \mu_i\tilde{Q}^{i,j} -\mu_j\tilde{Q}^{j,i}= \sum_{n,m=0}^{\infty} \frac{Y_{nm}}{n!m!}(\mu_i^n -\mu_j^n)(\mu_j \mu_i^m -\mu_i \mu_j^m) \,\,,  \label{G-40-4decoys} 
	\end{eqnarray}
	where $i,j \in\{0,1,2,3\} $. Since $G_{40}^{i,i}=0$ and $G_{40}^{i,j}=G_{40}^{j,i}$, we only have six distinct combinations that read (for $j>i$): $G_{40}^{0,1},G_{40}^{0,2},G_{40}^{0,3},G_{40}^{1,2},G_{40}^{1,3},G_{40}^{2,3}$.\\
	We now take the linear combination of the $G_{40}^{i,j}$ such that even the yields $Y_{1m},Y_{2m},Y_{n2}$ and $Y_{n3}$ are removed:
	\begin{eqnarray}
	\sum_{j>i} c_{i,j} G_{40}^{i,j} = \sum_{n,m=0}^{\infty} \frac{Y_{nm}}{n!m!} \sum_{j>i} c_{i,j} (\mu_i^n -\mu_j^n)(\mu_j \mu_i^m -\mu_i \mu_j^m)   \label{combG-40} \,\,,
	\end{eqnarray}
	where we implicitly assume that both indexes $i,j$ run over the set $\{0,1,2,3\}$.
	For $Y_{1m},Y_{2m},Y_{n2}$ and $Y_{n3}$ to be removed, the real coefficients $c_{i,j}$ must satisfy:
	\begin{eqnarray}
	\left\{
	{\begin{array}{lcl}
	 \sum_{j>i} c_{i,j} (\mu_i^n -\mu_j^n)(\mu_j \mu_i^2 -\mu_i \mu_j^2) & = &0  \quad\forall\,n \\
	 \sum_{j>i} c_{i,j} (\mu_i^n -\mu_j^n)(\mu_j \mu_i^3 -\mu_i \mu_j^3) & = &0  \quad\forall\,n \\
	 \sum_{j>i} c_{i,j} (\mu_i -\mu_j)(\mu_j \mu_i^m -\mu_i \mu_j^m) & = &0  \quad\forall\,m \\
	 \sum_{j>i} c_{i,j} (\mu_i^2 -\mu_j^2)(\mu_j \mu_i^m -\mu_i \mu_j^m) & = &0  \quad\forall\,m\,. \\
	\end{array}}
	 \right.  \label{system-40}
	\end{eqnarray}
	We now notice that the system (\ref{system-40}) is exactly the same system solved in \ref{bound_on_Y40-4decoys} while bounding $Y_{04}$, thus the solution for the coefficients $c_{i,j}$ is given in (\ref{cij}).
	By substituting the solution (\ref{cij}) back into (\ref{combG-40}), one gets:
	\begin{eqnarray}
	\fl \sum_{j>i} c_{i,j} G_{40}^{i,j} = \sum_{n=3}^{\infty}\frac{Y_{n0}}{n!} A_{04}(\mu_0,\mu_1,\mu_2,\mu_3,n) + \sum_{\stackrel[m=4]{n=3}{}}^{\infty} \frac{Y_{nm}}{n!m!} B_{04}(\mu_0,\mu_1,\mu_2,\mu_3,m,n)    \label{comb2G-40} \,\,,
	\end{eqnarray}
	where $A_{04}$ and $B_{04}$ are the coefficients defined in (\ref{A0m},\ref{Bnm}) while bounding $Y_{04}$. Hence we can adopt the observations made on the sign of $A_{04}$ and $B_{04}$ from \ref{bound_on_Y04-4decoys} and conclude that a valid upper bound on $Y_{40}$ is obtained by setting to zero all the other yields $Y_{n0}$ and to 1 the yields $Y_{nm}$ with $n\geq 3$ and $m\geq 4$ in (\ref{comb2G-40}). The upper bound on $Y_{40}$ then reads:
	\begin{eqnarray}
	 Y^U_{40} = \frac{4!}{A_{04}(\mu_0,\mu_1,\mu_2,\mu_3,4)} \left[\sum_{j>i} c_{i,j} G_{40}^{i,j} - \sum_{\stackrel[m=4]{n=3}{}}^{\infty} \frac{B_{04}(\mu_0,\mu_1,\mu_2,\mu_3,m,n)}{n!m!} \right]  \label{Y40-upperbound-4decoys}\,\,,
	\end{eqnarray}
	where $c_{i,j}$, $G_{40}^{i,j}$, $A_{04}(\mu_0,\mu_1,\mu_2,\mu_3,4)$ and the sum over $B_{04}$ are given in (\ref{cij}), (\ref{G-40-4decoys}), (\ref{A04}) and (\ref{sum-Bnm}), respectively.

	\subsection{Upper bound on $Y_{13}$} \label{bound_on_Y13-4decoys}
	We consider the most general combination of all sixteen gains:
	\begin{eqnarray}
		\sum_{i,j=0}^{3} c_{i,j} \tilde{Q}^{i,j} = \sum_{n,m=0}^{\infty} \frac{Y_{nm}}{n!m!}\left[\sum_{i,j=0}^{3} c_{i,j} \mu_i^n \mu_j^m \right]\,\,, \label{G-13-4decoys}
	\end{eqnarray}
	and require that the terms $Y_{n0},Y_{n1},Y_{n2},Y_{0m},Y_{2m}$ and $Y_{3m}$ are removed, by imposing proper conditions on the real coefficients $c_{i,j}$:
	\begin{eqnarray}
	\fl &Y_{n0}\mbox{ removed:}\quad \sum_{i,j=0}^{3} c_{i,j} \mu_i^n =0 \quad\forall\, n \quad\Leftarrow\quad \sum_{j=0}^{3} c_{i,j}=0  \quad \mbox{for} \quad i=0,1,2,3  \label{Yn0removed-13-4decoys} \\
	\fl &Y_{n1}\mbox{ removed:}\quad \sum_{i,j=0}^{3} c_{i,j} \mu_i^n \mu_j =0 \quad\forall\, n \quad\Leftarrow\quad \sum_{j=0}^{3} c_{i,j} \mu_j =0 \quad \mbox{for} \quad i=0,1,2,3  \label{Yn1removed-13-4decoys} \\
	\fl &Y_{n2}\mbox{ removed:}\quad \sum_{i,j=0}^{3} c_{i,j} \mu_i^n \mu_j^2 =0 \quad\forall\, n \quad\Leftarrow\quad \sum_{j=0}^{3} c_{i,j} \mu_j^2 =0 \quad \mbox{for} \quad i=0,1,2,3  \label{Yn2removed-13-4decoys} \\
	\fl &Y_{0m}\mbox{ removed:}\quad \sum_{i,j=0}^{3} c_{i,j} \mu_j^m=0 \quad\forall\, m \quad\Leftarrow\quad \sum_{i=0}^{3} c_{i,j}=0 \quad \mbox{for} \quad j=0,1,2,3  \label{Y0mremoved-13-4decoys} \\
	\fl &Y_{2m}\mbox{ removed:}\quad \sum_{i,j=0}^{3} c_{i,j} \mu_i^2 \mu_j^m=0 \quad\forall\, m \quad\Leftarrow\quad \sum_{i=0}^{3} c_{i,j} \mu_i^2 =0  \quad \mbox{for} \quad j=0,1,2,3  \label{Y2mremoved-13-4decoys} \\
	\fl &Y_{3m}\mbox{ removed:}\quad \sum_{i,j=0}^{3} c_{i,j} \mu_i^3 \mu_j^m=0 \quad\forall\, m \quad\Leftarrow\quad \sum_{i=0}^{3} c_{i,j} \mu_i^3 =0  \quad \mbox{for} \quad j=0,1,2,3  \label{Y3mremoved-13-4decoys} \,\,.
	\end{eqnarray}
	The twenty-four conditions given by (\ref{Yn0removed-13-4decoys} - \ref{Y3mremoved-13-4decoys}) form an over-determined system of equations for the sixteen variables $c_{i,j}$. However, thanks to the symmetries of the problem, a unique solution for $c_{i,j}$ exists and reads (we rescale every coefficient by requiring $c_{0,0}=1$):
	\begin{eqnarray}
		&c_{0,0}=1 \,\,,\nonumber\\
		&c_{0,1}=\frac{(\mu_0-\mu_2) (\mu_0-\mu_3)}{(\mu_2-\mu_1) (\mu_1-\mu_3)}  \,\,,\nonumber\\
		&c_{0,2}=\frac{(\mu_0-\mu_1) (\mu_0-\mu_3)}{(\mu_1-\mu_2) (\mu_2-\mu_3)} \,\,, \nonumber\\
		&c_{0,3}=\frac{(\mu_0-\mu_1) (\mu_0-\mu_2)}{(\mu_1-\mu_3) (\mu_3-\mu_2)} \,\,,\nonumber\\
		&c_{1,0}=-\frac{(\mu_0-\mu_2) (\mu_0-\mu_3) [\mu_0 (\mu_2+\mu_3)+\mu_2 \mu_3]}{(\mu_1-\mu_2)(\mu_1-\mu_3) [\mu_1 (\mu_2+\mu_3)+\mu_2 \mu_3]} \,\,,\nonumber\\
		&c_{1,1}=\frac{(\mu_0-\mu_2)^2 (\mu_0-\mu_3)^2 [\mu_0 (\mu_2+\mu_3)+\mu_2 \mu_3]}{(\mu_1-\mu_2)^2 (\mu_1-\mu_3)^2 [\mu_1 (\mu_2+\mu_3)+\mu_2 \mu_3]} \,\,,\nonumber\\
		&c_{1,2}=-\frac{(\mu_0-\mu_1) (\mu_0-\mu_2) (\mu_0-\mu_3)^2 [\mu_0 (\mu_2+\mu_3)+\mu_2 \mu_3]}{(\mu_1-\mu_2)^2 (\mu_1-\mu_3) (\mu_2-\mu_3) [\mu_1 (\mu_2+\mu_3)+\mu_2 \mu_3]} \,\,,\nonumber\\
		&c_{1,3}=\frac{(\mu_0-\mu_1) (\mu_0-\mu_2)^2 (\mu_0-\mu_3) [\mu_0 (\mu_2+\mu_3)+\mu_2 \mu_3]}{(\mu_1-\mu_2) (\mu_1-\mu_3)^2 (\mu_2-\mu_3) [\mu_1 (\mu_2+\mu_3)+\mu_2 \mu_3]} \,\,,\nonumber\\
		&c_{2,0}= \frac{(\mu_0-\mu_1) (\mu_0-\mu_3) [\mu_0 (\mu_1+\mu_3)+\mu_1 \mu_3]}{(\mu_1-\mu_2) (\mu_2-\mu_3) [\mu_1 (\mu_2+\mu_3)+\mu_2 \mu_3]}\,\,,\nonumber\\
		&c_{2,1}= -\frac{(\mu_0-\mu_1) (\mu_0-\mu_2) (\mu_0-\mu_3)^2 [\mu_0 (\mu_1+\mu_3)+\mu_1 \mu_3]}{(\mu_1-\mu_2)^2 (\mu_1-\mu_3) (\mu_2-\mu_3) [\mu_1 (\mu_2+\mu_3)+\mu_2 \mu_3]} \,\,,\nonumber\\
		&c_{2,2}= \frac{(\mu_0-\mu_1)^2 (\mu_0-\mu_3)^2 [\mu_0 (\mu_1+\mu_3)+\mu_1 \mu_3]}{(\mu_1-\mu_2)^2 (\mu_2-\mu_3)^2 [\mu_1 (\mu_2+\mu_3)+\mu_2 \mu_3]} \,\,,\nonumber\\
		&c_{2,3}=-\frac{(\mu_0-\mu_1)^2 (\mu_0-\mu_2) (\mu_0-\mu_3) [\mu_0 (\mu_1+\mu_3)+\mu_1 \mu_3]}{(\mu_1-\mu_2) (\mu_1-\mu_3) (\mu_2-\mu_3)^2 [\mu_1 (\mu_2+\mu_3)+\mu_2 \mu_3]} \,\,,\nonumber\\
		&c_{3,0}=\frac{(\mu_0-\mu_1) (\mu_0-\mu_2) [\mu_0 (\mu_1+\mu_2)+\mu_1 \mu_2]}{(\mu_1-\mu_3) (\mu_3-\mu_2) [\mu_1 (\mu_2+\mu_3)+\mu_2 \mu_3]} \,\,,\nonumber\\
		&c_{3,1}=\frac{(\mu_0-\mu_1) (\mu_0-\mu_2)^2 (\mu_0-\mu_3) [\mu_0 (\mu_1+\mu_2)+\mu_1 \mu_2]}{(\mu_1-\mu_2) (\mu_1-\mu_3)^2 (\mu_2-\mu_3) [\mu_1 (\mu_2+\mu_3)+\mu_2 \mu_3]} \,\,,\nonumber\\
		&c_{3,2}=-\frac{(\mu_0-\mu_1)^2 (\mu_0-\mu_2) (\mu_0-\mu_3) [\mu_0 (\mu_1+\mu_2)+\mu_1 \mu_2]}{(\mu_1-\mu_2) (\mu_1-\mu_3) (\mu_2-\mu_3)^2 [\mu_1 (\mu_2+\mu_3)+\mu_2 \mu_3]} \,\,,\nonumber\\
		&c_{3,3}=\frac{(\mu_0-\mu_1)^2 (\mu_0-\mu_2)^2 [\mu_0 (\mu_1+\mu_2)+\mu_1 \mu_2]}{(\mu_1-\mu_3)^2  (\mu_2-\mu_3)^2 [\mu_1 (\mu_2+\mu_3)+\mu_2 \mu_3]}    \label{cij-13-4decoys}\,\,.
	\end{eqnarray}
	By substituting these expressions back into (\ref{G-13-4decoys}) and by making some simplifications, one gets:
	\begin{eqnarray}
	 \fl	\sum_{i,j=0}^{3} c_{i,j} \tilde{Q}^{i,j} = \sum_{m=3}^{\infty}\frac{Y_{1m}}{m!} A_{13}(\mu_0,\mu_1,\mu_2,\mu_3,m) + \sum_{\stackrel[m=3]{n=4}{}}^{\infty} \frac{Y_{nm}}{n!m!} A_{13}(\mu_0,\mu_1,\mu_2,\mu_3,m)\cdot C_n   \,\,,\label{combG-13-4decoys} 
	\end{eqnarray}
	where:
	\begin{eqnarray}
	\fl  A_{13}(\mu_0,\mu_1,\mu_2,\mu_3,m) &= \frac{(\mu_0 -\mu_1)^2 (\mu_0 -\mu_2)^2(\mu_0 -\mu_3)^2 }{\mu_2 \mu_3+ \mu_1 \mu_2+\mu_1 \mu_3} \left(\sum_{i_1 \leq i_2 \leq \dots \leq i_{m-3}} \mu_{i_1}\mu_{i_2} \cdot \dots \cdot \mu_{i_{m-3}}\right) \,\,, \label{A1m}
	\end{eqnarray}
	and $C_n$ ($n \geq 5$) is defined recursively as:
	\begin{eqnarray}
	 \fl  \left\{
	 	{\begin{array}{lcl}
	 	C_{n} &=& \left[\sum_{j=1}^{n-4} (\mu^j_0+\mu^j_1+\mu^j_2+\mu^j_3) C_{n-j} -\mu_0 \mu_1 \mu_2 \mu_3 \left(\sum_{i_1 \leq i_2 \leq \dots \leq i_{n-5}} \mu_{i_1}\mu_{i_2} \cdot \dots \cdot
	    \mu_{i_{n-5}}\right)\right]/(n-4)   \nonumber \\
	  C_4 &=& \mu_0 \mu_1 \mu_2 +\mu_0 \mu_1 \mu_3 +\mu_0 \mu_2 \mu_3 + \mu_1 \mu_2 \mu_3  \,\,.
	  	\end{array}}
	  	 \right.  \nonumber\\ \fl  \label{Cn}  
	\end{eqnarray}
	In (\ref{A1m},\ref{Cn}) we assume that the indexes $i_j$ in the sums run over the set $\{0,1,2,3\}$ and we define $\sum_{i_1 \leq i_2 \leq \dots \leq i_{m-3}} \mu_{i_1}\mu_{i_2} \cdot \dots \cdot \mu_{i_{m-3}}|_{m=3} =1$. From (\ref{A1m}) we deduce that the sign of $Y_{1m}$'s coefficient is always positive, while from (\ref{Cn}) we deduce that $Y_{nm}$'s coefficient has always equal sign to that of $Y_{1m}$, since $C_n$ is always a positive quantity. Therefore a valid upper bound on $Y_{13}$ is obtained by setting to zero all the other yields  in (\ref{combG-13-4decoys}). The upper bound on $Y_{13}$ then reads:
	\begin{eqnarray}
	 Y_{13}^U = \frac{6}{A_{13}(\mu_0,\mu_1,\mu_2,\mu_3,3)} \left(	\sum_{i,j=0}^{3} c_{i,j} \tilde{Q}^{i,j}\right)\,\,, \label{Y13-upperbound-4decoys}
	\end{eqnarray}
	where $c_{i,j}$ are defined in (\ref{cij-13-4decoys}) and $A_{13}(\mu_0,\mu_1,\mu_2,\mu_3,3)$ reads:
	\begin{eqnarray}
	 A_{13}(\mu_0,\mu_1,\mu_2,\mu_3,3)=  \frac{(\mu_0 -\mu_1)^2 (\mu_0 -\mu_2)^2(\mu_0 -\mu_3)^2 }{\mu_2 \mu_3+ \mu_1 \mu_2+\mu_1 \mu_3} \,\,. \label{A13}
	\end{eqnarray}

	\subsection{Upper bound on $Y_{31}$} \label{bound_on_Y31-4decoys}
	We consider the most general combination of all sixteen gains:
	\begin{eqnarray}
		\sum_{i,j=0}^{3} c_{i,j} \tilde{Q}^{i,j} = \sum_{n,m=0}^{\infty} \frac{Y_{nm}}{n!m!}\left[\sum_{i,j=0}^{3} c_{i,j} \mu_i^n \mu_j^m \right] \label{G-31-4decoys}\,\,,
	\end{eqnarray}
	and require that the terms $Y_{n0},Y_{n2},Y_{n3},Y_{0m},Y_{1m}$ and $Y_{2m}$ are removed, by imposing proper conditions on the real coefficients $c_{i,j}$:
	\begin{eqnarray}
	\fl &Y_{n0}\mbox{ removed:}\quad \sum_{i,j=0}^{3} c_{i,j} \mu_i^n =0 \quad\forall\, n \quad\Leftarrow\quad \sum_{j=0}^{3} c_{i,j}=0  \quad \mbox{for} \quad i=0,1,2,3  \label{Yn0removed-31-4decoys} \\
	\fl &Y_{n2}\mbox{ removed:}\quad \sum_{i,j=0}^{3} c_{i,j} \mu_i^n \mu_j^2 =0 \quad\forall\, n \quad\Leftarrow\quad \sum_{j=0}^{3} c_{i,j} \mu_j^2 =0 \quad \mbox{for} \quad i=0,1,2,3  \label{Yn2removed-31-4decoys} \\
	\fl &Y_{n3}\mbox{ removed:}\quad \sum_{i,j=0}^{3} c_{i,j} \mu_i^n \mu_j^3 =0 \quad\forall\, n \quad\Leftarrow\quad \sum_{j=0}^{3} c_{i,j} \mu_j^3 =0 \quad \mbox{for} \quad i=0,1,2,3  \label{Yn3removed-31-4decoys} \\
	\fl &Y_{0m}\mbox{ removed:}\quad \sum_{i,j=0}^{3} c_{i,j} \mu_j^m=0 \quad\forall\, m \quad\Leftarrow\quad \sum_{i=0}^{3} c_{i,j}=0 \quad \mbox{for} \quad j=0,1,2,3  \label{Y0mremoved-31-4decoys} \\
	\fl &Y_{1m}\mbox{ removed:}\quad \sum_{i,j=0}^{3} c_{i,j} \mu_i \mu_j^m=0 \quad\forall\, m \quad\Leftarrow\quad \sum_{i=0}^{3} c_{i,j} \mu_i =0  \quad \mbox{for} \quad j=0,1,2,3  \label{Y1mremoved-31-4decoys} \\
	\fl &Y_{2m}\mbox{ removed:}\quad \sum_{i,j=0}^{3} c_{i,j} \mu_i^2 \mu_j^m=0 \quad\forall\, m \quad\Leftarrow\quad \sum_{i=0}^{3} c_{i,j} \mu_i^2 =0  \quad \mbox{for} \quad j=0,1,2,3  \label{Y2mremoved-31-4decoys} \,\,.
	\end{eqnarray}
	The twenty-four conditions (\ref{Yn0removed-31-4decoys} - \ref{Y2mremoved-31-4decoys}) form an over-determined system of equations for the sixteen variables $c_{i,j}$. However, thanks to the symmetries of the problem, a unique solution for $c_{i,j}$ exists and reads (we rescale every coefficient by requiring $c_{0,0}=1$):
	\begin{eqnarray}
		&c_{0,0}=1 \,\,,\nonumber\\
		&c_{0,1}=-\frac{(\mu_0-\mu_2) (\mu_0-\mu_3) [\mu_0 (\mu_2+\mu_3)+\mu_2 \mu_3]}{(\mu_1-\mu_2)(\mu_1-\mu_3) [\mu_1 (\mu_2+\mu_3)+\mu_2 \mu_3]}  \,\,,\nonumber\\
		&c_{0,2}=\frac{(\mu_0-\mu_1) (\mu_0-\mu_3) [\mu_0 (\mu_1+\mu_3)+\mu_1 \mu_3]}{(\mu_1-\mu_2) (\mu_2-\mu_3) [\mu_1 (\mu_2+\mu_3)+\mu_2 \mu_3]} \,\,,\nonumber\\
		&c_{0,3}=\frac{(\mu_0-\mu_1) (\mu_0-\mu_2) [\mu_0 (\mu_1+\mu_2)+\mu_1 \mu_2]}{(\mu_1-\mu_3) (\mu_3-\mu_2) [\mu_1 (\mu_2+\mu_3)+\mu_2 \mu_3]} \,\,,\nonumber\\
		&c_{1,0}=\frac{(\mu_0-\mu_2) (\mu_0-\mu_3)}{(\mu_2-\mu_1) (\mu_1-\mu_3)} \,\,,\nonumber\\
		&c_{1,1}=\frac{(\mu_0-\mu_2)^2 (\mu_0-\mu_3)^2 [\mu_0 (\mu_2+\mu_3)+\mu_2 \mu_3]}{(\mu_1-\mu_2)^2 (\mu_1-\mu_3)^2 [\mu_1 (\mu_2+\mu_3)+\mu_2 \mu_3]} \,\,,\nonumber\\
		&c_{1,2}=-\frac{(\mu_0-\mu_1) (\mu_0-\mu_2) (\mu_0-\mu_3)^2 [\mu_0 (\mu_1+\mu_3)+\mu_1 \mu_3]}{(\mu_1-\mu_2)^2 (\mu_1-\mu_3) (\mu_2-\mu_3) [\mu_1 (\mu_2+\mu_3)+\mu_2 \mu_3]} \,\,,\nonumber\\
		&c_{1,3}=\frac{(\mu_0-\mu_1) (\mu_0-\mu_2)^2 (\mu_0-\mu_3) [\mu_0 (\mu_1+\mu_2)+\mu_1 \mu_2]}{(\mu_1-\mu_2) (\mu_1-\mu_3)^2 (\mu_2-\mu_3) [\mu_1 (\mu_2+\mu_3)+\mu_2 \mu_3]} \,\,,\nonumber\\
		&c_{2,0}=\frac{(\mu_0-\mu_1) (\mu_0-\mu_3)}{(\mu_1-\mu_2) (\mu_2-\mu_3)} \,\,,\nonumber\\
		&c_{2,1}=-\frac{(\mu_0-\mu_1) (\mu_0-\mu_2) (\mu_0-\mu_3)^2 [\mu_0 (\mu_2+\mu_3)+\mu_2 \mu_3]}{(\mu_1-\mu_2)^2 (\mu_1-\mu_3) (\mu_2-\mu_3) [\mu_1 (\mu_2+\mu_3)+\mu_2 \mu_3]}  \,\,,\nonumber\\
		&c_{2,2}= \frac{(\mu_0-\mu_1)^2 (\mu_0-\mu_3)^2 [\mu_0 (\mu_1+\mu_3)+\mu_1 \mu_3]}{(\mu_1-\mu_2)^2 (\mu_2-\mu_3)^2 [\mu_1 (\mu_2+\mu_3)+\mu_2 \mu_3]} \,\,,\nonumber\\
		&c_{2,3}=-\frac{(\mu_0-\mu_1)^2 (\mu_0-\mu_2) (\mu_0-\mu_3) [\mu_0 (\mu_1+\mu_2)+\mu_1 \mu_2]}{(\mu_1-\mu_2) (\mu_1-\mu_3) (\mu_2-\mu_3)^2 [\mu_1 (\mu_2+\mu_3)+\mu_2 \mu_3]} \,\,,\nonumber\\
		&c_{3,0}=\frac{(\mu_0-\mu_1) (\mu_0-\mu_2)}{(\mu_1-\mu_3) (\mu_3-\mu_2)} \,\,,\nonumber\\
		&c_{3,1}=\frac{(\mu_0-\mu_1) (\mu_0-\mu_2)^2 (\mu_0-\mu_3) [\mu_0 (\mu_2+\mu_3)+\mu_2 \mu_3]}{(\mu_1-\mu_2) (\mu_1-\mu_3)^2 (\mu_2-\mu_3) [\mu_1 (\mu_2+\mu_3)+\mu_2 \mu_3]} \,\,,\nonumber\\
		&c_{3,2}=-\frac{(\mu_0-\mu_1)^2 (\mu_0-\mu_2) (\mu_0-\mu_3) [\mu_0 (\mu_1+\mu_3)+\mu_1 \mu_3]}{(\mu_1-\mu_2) (\mu_1-\mu_3) (\mu_2-\mu_3)^2 [\mu_1 (\mu_2+\mu_3)+\mu_2 \mu_3]} \,\,,\nonumber\\
		&c_{3,3}=\frac{(\mu_0-\mu_1)^2 (\mu_0-\mu_2)^2 [\mu_0 (\mu_1+\mu_2)+\mu_1 \mu_2]}{(\mu_1-\mu_3)^2  (\mu_2-\mu_3)^2 [\mu_1 (\mu_2+\mu_3)+\mu_2 \mu_3]}    \label{cij-31-4decoys}\,\,.
	\end{eqnarray}
	By substituting these expressions back into (\ref{G-31-4decoys}) and by making some simplifications, one gets:
	\begin{eqnarray}
	 \fl \sum_{i,j=0}^{3} c_{i,j} \tilde{Q}^{i,j} = \sum_{n=3}^{\infty}\frac{Y_{n1}}{n!} A_{13}(\mu_0,\mu_1,\mu_2,\mu_3,n) + \sum_{\stackrel[m=4]{n=3}{}}^{\infty} \frac{Y_{nm}}{n!m!} A_{13}(\mu_0,\mu_1,\mu_2,\mu_3,n) \cdot C_m   \,\,,\label{combG-31-4decoys} 
	\end{eqnarray}
	where $A_{13}$ and $C_m$ also appear in \ref{bound_on_Y13-4decoys} when bounding $Y_{13}$ and are defined as (\ref{A1m}) and (\ref{Cn}), respectively. Thus, following the same lines of \ref{bound_on_Y13-4decoys}, we conclude that all yields in (\ref{combG-31-4decoys}) are multiplied by a positive factor. A valid upper bound on $Y_{31}$ is then obtained by setting to zero all the other yields in (\ref{combG-31-4decoys}). We obtain:
	\begin{eqnarray}
	 Y_{31}^U = \frac{6}{A_{13}(\mu_0,\mu_1,\mu_2,\mu_3,3)} \left(	\sum_{i,j=0}^{3} c_{i,j} \tilde{Q}^{i,j}\right) \,\,,\label{Y31-upperbound-4decoys}
	\end{eqnarray}
	where $c_{i,j}$ and $A_{13}(\mu_0,\mu_1,\mu_2,\mu_3,3)$ are defined in (\ref{cij-31-4decoys}) and (\ref{A13}), respectively.

\end{document}